\documentclass[showpacs,prl,onecolumn,aps,superscriptaddress,preprintnumbers,nofootinbib]{revtex4}
\usepackage[T1]{fontenc}
\usepackage[latin9]{inputenc}
\usepackage{graphicx,hyperref}

\usepackage{amsmath,amssymb}
\usepackage{epsfig}
\usepackage{graphicx}
\usepackage{amsmath}
\usepackage{amsfonts}
\usepackage{mathrsfs}
\usepackage{pifont}
\usepackage{relsize}
\usepackage{mathtools}
\def\slashchar#1{\setbox0=\hbox{$#1$}     		
   \dimen0=\wd0                                 	
   \setbox1=\hbox{/} \dimen1=\wd1               	
   \ifdim\dimen0>\dimen1                        	
      \rlap{\hbox to \dimen0{\hfil/\hfil}}      	
      #1                                        	
   \else                                        	
      \rlap{\hbox to \dimen1{\hfil$#1$\hfil}}   	
      /                                         	
   \fi}

\renewcommand{\vec}{\boldsymbol}
\newcommand{\beq}{\begin{equation}}
\newcommand{\eeq}{\end{equation}}
\newcommand{\bea}{\begin{eqnarray}}
\newcommand{\eea}{\end{eqnarray}}
\newcommand{\baa}{\begin{array}}
\newcommand{\eaa}{\end{array}}

\def\eq#1{{Eq.~(\ref{#1})}}

\def\fig#1{{Fig.~\ref{#1}}}
\newcommand{\intl}{\int\limits}

\newcommand{\bas}{\bar{\alpha}_S}
\newcommand{\as}{\alpha_S}

\newcommand{\nn}{\nonumber}

\newcommand{\h}{\frac{1}{2}}

\newcommand{\x}{\vec{x}}
\newcommand{\vb}{\vec{b}}

\newcommand{\Lb}{\left(}
\newcommand{\Rb}{\right)}

\newcommand{\pp}{\partial}

\renewcommand{\vec}[1]{\boldsymbol{#1}}

\newcommand{\dY}{\delta \tilde{Y}}

\numberwithin{equation}{section}

\begin{document}
\title{Homotopy approach for scattering amplitude  for running QCD coupling }
\author{Carlos Contreras}
\email{carlos.contreras@usm.cl}
\affiliation{Departamento de F\'isica, Universidad T\'ecnica Federico Santa Mar\'ia,  Avda. Espa\~na 1680, Casilla 110-V, Valpara\'iso, Chile}
\author{Jos\'e Garrido}
\email{jose.garridom@sansano.usm.cl}
\affiliation{Departamento de F\'isica, Universidad T\'ecnica Federico Santa Mar\'ia,  Avda. Espa\~na 1680, Casilla 110-V, Valpara\'iso, Chile}
\author{Eugene Levin}
\email{leving@tauex.tau.ac.il}
\affiliation{Department of Particle Physics, School of Physics and Astronomy,
Raymond and Beverly Sackler
 Faculty of Exact Science, Tel Aviv University, Tel Aviv, 69978, Israel}
\date{\today}


\begin{abstract}
In this paper we proposed the homotopy approach for solving the nonlinear Balitsky-Kovchegov (BK) evolution equation with running QCD coupling. The approach consists of two steps. First, is the analytic solution to the nonlinear evolution equation for the simplified, leading twist  kernel. Second, is the iteration procedure that allow us to calculate corrections analytically or semi-numerically. For the leading twist kernel it is shown that the first iteration leads to $\leq 1\%$ accuracy.  The $\zeta = -\frac{4N_c}{b_0}Y \ln\Lb  \bas (1/Q^2_s(Y))/\bas\Lb r^2\Rb\Rb$  ($r$ is the dipole size, $Q_s$ is the saturation scale) and geometric scaling behaviour of the scattering amplitude are discussed as well as  the dependence on the value of the infrared cutoff.

 \end{abstract}
\maketitle

\vspace{-0.5cm}
\tableofcontents






\section{Introduction}

In this paper we are going to continue our search for the regular iteration procedure  of solving non-linear equations that govern QCD dynamics in the saturation region. Our general method is based on homotopy approach\cite{HE1,HE2} which we developed in our previous papers for the  Balitsky-Kovchegov\cite{BK} equation for the scattering amplitude\cite{CLMNEW}
and for the equation for the cross section of diffraction production\cite{KOLE,CGLM} .

Basically, our approach consists of two stages. In the first stage we find the analytic or almost analytic solution to the non-linear equation with  the simplified BFKL\footnote{BFKL stands for Balitsky, Fadin, Kuraev and Lipatov equation.}\cite{BFKL} kernel which collects the main features of the saturation. In the second stage we develop the numerical computation to estimate the corrections to the first iteration.

In this paper we wish to expand our procedure to the case of running QCD coupling. The good news is that we know how to introduce the running QCD coupling into the non-linear equation\cite{RAL}; and that we know the main difficulties of the searching of the solution related to the violation of the geometric behaviour of the scattering amplitude\cite{RUNAL}.

 \begin{figure}
 	\begin{center}
 	\leavevmode
 		\includegraphics[width=10cm]{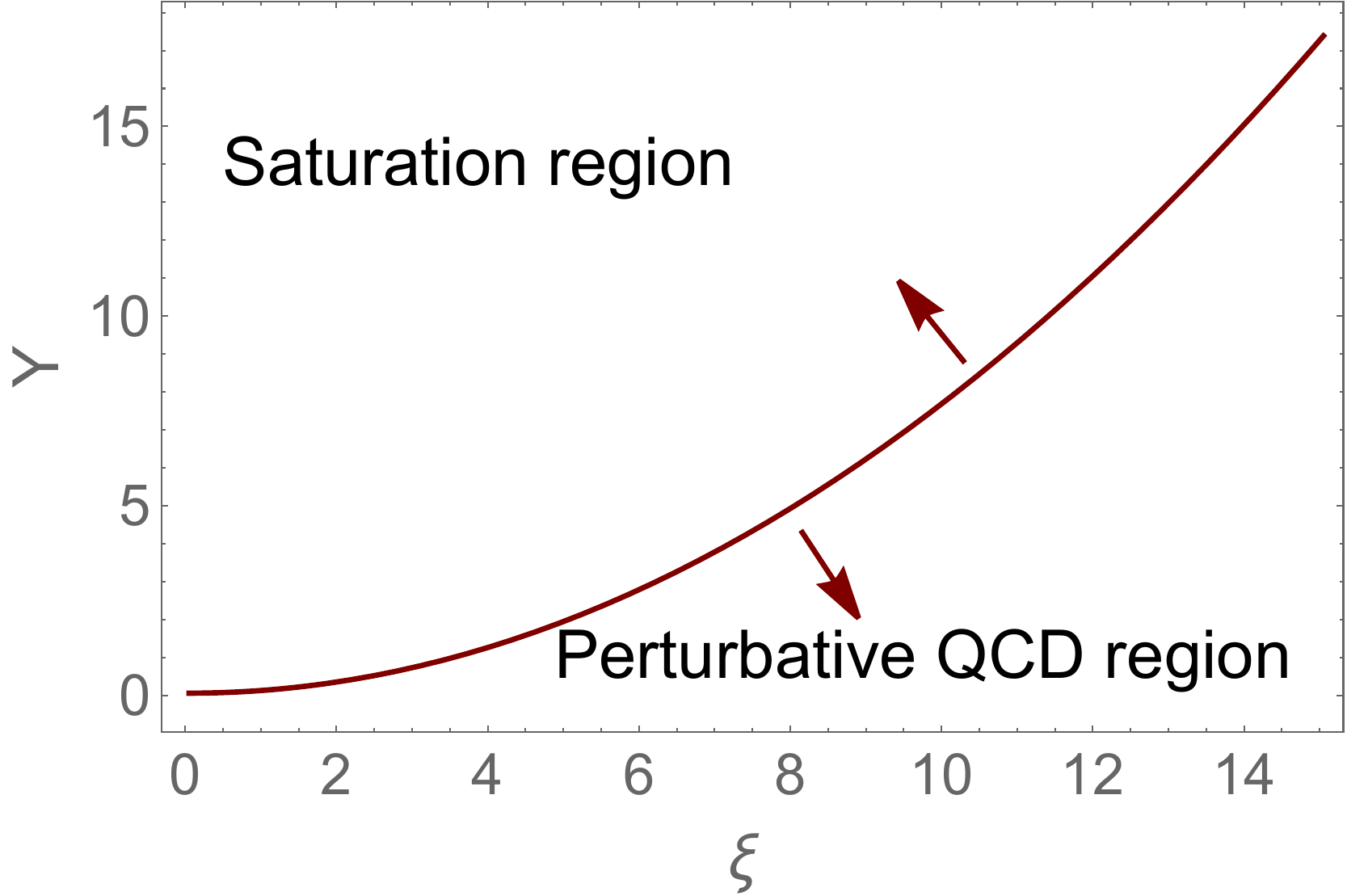}
 	\end{center}
 	\caption{Saturation region of QCD for the elastic scattering amplitude. The critical line: $ \frac{8 N_c}{b}\,\,\frac{\chi\Lb \gamma_{cr} \Rb}{ 1 \,-\,\gamma_{cr}}\,Y\,\equiv\,\,\xi^2_s\,\,=\,\,\xi^2\,$	
is shown in red. The initial condition for scattering with the dilute system of partons (with proton) is given at $\xi_s = 0$. The variable $\xi_s$ is defined as $\xi_s \,=\,\ln\Lb Q^2_s(Y,b)/Q_s^2(Y=Y_0,b)\Rb$.}
 	\label{sat}
 \end{figure}
 
 
 The detailed description of our approach is included in  the next sections. Here we wish 
 to discuss the kinematic region where we are looking for the solution and general assumptions that we make in fixing initial and boundary conditions. 
The nonlinear Balitsky-Kovchegov equation has the following form\cite{BK} for the scattering amplitude of the dipole with size $r$: 
\bea 
\frac{\partial N\Lb r,Y;\vec{b} \Rb}{\partial\,Y}\,
\,\,&
=&\,\,\,\,\int\,\frac{d^2 r_1}{2 \pi}\,K\Lb r; r_1,r_2\Rb \times 
\Big\{N\Lb r_1,Y;\vec{b} \,-\, 
\frac{1}{2}\,\vec{r}_2\Rb \,+\, N\Lb r_2,Y;\vec{b} \,-\, 
\frac{1}{2}\,\vec{r}_1\Rb\,-\, \,\, N\Lb r,Y;\vec{b}\Rb\,\,\nn\\
 &-& \,\,
 N\Lb r_1,Y;\vec{b} - \frac{1}{2}\,\vec{r}_2 \Rb\, N\Lb r_2 ,Y;\vec{b}- \frac{1}{2} \vec{r}_1\Rb \Big\} \label{GA1}
\eea
where $Y = \ln(1/x)$ is the rapidity of the incoming dipole; $N$ is the imaginary part of the scattering amplitude and $b$ is the impact parameter of this scattering process and $\vec{r}_2 = \vec{r} - \vec{r}_1$.  The BFKL kernel
with running QCD coupling  $K\Lb r;r_1,r_2\Rb$ in the Balitsky prescription\cite{RAL} has the following form 
\beq \label{K}
 K_{rc}^{Bal}\Lb r; r_1,r_2\Rb\,\,=\,\,\bas\Lb r^2\Rb \left\{ \frac{r^2}{r^2_1\,r^2_2}\,\,+\,\,\frac{1}{r_1^2}\Lb\frac{\bas\Lb r^2_1\Rb}{\bas\Lb r^2_2\Rb}\,-\,1\Rb
\,+\,\frac{1}{r_2^2}\Lb \frac{\bas\Lb r^2_2\Rb}{\bas\Lb r^2_1\Rb}\,-\,1\Rb\right\}
\eeq
while in the Kovchegov-Weigert prescription\cite{RAL2} it has the following form
\beq \label{KW}
 K_{rc}^{KW}\Lb r; r_1,r_2\Rb\,\,=\,\,\bas\Lb r^2_1\Rb \left\{ \frac{1}{r^2_1 }\,\,-2\,\frac{\bas\Lb r^2_2\Rb}{\bas\Lb R^2_0\Rb}
\,\frac{\vec{r}_1\cdot\vec{r}_2}{r_1^2 r_2^2}\,+\,\frac{1}{r_2^2} \frac{\bas\Lb r^2_2\Rb}{\bas\Lb r^2_1\Rb}\right\}
\eeq
with
\beq \label{Ro}
R_0^2\Lb r; r_1,r_2\Rb\,\,=\, r_1  r_2   \Lb \frac{r_2}{r_1 }\Rb ^{  \frac{r_1^2+r_2^2}{r_1^2-r_2^2} -2 \frac{r_1^2r_2^2}{\vec{r}_1\cdot \vec{r}_2}\,\frac{1}{r_1^2-r_2^2}} 
\eeq

Both prescriptions neglect different contributions. However, numerical studies \cite{KOVCUT} indicate that the Balitsky prescription provides a closer result to the full answer, where no  approximations are made. Therefore, we adopt the Balitsky prescription in our analysis.

In \eq{K} and \eq{KW}
$\as $ is the QCD coupling 
\beq \label{AL}
\as\Lb r^2\Rb\,\,=\,\,\frac{\as\Lb R^2\Rb}{1\,\,+\,\,\frac{\as\Lb R^2\Rb}{4 \pi b_0}\ln\Lb R^2/r^2\Rb}\,\,=\,\,\frac{4 \pi}{b_0 \,\ln\Lb 1/\Lb r^2\,\Lambda^2_{QCD}\Rb\Rb}
\eeq
 and  $\bas = N_c \as/\pi$  with  $b_0 = 11 N_c /3 - 2 N_f/3$ for number of colours $N_c$ and the number of flavours $N_f$\footnote{In our numerical estimates we use $N_c=N_f=3$.}.  
 $R$ is the arbitrary size (so called the renormalization point)  which the physical observables do not depend on. 

 First we need to find the solution to the nonlinear equation in the vicinity of the saturation scale to specify the kinematic regions where we will attempt to solve the nonlinear equations.
 
 In the vicinity of the saturation scale where $r^2 \, \approx r^2_1\, \approx r^2_2\, \approx 1/Q^2_s$,  we can consider  that $\bas\Lb r^2\Rb \,=\,\bas\Lb r^2_1\Rb = \bas \Lb r^2_2\Rb$ . We can see  that  for
 $R = r$\beq \label{GA2}
\as\Lb r^2_i \Rb\,\,=\,\,\frac{\as\Lb r^2\Rb}{1\,\,+\,\,\frac{\as\Lb r^2\Rb}{4 \pi b}\ln\Lb r^2/r^2_i\Rb}\,\,\,\,\xrightarrow{ \ln\Lb r^2/r^2_i\Rb\,\ll\,ln\Lb r^2 \,\Lambda^2_{QCD}\Rb }\,\,\,\,\as\Lb r^2 \Rb
\eeq
For $r^2 \,\varpropto \,1/Q^2_s$  condition  $|\ln\Lb r^2_i\,Q^2_s\Rb|\,\ll\,\ln \Lb Q^2_s/\Lambda^2_{QCD}\Rb$ determines the  kinematic region which we call vicinity of the saturation scale. Using this simplification the kernel of \eq{GA1} takes the form:
\beq \label{KK}
K\Lb r; r_1,r_2\Rb\,\,\,=\,\,\bas\Lb r^2\Rb \,\frac{r^2}{r^2_1\,r^2_2}
\eeq

The second simplification stems from the observation that for the equation for the saturation scale $Q_s$  we do not need to know the precise  form of the  non-linear term \cite{GLR,MUT,MUPE} (see also Ref.\cite{KOLEB}). Therefore, to find $Q_s$ as well as the behaviour of the amplitude in the vicinity of the saturation scale we need to solve the linear BFKL equation, but using the procedure
 which will be suitable for the solution of the non-linear equation with a general non-linear term. It is enough to use the semiclassical approximation for the amplitude $N\Lb r,Y;b \Rb$, which has the form
\beq \label{SCSOL}
N_A\Lb Y, \xi\Rb\,\,= \,\,e^{S\Lb Y, \xi\Rb}\,\,=\,\,e^{\omega\Lb Y,\xi\Rb\,Y\,+\,\Lb 1 - \gamma\Lb Y; \xi\Rb\Rb\,\xi\,+\,S_0} 
\eeq
where $\xi \,=\,\ln\Lb r^2 Q^2_s\Lb Y = Y_0; b \Rb\Rb$. In \eq{SCSOL} we are searching  for functions  $\omega\Lb Y,\xi\Rb$ and $\gamma\Lb Y,\xi\Rb$
which are smooth functions of  both arguments (see Ref.\cite{KOLEB} for details).  Plugging in the linear BFKL equation the solution of \eq{SCSOL} and taking into account that function $(r^2)^f \equiv \exp\Lb f \,\xi\Rb$ is the eigenfunction of the BFKL equation,  viz.:
\bea \label{GA5}
&&\bas\Lb r^2\Rb \int\,\frac{d^2 r_1}{2 \pi}\,K\Lb r; r_1,r_2\Rb\,(r^2_1)^f \,\,=\,\, \bas\Lb r^2\Rb\chi\Lb f \Rb\,(r^2)^f\,\,
\mbox{with}\,\,\,
\chi\Lb f \Rb \,\,=\,\,2 \,\psi(1) - \psi(f) - \psi(1 - f) \nn\\
&&\,\,\mbox{where} \,\, \psi(z) = d \ln \Gamma(z)/d z\, \mbox{ and } \,\Gamma(z) \,\mbox{ is Euler gamma function}
\eea
we obtain that
\beq \label{GA6}
\omega\Lb Y,\xi\Rb\,\,\,=\,\,\bas\Lb \xi\Rb\,\chi\Lb \gamma\Lb Y, \xi\Rb\Rb
\eeq

This solution has a form of wave-package and the critical line is the specific trajectory for this wave-package which coincides with the its front line. In other words, it is the  trajectory on which the phase velocity ($v_{ph}$) for the wave-package is the same as
the group velocity ($v_{gr}$). The equation $v_{gr} \,=\, v_{ph}$ has the folowing form for \eq{GA6}
\beq \label{CRL}
v_{ph}\,\,=\,\,\bas
\Lb r^2\Rb \frac{\chi\Lb \gamma_{cr} \Rb}{ 1 \,-\,\gamma_{cr}}\,\,\,=\,\,\,- \bas\Lb r^2 \Rb \chi'\Lb \gamma_{cr}\Rb\,\,=\,\,v_{gr}
\eeq
with the solution $\gamma_{cr} \,=\,0.37$.
\eq{CRL} can be translated into the following equation for the critical trajectory
\beq \label{GA7}
\frac{d \xi\Lb Y\Rb}{ d Y}\,\,=\,\,v_{ph}\,\,=\,\,\bas\Lb \xi\Rb\,\frac{\chi\Lb \gamma_{cr} \Rb}{ 1 \,-\,\gamma_{cr}}
\eeq
with the solution
\beq \label{GA8}
\frac{8 N_c}{b_0}\,\,\frac{\chi\Lb \gamma_{cr} \Rb}{ 1 \,-\,\gamma_{cr}}\,Y\,\equiv\,\,\xi^2_s\,\,=\,\,\xi^2\,\,-\,\,\xi^2_0\,\,\
\eeq
where $\xi_0\,=\,\ln\Lb  Q^2_s\Lb Y = Y_0; b \Rb/\Lambda^2_{QCD}\Rb$   and   $\xi_s \,=\,\ln\Lb Q^2_s(Y,b)/Q_s^2(Y=Y_0,b)\Rb$. For simplicity we  choose $ Q_s^2(Y=Y_0,b) = \Lambda^2_{QCD}$. It means that we model the DIS with the proton by the scattering of dipole $r$ with the dipole  size is equal to $R = 1/\Lambda_{QCD}$.

\eq{GA8} results in 
the saturation moment $Q_s^2(Y)$ which is equal to
 \beq \label{QS}
 Q^2_s\Lb Y, b\Rb\,\,=\,\,Q^2_s\Lb Y=Y_0, b\Rb \,e^{\sqrt{\frac{8 N_c}{b_0}\,\frac{\chi\Lb \gamma_{cr} \Rb}{ 1 \,-\,\gamma_{cr}}\,Y}}\
 \eeq 
We note that the dependence of $Q_s$ on $Y$ changes from a linear behavior in the fixed coupling case to the form given by \eq{QS} in the running coupling case. This change  slows down the evolution of the scattering amplitude.

The behaviour of the scattering amplitude in the vicinity of the saturation scale has been found in Refs.\cite{IIM,MUT,KOLEB} and it has the form:
\beq \label{SCAQS} 
N\Lb Y, \xi; b\Rb\,\,=\,\,N_0 \Lb r^2\,Q^2_s\Lb Y,b\Rb\Rb^{\bar{\gamma}}~~~~\mbox{with}~~~\bar{\gamma} = 1 - \gamma_{cr}
\eeq
One can see that we have a geometric scaling behaviour for the scattering amplitude in the vicinity of the saturation scale. \eq{SCAQS} gives us the boundary condition for the nonlinear equation, viz.:
\beq 
\label{IBC}
N\Lb Y,\xi = -\xi_s; b\Rb\,=\,N_0;\,\,\,\,\,\,\,\,\,\,\,\frac{ \partial N\Lb Y, \xi = -\xi_s; b\Rb}{\partial \xi}\,\,=\,\,\,\bar{\gamma}\,\,N_0
\eeq

  We wish to note that in the region of perturbative QCD  (see \fig{sat}) the scattering amplitude is the solution to the linear BFKL equation:
  \beq  \label{BFKL}
\frac{\partial N^{\tiny{\rm BFKL}}\Lb r,Y;\vec{b} \Rb}{\partial\,Y}\,
= \int\,\frac{d^2 r_1}{2 \pi}\,K\Lb r; r_1,r_2\Rb 
\Big\{N^{\tiny{\rm BFKL}}\Lb r_1,Y;\vec{b} \,-\, 
\frac{1}{2}\,\vec{r}_2\Rb \,+\, N^{\tiny{\rm BFKL}}\Lb r_2,Y;\vec{b} \,-\, 
\frac{1}{2}\,\vec{r}_1\Rb\,-\, \,\, N^{\tiny{\rm BFKL}}\Lb r,Y;\vec{b}\Rb\Big\}\eeq 

The initial condition for this equation is the Born Approximation  (the exchange of two gluons) for the scattering of the dipole with the  size $r$ with the dipole with the size $R=1/\Lambda_{QCD}$.

 We have to specify the region of $r_{\perp}$ which we are dealing with in the saturation region $ r^2_{\perp}\,Q^2_s\Lb Y,b\Rb\,>\,1$. As has been noted in 
 Refs.\cite{GOST,BEST}  actually for very large $r_{\perp} $ the non-linear corrections are not important and we have to solve linear BFKL equation. This feature can be seen  directly
 from the eigenfunction of this equation.  Indeed, the eigenfunction
 (the scattering amplitude of two dipoles with sizes $r _{\perp}\equiv  x_{10}$ and $R$) has the
 following form \cite{LIP}
\beq \label{EIGENF}
\phi_\gamma\Lb \vec{r}_{\perp} , \vec{R}, \vec{b}\Rb\,\,\,=\,\,\,\Lb \frac{
 r^2\,R^2}{\Lb \vec{b}  + \h(\vec{r}_{\perp} - \vec{R})\Rb^2\,\Lb \vec{b} 
 -  \h(\vec{r}_{\perp} - \vec{R})\Rb^2}\Rb^\gamma  \,\xrightarrow{ b \gg\,r, R} \Lb\frac{R^2 r^2}{b^4}\Rb^\gamma~~\mbox{with}\,\,0 \,<\,Re \,\gamma\,<\,1
 \eeq 

One can see that for $r_{\perp}=x_{10} \,>\,min[R, b]$,  $\phi_\gamma$ starts to be smaller and the non-linear term in the BK equation 
could be neglected.
The typical process, that we bear in mind, is the deep inelastic scattering(DIS) with a nucleus at $Q^2\, \geq \,1 GeV^2$ and at small values of $x$. In our previous papers\cite{CLMNEW,CGLM} we have commented on impact parameter behaviour for the scattering with nucleus.

In our paper we can distinguish three parts. The first one is the brief discussion  of the solution to the nonlinear evolution equation for the model leading twist BFKL kernel (section II and III).  In these sections we discuss the main results of Ref.\cite{RUNAL} and present the detailed analysis of self-similar solution to this equation. These sections contain  the main theoretical basis for the first stage of our approach: finding  an analytic or almost analytic solution to the non-linear equation. 
  In section IV-VI we proposed the homotopy approach for the leading twist BFKL kernel. We demonstrated that the first iteration can be done based on the results of  section III and estimates the next corrections. It turns out that the second iteration gives the solution within accuracy of several percents. Sections VIII and IX present our homotopy approach to the nonlinear equation with the general BFKL kernel. We observe that we need to make two iterations to reach a several percents accuracy.  We show that we can suggest a pure analytic solution at the first iteration. In the conclusion we summarize our results and outline the problems that we faced developing our approach.


\section{The main features of the nonlinear equation for the leading twist BFKL kernel} 

In this section we briefly review the main result of Ref.\cite{RUNAL}  on the solution to the nonlinear equation with the simplified BFKL kernel.
 Following Ref. \cite{LETU} we simplify the kernel by taking into account only log contributions. In other words, we would like to consider only leading twist contribution   to the BFKL kernel, which contains all twists.
Actually we have two types of the logarithmic contributions:  $\ln\Lb r^2 \Lambda^2_{QCD}\Rb$ for $r^2 \,\ll\, 1/Q^2_s$ and $  \ln\Lb r^2\,Q^2_s\Rb$ for   $r^2 \,>\,1/Q^2_s$. In the saturation region we are dealing with the second kind of logs (see Refs.\cite{CLMNEW,CGLM}).  They come
 from the decay of the large size dipole into one small size dipole  and one large size dipole.  However, the size of the small dipole is still larger than $1/Q_s$.  It turns out that
$\bas$
 depends on the size of produced dipole, if this size is the smallest one. It follows directly from \eq{K} in the kinematic regions: 
$ r \approx r_2 \,\gg\,r_1\,\,\gg\,1/Q_s$  and $ r \approx r_1 \,\gg\,r_2\,\,\gg\,1/Q_s$
 (see
Ref. \cite{MURC} for additional arguments). This observation can be translated in the following form of the kernel
\beq \label{K2}
 \int d^2 r' \,K\Lb r, r'\Rb\,\,\rightarrow\,\pi\, \int^{r^2}_{1/Q^2_s(Y,b)} \frac{\bas\Lb 
r^2_1\Rb d r^2_1}{r^2_1}\,\,+\,\,
\pi\, \int^{r^2}_{1/Q^2_s(Y,b)} \frac{ \bas\Lb r^2_2\Rb d r^2_2}{r^2_2}
\eeq

One can see that this kernel leads to the $\Lb  \int^{r^2}_{1/Q^2_s(Y,b)} \frac{\bas\Lb r^2_1\Rb d r^2_1}{r^2_1}\Rb^n$-contributions. Introducing a new function
\beq \label{TN}
 \tilde{N} \Lb r,Y;b\Rb\,\,=\,\,\intl^{r^2}_{1/Q^2_s} d r'^2\,\frac{\bas\Lb r'^2\Rb}{r'^2} \,N\Lb r',Y;b\Rb
\eeq
one obtain the following equation
\beq \label{SK2}
\frac{\partial N\Lb r, Y; b\Rb}{\partial Y}\,\,=\,\, \tilde{N} \Lb r,Y;b\Rb\,\Big( 1\,\,-\,\,N\Lb r, Y; b\Rb \Big)
\eeq
Introducing a new variable
\beq \label{L}
l\,\,=\,\,\int^{r^2}\,d r'^2\frac{\bas\Lb r'^2\Rb}{r'^2}\,\,=\,\,-\frac{4 N_c}{b_0}\,\ln\Big(4N_c/ \Lb b_0\,\bas\Lb r^2 \Rb\Rb\Big)\,\,=\,\,-\frac{4 N_c}{b_0}\,\ln \Lb \bar{\xi}\Rb
\eeq
with $\bar{\xi}\,=\,-\ln\Lb r^2\,\Lambda^2_{QCD}\Rb\,\equiv\,- \xi$
and new function $\Omega\Lb r, Y; b \Rb$
\beq \label{PHI}
N\Lb r, Y; b\Rb\,\,=\,\,1\,\,\,-\,\,e^{- \Omega\Lb r,Y;b\Rb}
\eeq
we obtain the following equation
\beq \label{SK3}
\frac{ \partial^2 \Omega\Lb r, Y; b \Rb}{\partial Y \partial l}\,\,=\,\,1\,\,\,-\,\,\,e^{ - \Omega\Lb r,Y;b\Rb}
\eeq

One can see that \eq{SK3} has the same form as the nonlinear equation with the fixed QCD coupling\cite{CLMNEW,CGLM} but $\xi$ is replaced by $l$.

Below we will often use the variable $l - l_s$ which is equal to
\beq \label{XIL}
l \,-\,l_s\,\,=\,\,\intl^{r^2}_{1/Q^2_s} d r'^2\frac{\bas\Lb r'^2\Rb}{r'^2}\,\,=\,\,-\frac{4 \,N_c}{b_0} \ln\Lb \frac{ - \xi}{\xi_s}\Rb
\eeq


\section{Solutions to the simplified equation}
\begin{boldmath}
\subsection{ Solution for $\Omega\, \gg\, 1$ }
\end{boldmath}

Searching for the solution to \eq{SK3} we start  with  finding the asymptotic behaviour of $\Omega$ at large values of $Y$ and $l$. In this kinematic region we expect that 
 that $\Omega$ will be large since the scattering amplitude approaches to unity.
  Hence, in this region \eq{SK3}  degenerates to a very simple equation
\beq \label{PL1}
\frac{\partial^2 \Omega\Lb Y,l ;b\Rb}{\partial Y \partial l }\,\,=\,\,1
\eeq
with obvious solution:
\beq \label{PL2}
 \tilde\Omega_{\infty}\Lb Y,l ;b\Rb\,\,=\,\,Y\,l  \,+\,F\Lb Y\Rb\,+\,G\Lb l  \Rb
\eeq
 Functions $F$ and $G$ should be found from the initial conditions of \eq{IBC} and they take the form 
\beq \label{PL5}
\tilde{\Omega}_{\infty}\Lb Y, l ;b\Rb\,\,=\,\,Y\,\Lb l  - l_s\Rb \,-\,\bar{\gamma}\,\Omega_0\, \Lb e^{-\frac{b_0}{4 N_c}  l} \,-\,e^{-\frac{b_0}{4 N_c} l_s}\Rb\,+\,\frac{1}{4\,\kappa} \Lb e^{-\frac{b_0}{2 N_c} l}\,-\,e^{-\frac{b_0}{2 N_c}l_s}\Rb\,\,+\,\,\Omega_0
\eeq
where $l_s\,\,=\,\,-\,\frac{4\, N_c}{b_0} \ln  \xi_s$ and $\kappa = \chi\Lb \gamma_{cr}\Rb/(1 - \gamma_{cr})$.

 \eq{PL5} at $\bar{\xi }\,\to\, \xi_s$ leads to 
\beq \label{PL7}
\tilde{\Omega}_{\infty}\Lb Y, l;  b\Rb\,\,\xrightarrow{ \bar{\xi} - \xi_s \,\ll\,\xi_s}\,\,\Omega_0 \,+\,\bar{\gamma} \Omega_0 \Lb\xi +\xi_s\Rb\,\,=\,\,\Omega_0\,\,+\,\,\,\bar{\gamma}\,\Omega_0\,z
\eeq
showing the geometric scaling behaviour.  Hence, we can hope that the solution will show the geometric scaling behaviour in the vicinity of the saturation scale.
Note, that \eq{PL5} does not depend on $l - l_s$ except the first term.

\subsection{Traveling wave solution }
\eq{SK3} has general traveling wave solution  (see Ref.\cite{MATH} formula {\bf 3.5.3})
which can be found noticing that  $\Omega\Lb Y, l; b\Rb \,\equiv\,\phi\Lb \eta \equiv  a \,Y + b \,l;b\Rb$ reduced the equation to 
\beq \label{TW1}
a\,b \frac{d^2 \phi\Lb \eta; b \Rb}{d \eta^2}\,\,=\,\,1 - e^{- \phi\Lb \eta; b \Rb}
\eeq
The general solution of \eq{TW1} has the form
\beq \label{TW2}
\int^\phi_{\phi_0}\frac{d \phi'}{\sqrt{c \,+\,\frac{1}{2 \,a\,b}\Big( \phi'  - 1 + e^{-\phi'}\Big)}}\,\,=\,\, \eta\,\,=\,\, a\,Y \,+\,b \,(l-l_s)
\eeq
where $c, \phi_0, a$ and $b$ are arbitrary constants that  should be found from the initial and boundary conditions.

The initial conditions of \eq{IBC} can be written in terms of $Y$ and $l$ variables as
\beq \label{TWIC}
\phi\Lb \eta 
\,=\,a Y ; b \Rb \,\,=\,\,\,\phi_0\,;\,\,\,\,\,\,\,\,\,\,\,\,\phi'_\eta\Lb  \eta \,=\,a Y + b(l-l_s); b \Rb\Bigg{|}_{l = ls}\,\,=\,\,-\h\,\phi_0 \,\xi_s
\eeq
It should be mentioned that the variable $\eta$ is not the scaling variable $z \,=\,\ln\Lb \tau\Rb \,=\,\xi_s - \bar{\xi}$ with $\xi_s = \sqrt{\frac{8N_c\kappa}{b_0} \,Y}$.  One can see that we cannot satisfy the initial conditions  of \eq{TWIC}. 
Indeed, even to satisfy the first of \eq{TWIC} we need to choose $\eta = 0$ on the critical line. As you see we cannot do this with $a$ and $b$ being constants. The second equation depends on $Y$, but not on $\eta$, making impossible to satisfy this condition in the framework of traveling wave solution.

If we try to find a solution  which depends on $z$ ($\Omega\Lb Y; r^2;b\Rb = \Omega\Lb z; b\Rb$) we obtain the following equation (using $\kappa\,=\,4$ and the variable $
 \tilde{z}\,=\,\sqrt{\frac{16 N_c}{b_0}}\,z$)
\beq \label{TW3}
 \sqrt{\frac{16 N_c}{b_0}}\frac{ \tilde{z}}{ \sqrt{2\,Y}}\,\frac{d^2 \Omega\Lb \tilde{z}; b \Rb }{d\, \tilde{z}^2}\,\,+\,\,\frac{d^2 \Omega\Lb \tilde{z}; b \Rb}{d\, \tilde{z}^2}\,\,=\,\,1 \,
\,-\,\,e^{ - \Omega\Lb \tilde{z}; b \Rb}
\eeq
Therefore, only in the vicinity of the critical line where $  \sqrt{\frac{16 N_c}{b_0}}\, \tilde{z}\,\ll\,\sqrt{2\,Y}$ we can expect the geometric   scaling behaviour of the scattering amplitude. It should be stressed that at large value of $Y$ the region where we have the geometric scaling behaviour becomes rather large. Neglecting the first term in \eq{TW3} we obtain the equation in the same form as for frozen $\as$.  It is easy to find the solution to this equation that satisfies the initial condition of \eq{IBC}.  Actually,  the condition  $  \sqrt{\frac{16 N_c}{b_0}}\, \tilde{z}\,\ll\,\sqrt{2\,Y}$ can be rewritten as $\as\Lb Q^2_s\Rb\,\ln\Lb r^2Q^2_s\Rb \,\ll\,1$ and it  shows the region in which we can consider the running QCD coupling as being frozen at $r^2=1/Q^2_s$.

\subsection{Self-similar solution}
Generally speaking (see Ref.\cite{MATH} formulae {\bf 3.4.1.1} and {\bf 3.5.2})
\eq{SK3} has a self similar (functional separable) solution $\Omega\Lb Y, l;b\Rb \,=\,\Omega\Lb \zeta ; b \Rb$ with 
\beq \label{SS1}
\zeta\,\,=\,\,Y\,l
\eeq
For function $\Omega\Lb \zeta ; b \Rb$ we can reduce \eq{SK3} to the ordinary differential equation
\beq \label{SS2}
 \zeta\,\,\frac{d^2 \Omega\Lb \zeta ; b \Rb}{d \zeta^2}\,\,+\,\,\frac{d\,\Omega\Lb \zeta ; b \Rb}{d \zeta}\,\,=\,\,1\,\,-\,\,e^{- \Omega\Lb \zeta ; b \Rb}
\eeq

The initial condition of \eq{IBC} can be rewritten in the form
\bea \label{SSIC}
\Omega\Lb \zeta = Y\,l_s ; b \Rb\,\,&=&\,\,\Omega_0;\,\,\,\,\,\,\,\,\,\,\,\\
\frac{d \Omega\Lb \zeta= Y\,l_s ; b \Rb}{ d \zeta}\,\,&=&\,\,2\kappa\,\bar\gamma\,\Omega_0/\xi_s \,\,=\,\,
2\kappa\,\bar\gamma\,\Omega_0\Big{/} \sqrt{\frac{8 N_c\kappa}{b_0} \,Y}\nn
\eea

Generally speaking we cannot satisfy  \eq{SSIC} using the solution of \eq{SS2} since these conditions depend  not only on $\zeta$ but on extra variable $Y$.
However, at large value of $Y$ one can see that  \eq{SSIC} degenerates to
\beq \label{SSIC1}
\Omega\Lb \zeta = 0 ; b \Rb\,\,=\,\,\Omega_0;\,\,\,\,\,\,\,\,\,\,\,\frac{d \Omega\Lb \zeta=0 ; b \Rb}{ d \zeta}\,\,=\,\,0;
\eeq

Solution to \eq{SS2} we can find starting from $\Omega \gg 1$. \eq{SS2} takes the form:

\beq \label{SS3}
 \zeta\,\,\frac{d^2 \Omega^{(0)}\Lb \zeta ; b \Rb}{d \zeta^2}\,\,+\,\,\frac{d  \Omega^{(0)}\Lb \zeta ; b \Rb}{d \zeta}\,\,=\,\,1\eeq 
with the solution:
\beq \label{SS4}
\frac{ d \Omega^{(0)}
\Lb\zeta ;b \Rb}{d \zeta} \,=\, 1 \,-\,\frac{c_1}{\zeta} ;~~~~~~\Omega^{(0)}\Lb\zeta ;b \Rb\,=\,
\zeta \,+c_2 - c_1\ln \zeta
\eeq
Constants $c_1$ and $c_2$ have to be found from the boundary conditions of \eq{SSIC1}. One can see that $c_1$  is equal to zero and $c_2 = \Omega_0  - Y \,l_s$ if $\Omega_0 \,\ll\,1$. Note that \eq{PL1} has as a solution the arbitrary function of Y. Hence that first iteration gives the solution which is the function of\footnote{Starting from this equation we will use notation $\zeta$ for expression of \eq{ZNEW}.}
\beq \label{ZNEW}
\zeta = Y \,\Lb l - l_s\Rb
\eeq
We assume  the iterations of \eq{SS2} lead to function which depends on these $\zeta$. Actually this feature follows from section II if we assume that 
\beq \label{TN1}
 \tilde{N} \Lb r,Y;b\Rb\,\,=\,\,\intl^{r^2}_{1/Q^2_s} d r'^2\,\frac{\bas\Lb r'^2\Rb}{r'^2} \,N\Lb r',Y;b\Rb\,\,=\,\,\intl^{l - l_s}d l' N\Lb r',Y;b\Rb\eeq
 Note that for $N \to 1$ \eq{TN1} is valid. Generally \eq{TN1} means that we are looking for the solution $\Omega\Lb Y, l-l_s; b \Rb$.

 Bearing \eq{TN1} in mind we can rewrite \eq{SK3} in the form
\beq \label{SK31}
\frac{ \partial^2 \Omega\Lb Y,l-l_s; b \Rb}{\partial Y \partial (l - l_s)}\,\,=\,\,1\,\,\,-\,\,\,e^{ - \Omega\Lb Y,l-l_s;b\Rb}
\eeq
\eq{SK31} can be rewritten in the form of \eq{SS2} for $\zeta = Y\,(l - l_s)$.

For  the second iteration we have $\Omega\,=\,\Omega^{(0)} \,\,+\,\,\Omega^{(1)}$ with the following equation for $\Omega^{(1)}$:
\beq \label{SS5}
 \zeta\,\,\frac{d^2 \Omega^{(1)}\Lb \zeta ; b \Rb}{d \zeta^2}\,\,+\,\,\frac{d  \Omega^{(1)}\Lb \zeta ; b \Rb}{d \zeta}\,\,=\,\,-\,\,e^{- \Omega^{(0)}\Lb \zeta ; b \Rb}
\eeq

Searching $\frac{d  \Omega^{(1)}\Lb \zeta ; b \Rb}{d \zeta}$ in the form:
\beq \label{SS6}
\frac{d  \Omega^{(1)}\Lb \zeta ; b \Rb}{d \zeta}=\frac{1}{\zeta} C\Lb \zeta\Rb
\eeq
we have
\beq \label{SS7}
\frac{d  \Omega^{(1)}\Lb \zeta ; b \Rb}{d \zeta}= - \frac{1}{\zeta}\int^\zeta_0 d \zeta' \,e^{- \Omega^{(0)}\Lb \zeta' ; b \Rb}\,=\,-\,\frac{1}{\zeta}e^{- \Omega_0}\Lb 1\,+\, \sinh (\zeta )\,-\,\cosh (\zeta )\Rb\eeq

From \eq{SS7} we have for $ \Omega^{(1)}\Lb \zeta ; b \Rb$:
\beq \label{SS8}
\Omega^{(1)}\Lb \zeta ; b \Rb= -\int^\zeta_0 d \zeta'\frac{d  \Omega^{(1)}\Lb \zeta' ; b \Rb}{d \zeta'}\,\,+\,\,c_3= - e^{- \Omega^{(0)}}\underbrace{\Lb -\text{Chi}(\zeta )+\log (\zeta )+\text{Shi}(\zeta )+\gamma\Rb}_{ \zeta - \frac{1}{4} \zeta^2 + \mathcal{O}\Lb\zeta^3\Rb} \,\,+\,\,c_3
\eeq
where $c_3$ is a constant which we need to find from the initial and boundary conditions.  In \eq{SS8}  Chi$\Lb \zeta\Rb$ and Shi$\Lb\zeta\Rb$ are  the hyperbolic sine  and cosine integrals (see Ref.\cite{RY} formula {\bf 8.22}) and $\gamma$ is the Euler's constant (see Ref.\cite{RY} formula {\bf 9.72}).  Using the expansion at small $\zeta$ we can see that $c_3=0$.

 \begin{figure}
 	\begin{center}
 	\leavevmode
 		\includegraphics[width=11cm]{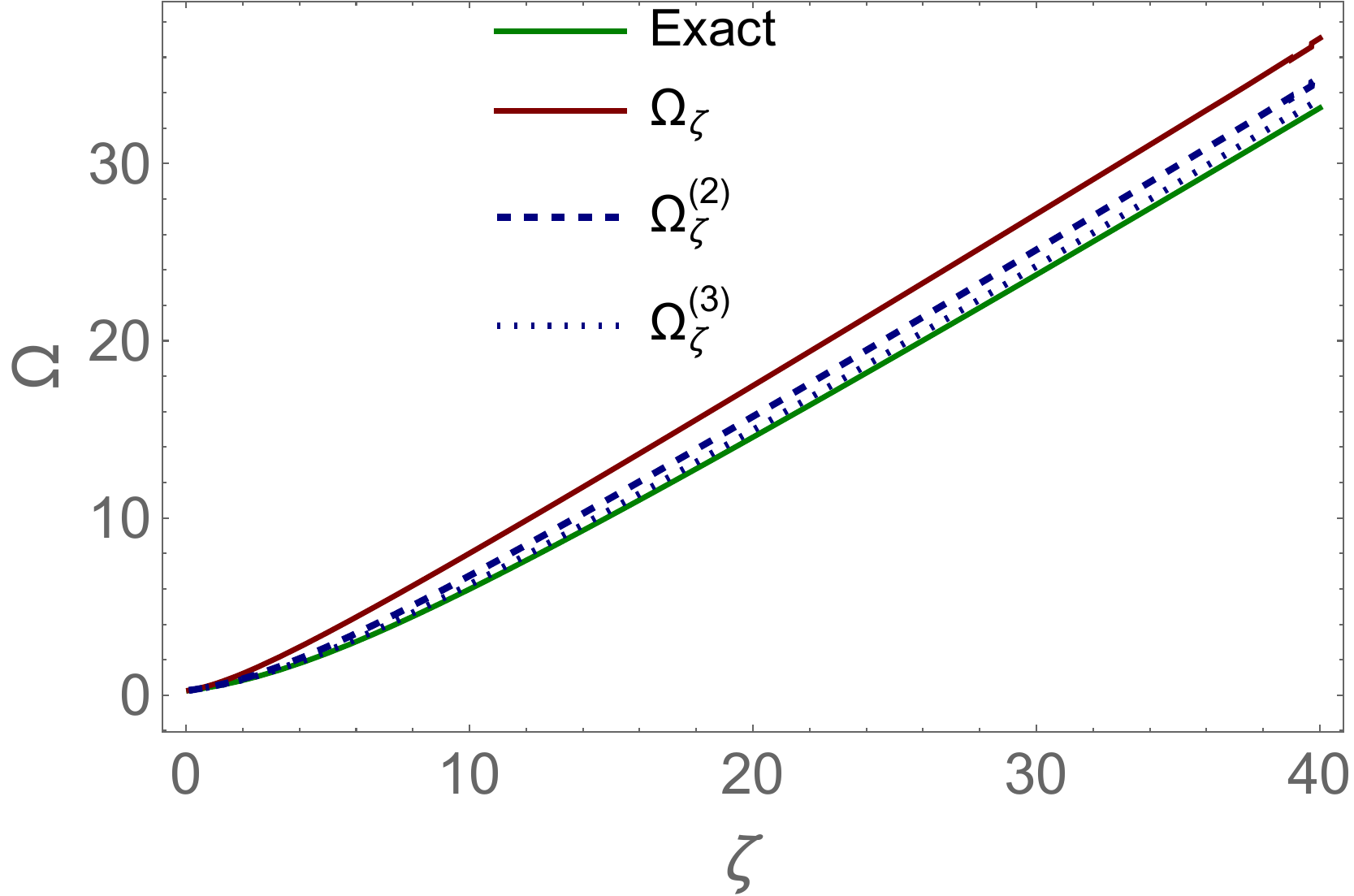}
 	\end{center}
 	\caption{$\Omega\Lb \zeta\Rb$ versus $\zeta$. The exact solution to \eq{SS2} with $\Omega_0=0.25$. $\Omega_\zeta = \Omega^{(0)} +\Omega^{(1)}$ from \eq{SS4} and \eq{SS5}.  $\Omega^{(2)}_\zeta \,=\,\Omega_\zeta + \Omega^{(2)}$, while  $\Omega^{(3)}_\zeta \,=\,\Omega^{(2)}_\zeta + \Omega^{(3)}$. $\Omega^{(2)}$ and $\Omega^{(3)}$ are given by \eq{NS1} and \eq{NS2}.	}
	\label{omegazeta}
 \end{figure}
 
\subsection{Numerical solution}
~
In \fig{omegazeta} we plot the exact solution to \eq{SS2} with the initial conditions of \eq{SSIC1} with
$\Omega_0 =0.25$ and $\Omega_\zeta = \Omega^{(0)} +\Omega^{(1)}$ taking from \eq{SS4} and \eq{SS5}. One sees that our analytical approximation is not quite good. We need to find the next iteration of \eq{SS5}: $ \Omega^{(2)}$. The equation for $ \Omega^{(2)}$ takes the form:
\beq \label{NS1}
 \frac{d }{d\,\zeta} \Lb \zeta \frac{d}{d \zeta} \Omega^{(2)}\Lb \zeta ; b \Rb\Rb\,\,=\,\,\,e^{- \Omega^{(0)}\Lb \zeta ; b \Rb}\Lb 1 - e^{ - \Omega^{(1)}\Lb \zeta ; b \Rb}\Rb
\eeq 
with the initial conditions: $\Omega^{(0)}\Lb \zeta=0 ; b \Rb =0$ and $\frac{d \Omega^{(0)}\Lb \zeta; b \Rb}{d \zeta}|_{\zeta=0} =0$. Note, that we keep in the exponent since from \fig{omegazeta} we see that $\Omega^{(1)} > 1$. The calculated $\Omega^{(2)}$ is shown in  \fig{omegazeta}.
One can see that we need to make more iteration.
 The general   expression for $i+1$ iteration is
 \beq \label{NS2}
 \frac{d }{d\,\zeta} \Lb \zeta \frac{d}{d \zeta} \Omega^{(i+1)}\Lb \zeta ; b \Rb\Rb\,\,=\,\,\,\exp\Lb- \sum^{i-1}_{l=0}\Omega^{(l)}\Lb \zeta ; b \Rb\Rb\Lb 1 - e^{ - \Omega^{(i)}\Lb \zeta ; b \Rb}\Rb
\eeq  

The solution to \eq{NS2} with zero initial conditions has the form:
 \beq \label{NS3}
 \Omega^{(i+1)}\Lb \zeta ; b \Rb\,\,=\,\,\intl^{\zeta}_0 d t \ln \Lb \frac{\zeta}{t}\Rb\, 
 \exp\Lb- \sum_{l=0}^{i-1}\Omega^{(l)}\Lb t ; b \Rb\Rb\,\Lb 1 - e^{ - \Omega^{(i)}\Lb t ; b \Rb} \Rb
 \eeq
 From \fig{omegazeta} one can see that the calculation of $\Omega^{(3)}$ provides a good description of the numerical solution for $\Omega_0 = 0.25$.


\section{Homotopy approach: first iteration. }
 
 The homotopy method we can use for the general equation:
 \beq \label{HOM1}
\mathscr{L}[u] +  \mathscr{N_{L}}[u]=0
\eeq 
 where the linear part $\mathscr{L}[u] $ is a differential  or integral-differential operator, but non-linear part $
 \mathscr{N_{L}}[u] $ has an arbitrary form. As a solution, we introduce  the following  equation for the homotopy function $ {\mathscr H}\Lb p,u\Rb$:
 \beq \label{HOM2}
 {\mathscr H}\Lb p,u\Rb\,\,=\,\,\mathscr{L}[u_p] \,+ \,  p\, \mathscr{N_{L}}[u_p] \,\,=\,\,0
 \eeq
 
 Solving \eq{HOM2} we reconstruct the function
 \beq \label{HOM3}
 u_p\Lb Y,  \x_{10},  \vb\Rb\,\,=\,\, u_0\Lb Y,  \x_{10},  \vb\Rb\,\,+\,\,p\, u_1\Lb Y,  \x_{10},  \vb\Rb \,+\,p^2\, u_2\Lb Y,  \x_{10},  \vb\Rb \,\,+\,\,\dots
 \eeq
 with $\mathscr{L}[u_0] = 0$. \eq{HOM3}  gives  the solution of the non-linear equation at $ p = 1$.  The hope is that several  terms in series of \eq{HOM3} will give a good  approximation in the solution of the non-linear  equation. 
  
  As we have mentioned in the introduction the main idea of the homotopy approach is to find the first iteration or  in other words, to introduce $\mathscr{L}[u_p]  $ as  the solution to the equation
  \beq \label{HOM4}
   \mathscr{L}[u_p]  \,\,=\,\,0  
  \eeq
  which will be analytical or almost analytical and which will absorbe the main features of \eq{HOM3}.  Our choice of $\mathscr{L}[u_p]$ is the following:
  \beq \label{HOM5}
 \mathscr{L}[\Omega] \,\,=\,\,\frac{ \partial^2 \Omega\Lb r, Y; b \Rb}{\partial Y \partial l}\,\,-\,\,1\,\,\,+\,\,\,e^{ - \Omega\Lb r,Y;b\Rb}
\eeq

The selfsimilar solution to \eq{HOM5}, which we have discussed in section III, 
  is not perfect since the boundary condition for \eq{HOM5} for this solution  has the form of \eq{SSIC1} but not \eq{IBC}. Therefore, we have to search for a different solution to \eq{HOM5}. Let us try to find it in the form:
  \beq \label{HOM6}
  \Omega\Lb \zeta, l - l_s;b\Rb\,\,=\,\,\underbrace{ \Omega^{(0)}\Lb \zeta, b\Rb \,\,+\,\,\Omega^{(1)}\Lb \zeta, b\Rb}_{\Omega_\zeta\Lb \zeta\Rb} \,\,+\,\,\Omega'\Lb \zeta, l - l_s;b\Rb
  \eeq
  assuming $  \Omega'\Lb \zeta, l - l_s; b\Rb\,\,\ll\,1$. Plugging \eq{HOM6} into \eq{HOM5} we reduce \eq{HOM5} to the following equation for $\Omega'\Lb \zeta, l-l_s;b\Rb$:
  \beq \label{HOM7}
  \dfrac{\pp^2\,\Omega'\Lb \zeta, l - l_s; b\Rb}{\pp Y\,\pp (l - l_s)}\,\,=\,\, e^{ - \Omega_\zeta\Lb \zeta\Rb}\, \Omega'\Lb \zeta, l - l_s; b\Rb 
  \eeq

 As we have mentioned our goal to find $\Omega'\Lb \zeta, l-l_s; b\Rb$ which is able to generate the initial condition of \eq{IBC}, viz.: $\frac{ \partial N\Lb Y, \xi =-\xi_s; b\Rb}{\partial \xi}\,\,=\,\,\,\bar{\gamma}\,\,N_0$.  Bearing this in mind we first will try to solve \eq{HOM7} for $\zeta \to 0$.
  $$
   \dfrac{\pp^2\,\Omega'\Lb \zeta, l -l_s; b\Rb}{\pp Y\,\pp (l - l_s)}\,\,=\,\, e^{- \Omega_0}\Omega'\Lb \zeta, l- l_s; b\Rb  $$
  Introducing $\dY = \exp\Lb - \Omega_0\Rb \,Y $ we reduce this equation to
  \beq \label{HOM71}
  \dfrac{\pp^2\,\Omega'\Lb \zeta, l - l_s; b\Rb}{\pp \dY\,\pp (l -  l_s)}\,\,=\,\, \Omega'\Lb \zeta, l -l_s; b\Rb 
  \eeq
  with the solution 
  
   \beq \label{HOM72}
  \Omega'\Lb \zeta, l- l_s ; b\Rb = \intl^{\epsilon + i \infty}_{\epsilon - i \infty} \frac{d \nu}{2 \pi\,i} e^{\frac{1}{\nu}\Lb l - l_s\Rb + \nu \,\dY} \phi\Lb \nu\Rb
  \eeq
  Calculating  $\frac{ \partial \Omega'\Lb Y, \xi = -\xi_s; b\Rb}{\partial \xi} $ we obtain
  \beq \label{HOM73}
  \frac{ \partial \Omega'\Lb Y, \xi = -\xi_s; b\Rb}{\partial \xi}\,\,=\,\,\dfrac{4\,N_c}{b_0}\frac{1}{\xi_s}  
  \intl^{\epsilon + i \infty}_{\epsilon - i \infty} \frac{d \nu}{2 \pi\,i} e^{ \nu \,\dY} \frac{\phi\Lb \nu\Rb}{\nu} 
  \eeq
  
    For $ \frac{\phi\Lb \nu\Rb}{\nu}   = \tilde\kappa\nu^{-m/2}\Theta\Lb \nu\Rb$ we obtain:
   \beq \label{HOM74}
  \frac{ \partial \Omega'\Lb Y, \xi =-\xi_s; b\Rb}{\partial \xi}\,\,=\,\,\frac{1}{2\pi}\frac{4N_c}{b_0\xi_s}\tilde\kappa
(-i e^{-\Omega_0}Y)^{\frac{1}{2} (\Re(m)-2)} \Gamma \left(1-\frac{\Re(m)}{2}\right)
  \xrightarrow{\Re(m)\, \to\, 3}\,\frac{4N_c e^{-\Omega_0/2}\tilde\kappa\sqrt{Y}}{b_0\sqrt{\pi}\xi_s} =\bar\gamma\Omega_0 
  \eeq
  The value of $\Omega'\Lb \zeta \to 0, l=l_s; b\Rb  \,\,\propto\,\,1/\sqrt{Y} \ll 1$.

For a general solution let us evaluate  $ \frac{ \partial \Omega'\Lb \zeta, \xi \to -\xi_s; b\Rb}{\partial \zeta}$ and   $ \frac{ \partial^2 \Omega'\Lb \zeta, \xi \to -\xi_s; b\Rb}{\partial \zeta^2 }$  using \eq{HOM72} which we rewrite as
   \beq \label{HOM75}
  \Omega'\Lb \zeta, l-l_s; b\Rb = \intl^{\epsilon + i \infty}_{\epsilon - i \infty} \frac{d \nu}{2 \pi\,i} e^{\frac{1}{\nu}\Lb l - l_s\Rb + \nu \frac{\tilde{\zeta}}{ l - l_s} } \phi\Lb \nu\Rb\,\,=\,\,\underbrace{\Lb \bar{\gamma}\,\Omega_0\,e^{\Omega_0/2}\frac{\sqrt{ \frac{8 \,N_c}{b_0}\frac{\chi\Lb \gamma_{cr}\Rb}{1 - \gamma_{cr}}}}{\sqrt{\pi}\,\frac{4 \,N_c}{b_0} }\Rb}_{\rm Const} \,\sqrt{l - l_s} K_{-\h} \Lb 2 \sqrt{\tilde{\zeta}}\Rb/( \tilde{\zeta})^{1/4}
  \eeq
  with $\tilde{\zeta} \,=\, \exp\Lb - \Omega_0\Rb\,\zeta \,=\, \exp\Lb - \Omega_0\Rb \,Y\,\Lb l - l_s\Rb$.
  
    The value of $\tilde\kappa$ ($ Const$) in \eq{HOM75} is chosen using \eq{IBC} and definitions of \eq{GA8} and \eq{L}.

  Coming back to solution of \eq{HOM7} we suggest to look for it in the form: \beq \label{HOM80}
 \Omega'\Lb \zeta, l-l_s; b\Rb= \mathcal{Z}\Lb \zeta\Rb \,\,\mathcal{L}\Lb l - l_s\Rb
 \eeq  
  Matching with \eq{HOM75} give us $\mathcal{L}\Lb l-l_s\Rb  \,=\,\Lb l \,-\,l_s\Rb^{\h} $ and for $\mathcal{Z}\Lb \zeta\Rb$ we obtain the following equation:
  \beq \label{HOM90}
\frac{3}{2}\mathcal{Z}_{\zeta}\Lb \zeta\Rb\,\,+\,\,\zeta\,\mathcal{Z}''_{\zeta,\zeta}\Lb \zeta\Rb  \,\,=\,\,e^{ - \Omega_\zeta\Lb \zeta\Rb}\mathcal{Z}\Lb \zeta\Rb\eeq

\begin{figure}[h]
 	\begin{center}
 	\leavevmode
 		\includegraphics[width=12cm]{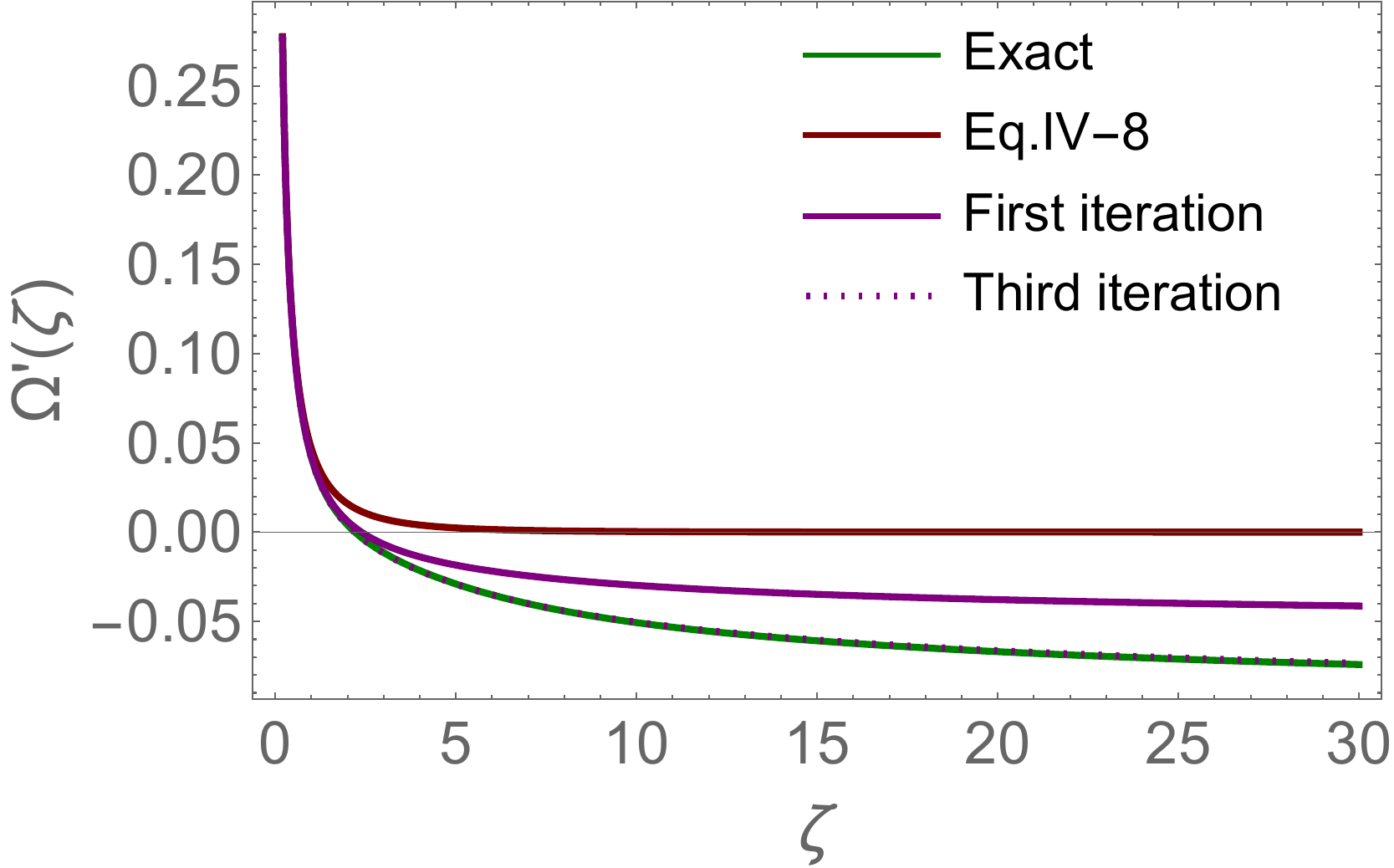}
 	\end{center}
 	\caption{$\Omega'\Lb \zeta\Rb$ versus $\zeta$ from \eq{HOM90}. $\Omega_0 = 1/4, \bar{\gamma} = 0.63$. The first iteration at small $\zeta$ is given by $ \mathcal{Z}\Lb \zeta ; \eq{HOM75}\Rb \,+\,\Delta   \mathcal{Z}\Lb \zeta; \eq{HOM120}\Rb$.	 The third iteration is calculated using \eq{HOM121}. 
}
 	\label{omegaprex}
 \end{figure}

  In \fig{omegaprex} we plot the exact solution to this equation with  $\Omega_\zeta\Lb \zeta\Rb$ given by \eq{SS4} and \eq{SS8} with $\Omega_0 = 1/4$.  
  For finding the analytical solution we suggest to search it as follows:
    \beq \label{HOM100}  
\Omega'\Lb \zeta,l-l_s\Rb\,\,=\,\,\Omega'\Lb \zeta,l -l_s;  \eq{HOM75}\Rb  \,\,+\,\,\Delta \Omega'\Lb \zeta,l-l_s\Rb
\eeq
with the equation for  $\Delta \Omega'\Lb \zeta,l-l_s\Rb = \Delta \mathcal{Z}'\,\Lb l\,-\,l_s\Rb^{\h}$ which has the form:
  \beq \label{HOM110}
\frac{3}{2}\Delta\mathcal{Z}'_{\zeta}\Lb \zeta\Rb\,\,+\,\,\zeta\,\Delta\mathcal{Z}''_{\zeta,\zeta}\Lb \zeta\Rb  \,\,=\,\,-\Lb e^{ - \Omega_0}\,-\,e^{ - \Omega_\zeta\Lb \zeta\Rb}\Rb\mathcal{Z}\Lb \zeta; \eq{HOM75}\Rb\eeq   
 Solution to this equation has the form:
 \beq \label{HOM120}  \Delta\mathcal{Z}\Lb \zeta\Rb =- \intl^\zeta_0 \frac{d t}{t ^{3/2}} \intl^t_0  d t' \sqrt{t'} \Lb e^{ - \Omega_0} \,-\,e^{ - \Omega_\zeta\Lb t'\Rb} \Rb\mathcal{Z}\Lb t'; \eq{HOM75}\Rb =  \,2\,\intl^\zeta_0 d t \Lb\sqrt{\frac{t}{\zeta}}-1\Rb \, \Lb e^{ - \Omega_0}\,-\,e^{ - \Omega_\zeta\Lb t\Rb} \Rb\mathcal{Z}\Lb t; \eq{HOM75}\Rb 
 \eeq
 
 For the second and higher iterations we have the following equations:
  \beq \label{HOM121} 
   \Delta^{(i)}\mathcal{Z}\Lb \zeta\Rb = \intl^\zeta_0 \frac{d t}{t ^{3/2}} \intl^t_0  d t' \sqrt{t'} \,e^{ - \Omega_\zeta\Lb t'\Rb}\Delta^{(i-1)}\mathcal{Z}\Lb t'\Rb =  -\,2\,\intl^\zeta_0 d t \Lb\sqrt{\frac{t}{\zeta}}-1\Rb \, e^{ - \Omega_\zeta\Lb t\Rb} \Delta^{(i-1)}\mathcal{Z}\Lb t\Rb 
 \eeq  
 In \fig{omegaprex} we plot  $ \Omega'\Lb \zeta\Rb = \mathcal{Z}\Lb \zeta; \eq{HOM75}\Rb\,\,+\,\,\Delta \mathcal{Z}\Lb \zeta\Rb $. One can see that this first iteration of \eq{HOM7} does not approach  $\Omega'\Lb \zeta\Rb$ with good accuracy. It turns out that only third iteration can describe the exact result for $Y \leq 25$.

\begin{figure}[h]
 	\begin{center}
 	\leavevmode
 		\includegraphics[width=11cm]{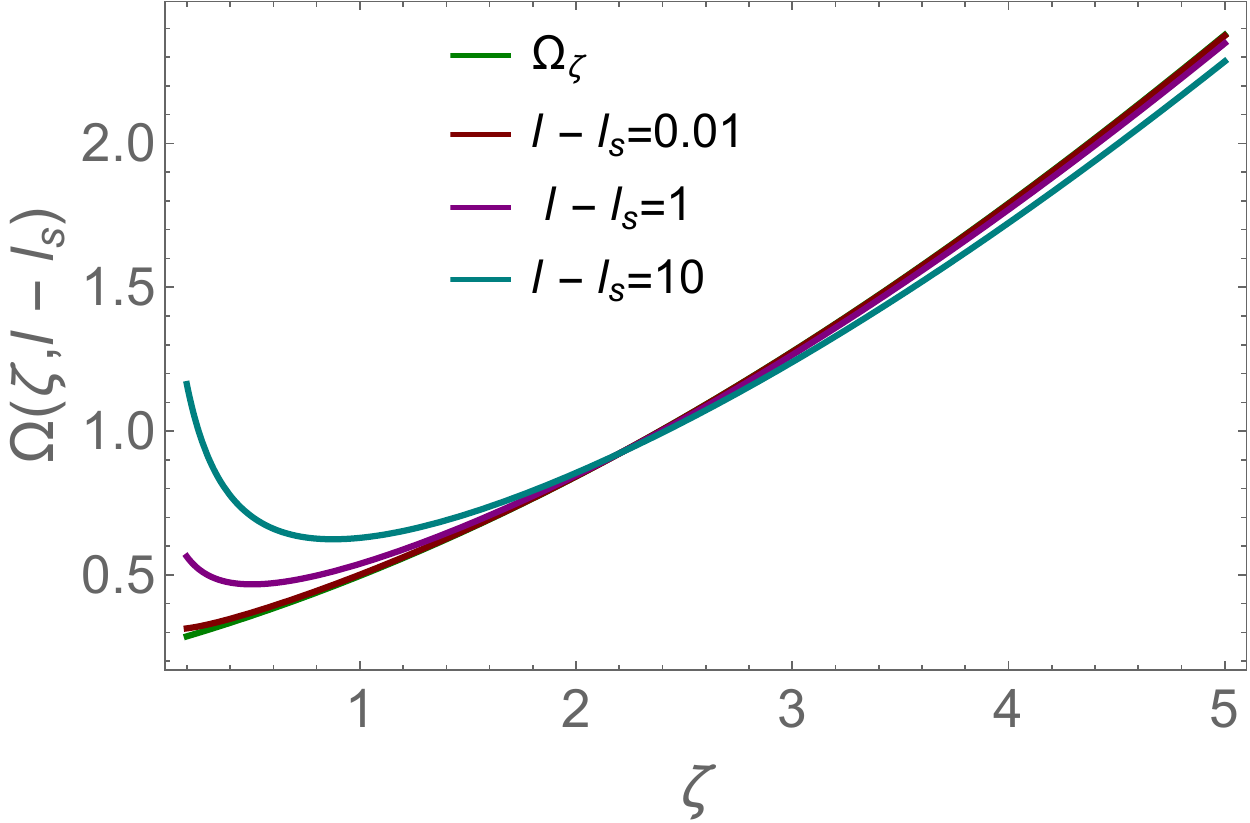}
 	\end{center}
 	\caption{$\Omega\Lb \zeta,l - l_s\Rb$ versus $\zeta$ at fixed $l - l_s$. $\Omega = \Omega_\zeta^{(3)} +\Omega'$ from \eq{NS3} and \eq{HOM80}. $\Omega'\Lb \zeta, l-l_s\Rb$ is the same as in \fig{omegaprex}.
}
 	\label{omegafull}
 \end{figure}

 In \fig{omegafull} the resulting $\Omega\Lb \zeta,l-l_s\Rb$ is shown. One can see that $\Omega'\Lb \zeta, l-l_s\Rb$ contribute only at rather small values of $\zeta$.
 
~

~

\section{Homotopy approach: second iteration  for the leading twist BFKL kernel}
 The next homotopy iteration  stems from \eq{HOM2}:
 \beq \label{SI1}
  {\mathscr H}\Lb p,\Omega^{(0)}+ p\Omega^{(1)} \Rb\,\,=\,\,\mathscr{L}[\Omega^{(0)}+ p\,\Omega^{(1)}] \,+ \,  p\, \mathscr{N_{L}}[\Omega^{(0)}] \,\,=\,\,0
  \eeq
  
  We need to account for the linear in $p$ terms in \eq{HOM2}.
  From \eq{SK2}-\eq{SK3} we obtain that \eq{SI1} takes the form:
 \beq \label{SI2}
 \dfrac{\pp^2 \Omega^{(1)}\Lb Y, l -l_s\Rb}{\pp Y\,\pp l}\,\,=\,\, \Lb 1 - e^{-\Omega^{(1)}\Lb Y, l-l_s\Rb}\Rb\,\,e^{ - \Omega^{(0)}\Lb Y,l-l_s\Rb} \,\,-\,\,\underbrace{\dfrac{\pp}{\pp\,l}\Lb  e^{ \Omega^{(0)}\Lb Y,l-l_s\Rb} \mathscr{N_{L}}[\Omega^{(0)}] \Rb}_{DH^{(0)}\Lb Y, l - l_s\Rb}
 \eeq
 with  $\Omega^{(0)}\Lb \zeta,l-l_s\Rb \equiv\,\Omega_\zeta^{(3)}\Lb \zeta\Rb+\Omega'\Lb\zeta,l-l_s\Rb$.

  In the derivation of \eq{SI2} we assumed that $\Lb \Omega^{(1)}\Lb Y, l-l_s\Rb\Rb^2 \,\ll\,\,\, \Omega^{(1)}\Lb Y, l - l_s\Rb $. The initial and boundary conditions for $\Omega^{(1)}\Lb Y, l-l_s\Rb$ takes the form:
  \beq \label{SIBC}   
  \Omega^{(1)}\Lb Y, l=l_s\Rb \,=\,0;~~~~~\dfrac{\pp \Omega^{(1)}\Lb Y, l-l_s\Rb}{\pp\,l}\Big{|}_{l = l_s} \,=\,0;
  \eeq  
    In this section we calculate $\mathscr{N_{L}}[\Omega^{(0)}] $ for the simplified BFKL kernel.
The BFKL kernel of \eq{GA1} includes the summation over all twist
 contributions. In the simplified approach we restrict ourselves 
 to the leading twist term only, which has the form\cite{LETU,CLMS} 
\bea \label{SIMKER}
\chi\Lb \gamma\Rb\,\,=\,\, \left\{\begin{array}{l}\,\,\,\,\,\,\,\frac{1}{\gamma}\,\,\,\,\,\,\,\,\,\mbox{for}\,\,\,\tau\,=\,r Q_s\,>\,1\,\,\,\,\,\mbox{summing} \Lb \ln\Lb r Q_s\Rb\Rb^n;\\ \\
\,\,\,\frac{1}{1 \,-\,\gamma}\,\,\,\,\,\mbox{for}\,\,\,\tau\,=\,r Q_s\,<\,1\,\,\,\,\,\mbox{summing}
\Lb \ln\Lb1/(r\,\Lambda_{\rm QCD})\Rb\Rb^n;\\  \end{array}
\right.
\eea
instead of the full expression of \eq{GA1}.  

For this simplified kernel \eq{K2} describe the contribution for $r_1 \leq r$, while
$N^{(0)}\Lb Y, l'\Rb$ in the region $r' > r$ in \eq{GA1} contributes to $ \mathscr{N_{L}}[N^{(0)}] $ which has the following form:
 \beq \label{SI3}
   \mathscr{N_{L}}[\Omega^{(0)}] = \intl_{r_1\,>\, r}\,\frac{d^2 r_1}{2 \pi}\,K\Lb r; r_1,r_2\Rb \exp\Lb \,-\,\,
 \Omega^{(0)}\Lb r_1,Y;\vec{b} - \frac{1}{2}\,\vec{r}_2 \Rb\,-\,  \Omega^{(0)}\Lb r_2 ,Y;\vec{b}- \frac{1}{2} \vec{r}_1\Rb \Rb \eeq
 where $ K\Lb r; r_1,r_2\Rb$ is given by \eq{K}. 
 
  From  \eq{AL} $\bas 
  \Lb r^2\Rb\,\,=\,\,- \frac{4 \,N_c}{b_0} \frac{1}{\xi}$. At $\xi=0 $, $\bas\Lb r^2\Rb$ has a singularity (Landau pole). The behaviour of  $\bas\Lb r^2\Rb$ for $\xi > 0$  is the problem of non-perturbative QCD which has not been solved. We assume that this region of integration does not contribute in our integrals, reducing \eq{SI31}
  to the expression:
  
  \beq \label{SI31}
   \mathscr{N_{L}}[\Omega^{(0)}] = \intl_{r_1\,>\, r}^{1/\Lambda_{QCD}}\,\frac{d^2 r_1}{2 \pi}\,K\Lb r; r_1,r_2\Rb \exp\Lb \,-\,\,
 \Omega^{(0)}\Lb r_1,Y;\vec{b} - \frac{1}{2}\,\vec{r}_2 \Rb\,-\,  \Omega^{(0)}\Lb r_2 ,Y;\vec{b}- \frac{1}{2} \vec{r}_1\Rb \Rb \eeq

 Taking $\Omega^{(0)}=\Omega_\zeta^{(3)} + \Omega' $,  we can obtain the estimate for  $ \mathscr{N_{L}}[N^{(0)}]$ for the leading twist BFKL kernel, viz.:
 \beq \label{SI4}
   \mathscr{N_{L}}[\Omega^{(0)}] = \h \bas\Lb r\Rb\, e^{\xi}\!\!\!\!\!\intl^0_{\xi  \geq -\xi_s}\!\!\!\!\!d \xi'   \, e^{-\xi'} 
\exp\Lb-\,\,
 2\,\Lb \Omega_\zeta^{(3)}\Lb \xi'\Rb + \Omega' \Lb Y, \xi'\Rb\, \Rb\Rb \, 
\eeq

\begin{figure}
 	\begin{center}
 	\leavevmode
	\begin{tabular}{c c }
 		\includegraphics[width=8.5cm]{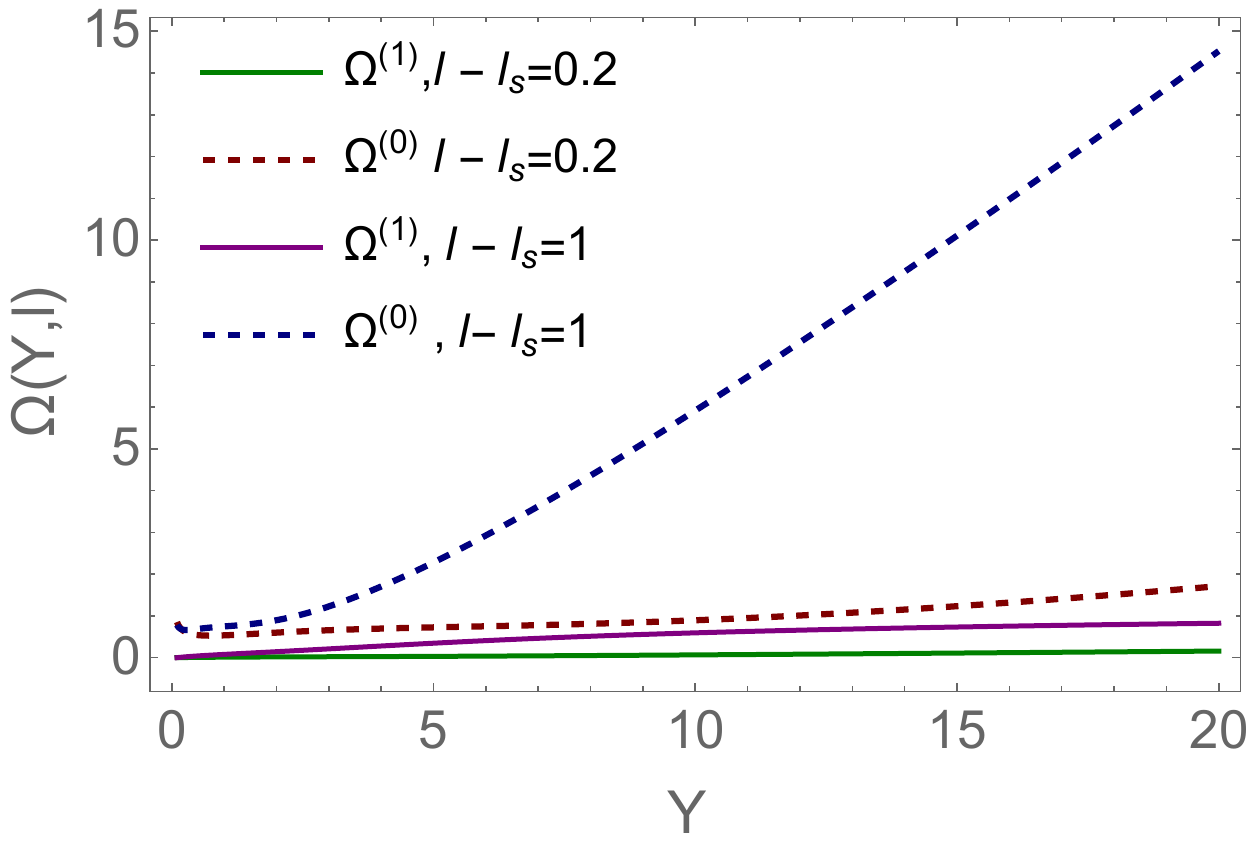}&\includegraphics[width=8.5cm]{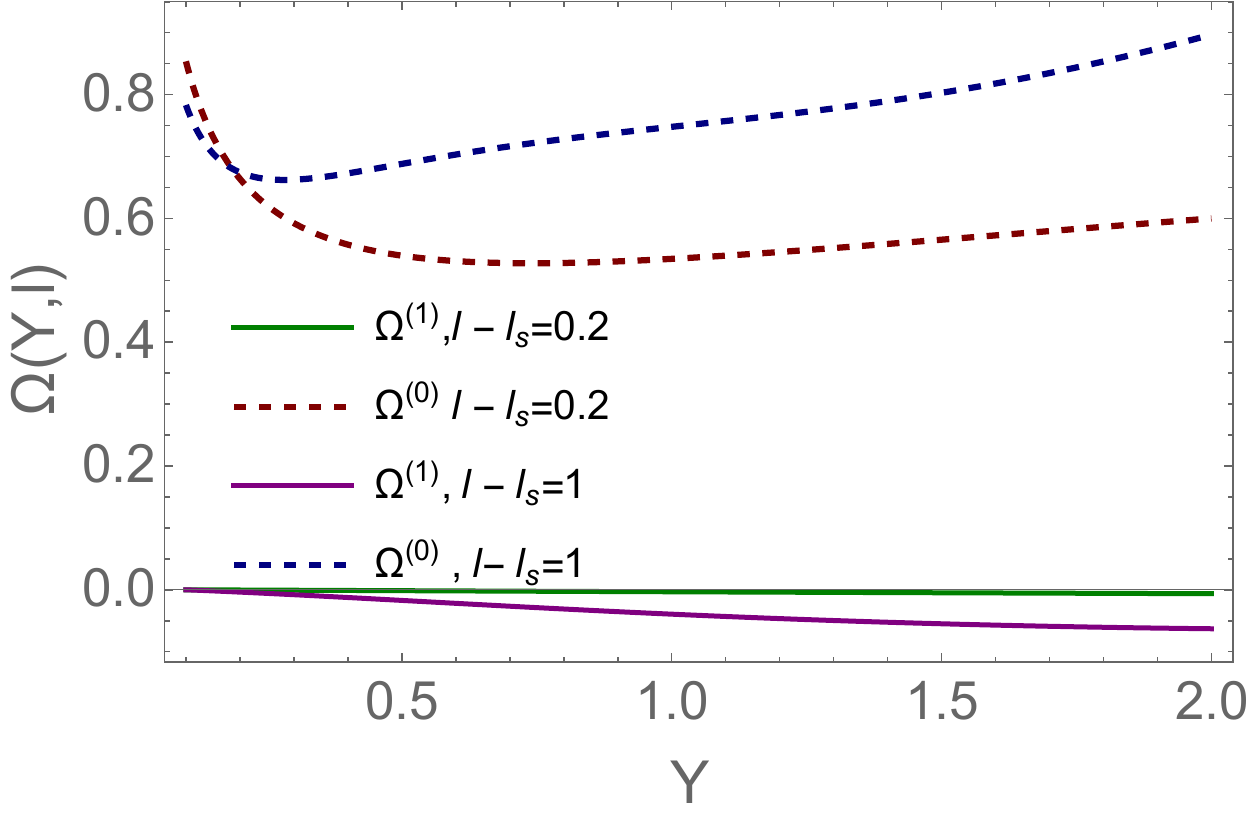} \\
		\fig{fin}-a & \fig{fin}-b\\
		\end{tabular}	\end{center}
	\caption{ $\Omega^{(1)}\Lb \zeta,l-l_s\Rb$ versus $\zeta$  at fixed $l -l_s$   in comparison with $\Omega^{(0)}\Lb \zeta.l-l_s\Rb$.  $\Omega_0 =0.25, \bar{\gamma}=0.63$.
}

 	\label{fin}
 \end{figure}

 Solving (numerically)  \eq{SI2} with the boundary and initial conditions of \eq{SIBC} we obtain $\Omega^{(1)}$. In \fig{fin} we plot $\Omega^{(0)}\Lb \zeta, l - l_s\Rb \equiv \Omega_\zeta\Lb \zeta, l - l_s\Rb $ and $\Omega^{(1)}\Lb \zeta, l - l_s\Rb$. One can see that $\Omega^{(1)}$ is much smaller that $\Omega^{(0)}$ as we expected. 
 
However, the corrections to the scattering amplitude
 \beq \label{SI5}
 \dfrac{N^{(1)}\Lb Y, l-l_s\Rb}{N^{(0)}\Lb Y, l-l_s\Rb} \,\,=\,\,\dfrac{ \Omega^{(1)}\Lb Y, l-l_s\Rb\exp\Lb - \Omega^{(0)}\Lb Y, l-l_s\Rb\Rb}{1\,\,-\,\,\exp\Lb - \Omega^{(0)}\Lb Y, l-l_s\Rb\Rb}
 \eeq 
turns out to be less that $5\%$   in the entire region of $Y$. Therefore, the first iteration of \eq{HOM4} gives the solution to our problem with accuracy which is not worse than $5\%$ at small values of $Y$ ($Y \leq 5$). Note, that for $Y > 5$ the accuracy is $< 2\%$. 
For better accuracy we have to include the second iteration of \eq{HOM2}, which is linear in $p$.
 
 ~

~

\section{Homotopy approach: third iteration  for the leading twist BFKL kernel}

\begin{figure}[h]
 	\begin{center}
 	\leavevmode
	\begin{tabular}{c c }
 		\includegraphics[width=8.5cm]{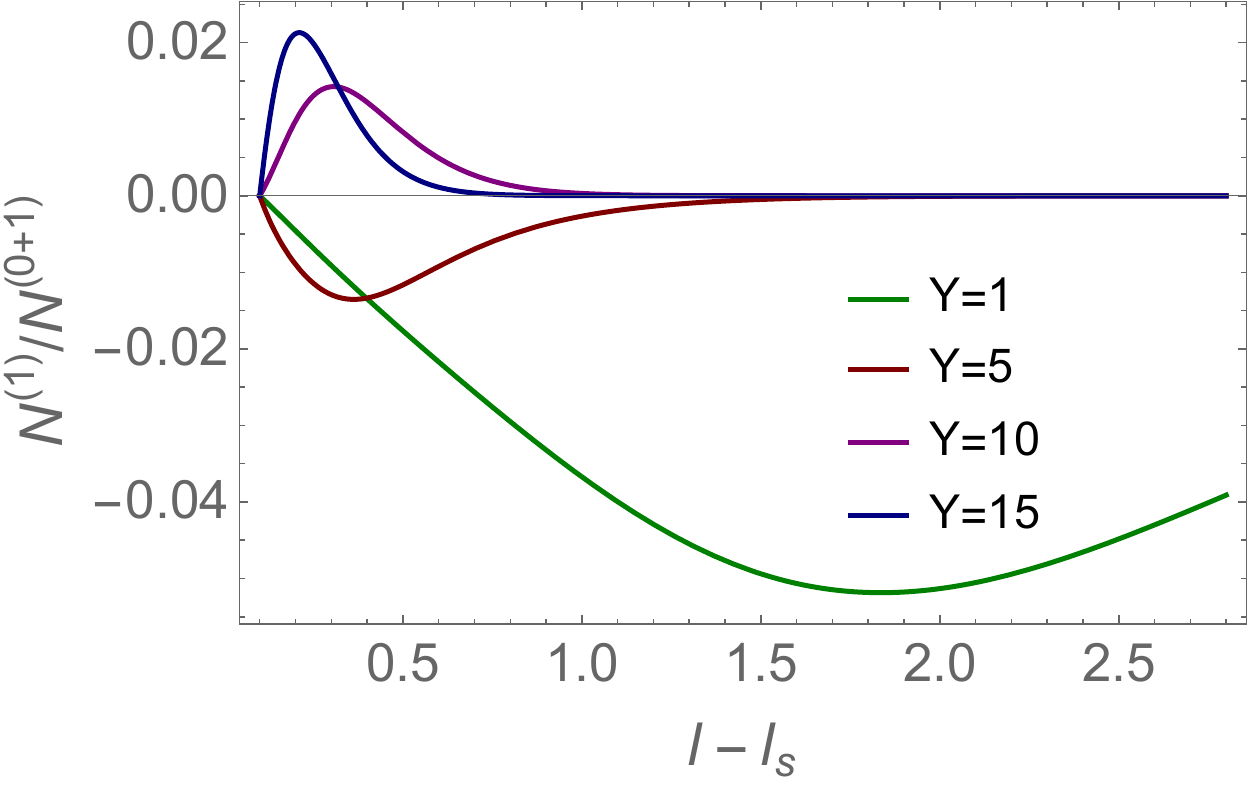}&\includegraphics[width=8.7cm]{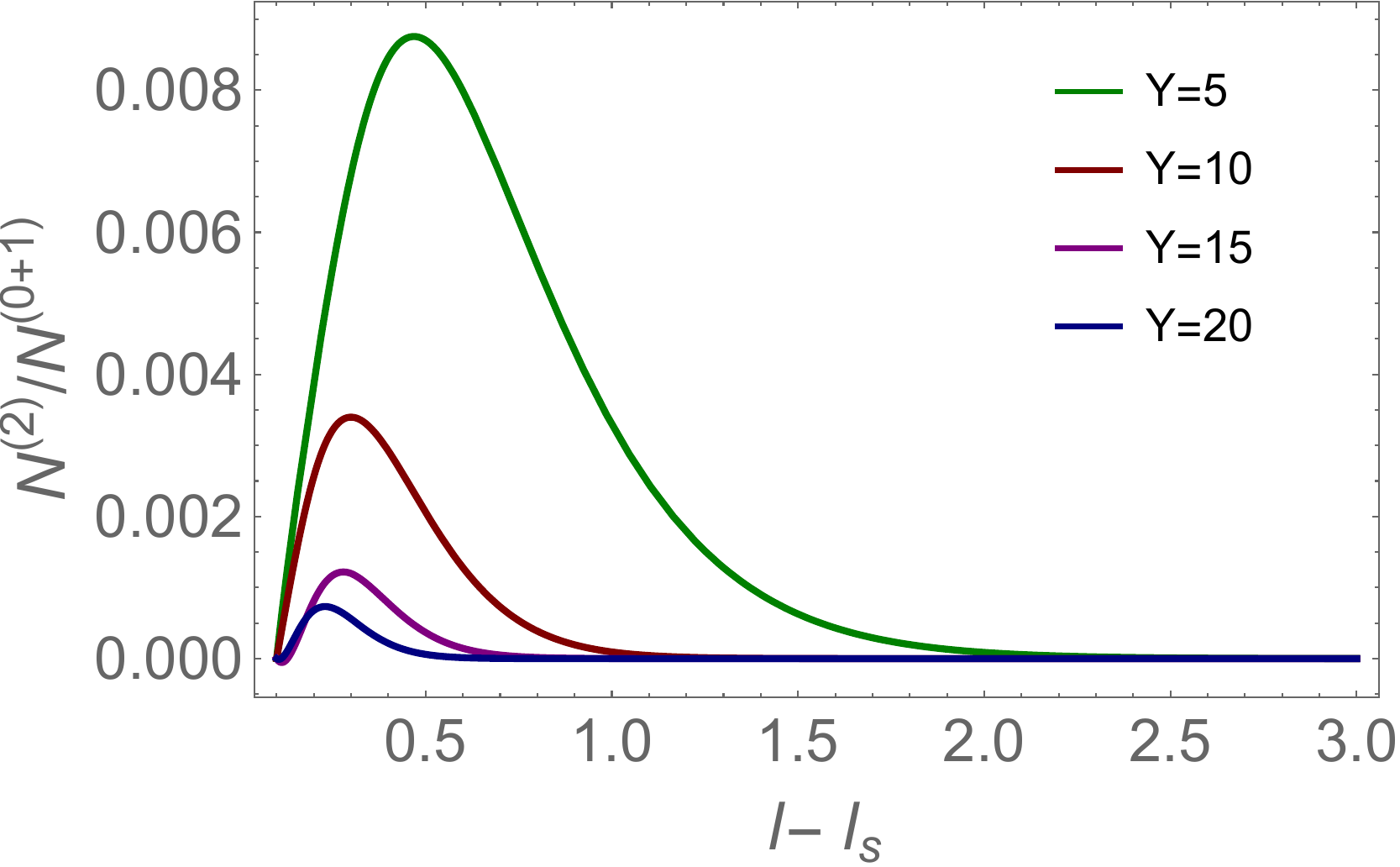} \\
		\fig{thi}-a & \fig{thi}-b\\
		\end{tabular}	\end{center}
 	\caption{ $\frac{N^{(1)}\Lb Y,l-l_s\Rb}{N^{(0+1)}\Lb Y,l-l_s\Rb}\,=\,\,\dfrac{ \Omega^{(1)}\Lb Y, l-l_s\Rb\exp\Lb - \Omega^{(0)}\Lb Y, l-l_s\Rb\Rb }{1\,\,-\,\,\exp\Lb - \Omega^{(0)}\Lb Y, l-l_s\Rb -\Omega^{(1)}\Lb Y, l-l_s\Rb\Rb}$	 
	  (\fig{thi}-a) and  $\frac{N^{(2)}\Lb Y,l-l_s\Rb}{N^{(0+1)}\Lb Y,l-l_s\Rb} =\,\,\dfrac{ \Omega^{(2)}\Lb Y, l-l_s\Rb\exp\Lb - \Omega^{(0)}\Lb Y, l-l_s\Rb -\Omega^{(1)}\Lb Y, l-l_s\Rb\Rb}{1\,\,-\,\,\exp\Lb - \Omega^{(0)}\Lb Y, l-l_s\Rb -\Omega^{(1)}\Lb Y, l-l_s\Rb\Rb}$
	(\fig{thi}-b)
	 versus $l-l_s$ at fixed $Y$  (see \eq{SI5}).  }	
	
\label{thi}
 \end{figure}

 ~
  For rather small $l - l_s$ the second iteration gives a sizable contribution  (about  $5\%$) to the scattering amplitude (see \fig{thi}-a).  Bearing this in mind we have to estimate the third iteration to demonstrate that  this iteration increases our accuracy.

 The general equation for this iteration has the following form:
\beq \label{THI1}
  {\mathscr H}\Lb p,\Omega^{(0)}+ p\,\Omega^{(1)} +  p\,\Omega^{(2)} \Rb\,\,=\,\,\mathscr{L}[\Omega^{(0)}+ p\,\Omega^{(1)} + p\,\Omega^{(2)} ] \,+ \,  p\, \mathscr{N_{L}}[\Omega^{(0)} + p\,\Omega^{(1)}] \,\,=\,\,0
  \eeq 
  \eq{THI1} is the expansion of the general equation (see \eq{HOM1} ) up to terms $\propto \,p^2$.
  From this equation we obtain:
  \beq \label{THI2}
 \dfrac{\pp^2 \Omega^{(2)}\Lb Y, l -l_s\Rb}{\pp Y\,\pp (l-l_s)}\,\,=\,\, \Omega^{(2)}\Lb Y, l-l_s\Rb\,\,e^{ - \Omega^{(0)}\Lb Y,l - l_s\Rb\,-\, \Omega^{(1)}\Lb Y,l-l_s\Rb} \,\,-\,\,\underbrace{\dfrac{\pp}{\pp\,(l-l_s)}\Lb  e^{ \Omega^{(0)}\Lb Y,l-l_s\Rb\,+\, \Omega^{(1)}\Lb Y,l-l_s\Rb} \mathscr{N_{L}}[\Omega^{(0)}\,+\, \Omega^{(1)}] \Rb}_{DH^{(1)}\Lb Y, l -l_s\Rb}
 \eeq  
   The initial and boundary conditions for $\Omega^{(2)}\Lb Y, l-l_s\Rb$  are the same as for  $\Omega^{(1)}\Lb Y, l-l_s\Rb$   and they are given by \eq{SIBC}.
  Repeating the same procedure as in the previous section we obtain that 
  $\mathscr{N_{L}}[\Omega^{(0)}\,+\, \Omega^{(1)}] $ has the form:  
 \bea \label{THI3}
  \mathscr{N_{L}}[\Omega^{(0)}+\Omega^{(1)}] & = & 
 \intl_{r_1\,>\, r}^{ 1/\Lambda_{QCD}}\,\frac{d^2 r_1}{2 \pi}\,K\Lb r; r_1,r_2\Rb \exp\Lb \,-\,\,
 \Omega^{(0)}\Lb r_1,Y;\vec{b} - \frac{1}{2}\,\vec{r}_2 \Rb\,-\,  \Omega^{(0)}\Lb r_2 ,Y;\vec{b}- \frac{1}{2} \vec{r}_1\Rb \Rb \nn\\
 &\times&~ \Lb \Omega^{(1)}\Lb r_1,Y;\vec{b} - \frac{1}{2}\,\vec{r}_2 \Rb\,+\,  \Omega^{(1)}\Lb r_2 ,Y;\vec{b}- \frac{1}{2} \vec{r}_1\Rb \Rb
  \eea
 where $ K\Lb r; r_1,r_2\Rb$ is given by \eq{K}.

 \begin{figure}[h]
 	\begin{center}
 	\leavevmode
	\begin{tabular}{c c }
 		\includegraphics[width=8.5cm]{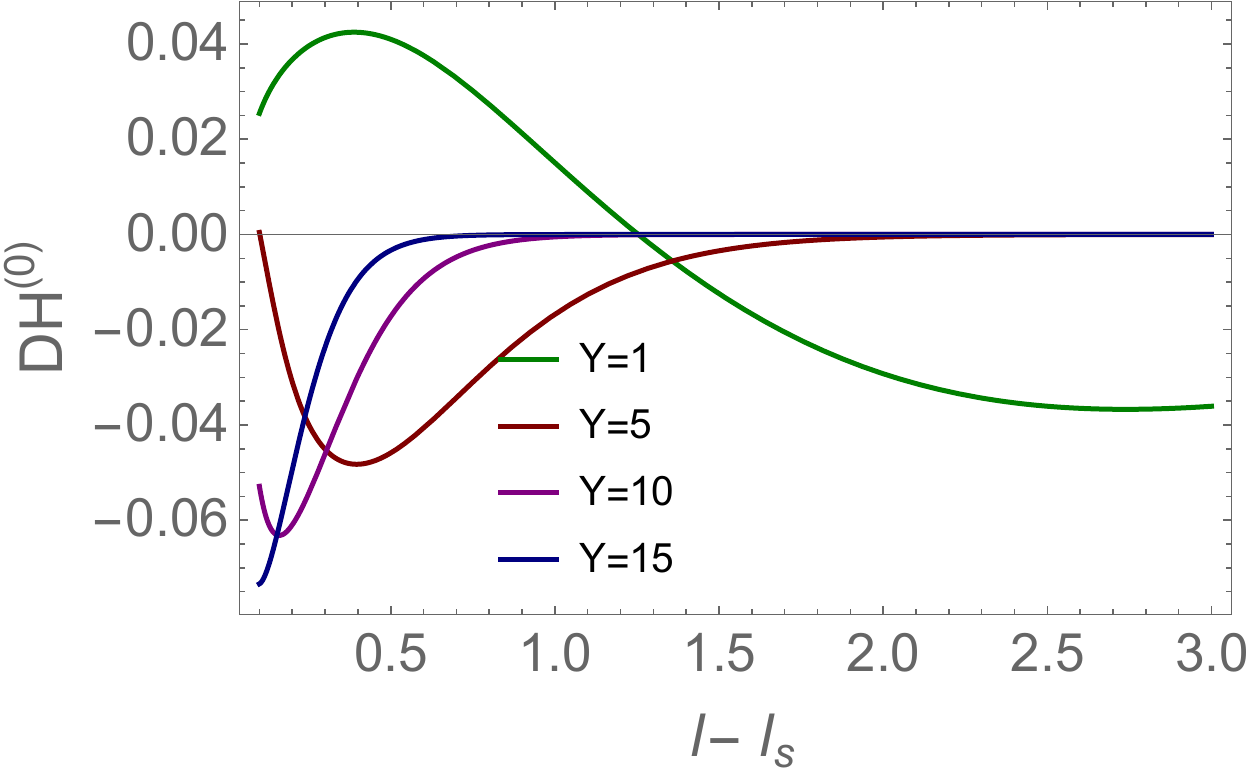}&\includegraphics[width=8.9cm]{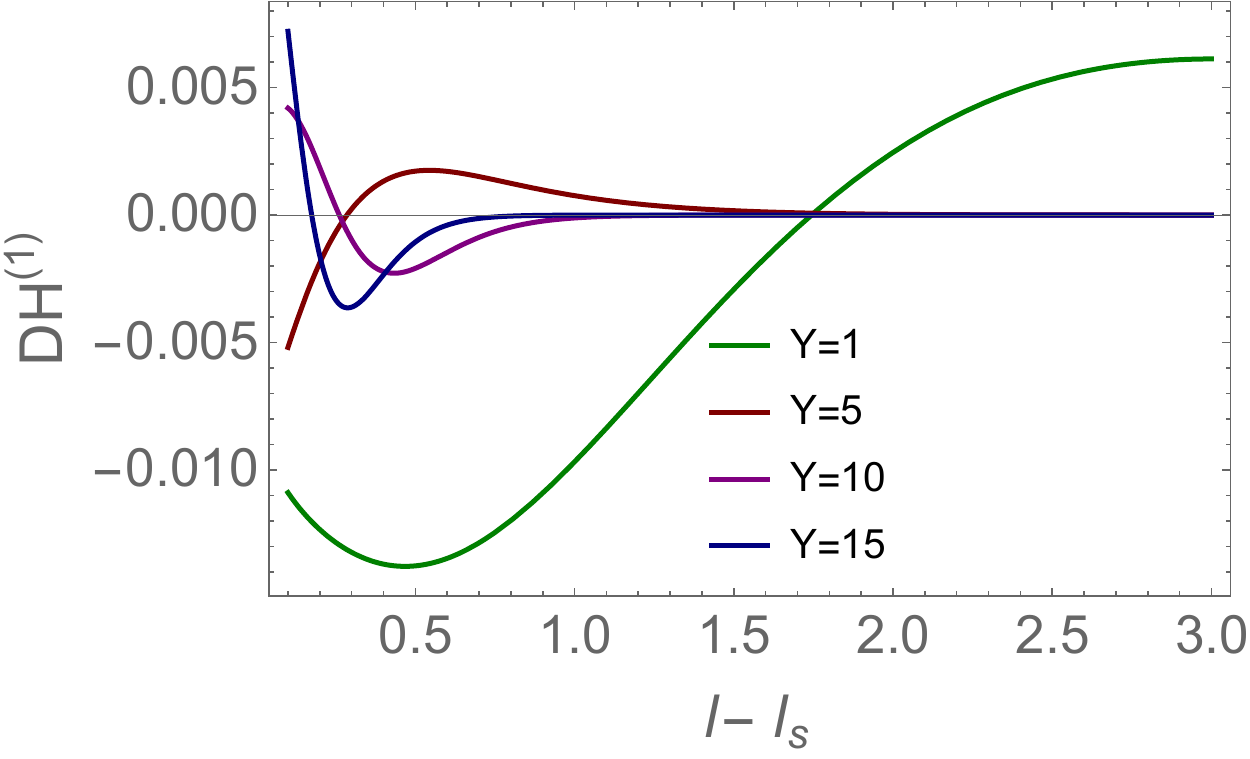} \\
		\fig{dh}-a & \fig{dh}-b\\
		\end{tabular}	\end{center}
 	\caption{ $DH^{(0)}\Lb Y, l - l_s\Rb$ (\fig{dh}-a) and $DH^{(1)}\Lb Y, l - l_s\Rb$ (\fig{dh}-b) from \eq{SI2} and \eq{THI2} respectively as  functions of
	$l-l_s$ at fixed $Y$.  $\Omega_0 =0.25, \bar{\gamma}=0.63$.	 
}	
	
\label{dh}
 \end{figure}
 

In \fig{dh} we plotted the nonhomogeneous terms of \eq{SI2} and \eq{THI2}. One can see that the nonhomogeneous term of \eq{SI2}  approximately is ten times larger than the same term on \eq{THI2}. Therefore, without solving \eq{THI2} we can conclude that  
$\Omega^{(2)}$ will be suppressed by one order in comparison with $\Omega^{(1)}$.  In \fig{thi} the solution to \eq{THI2} is plotted.
 One can see that the third iteration leads to small corrections in the entire region of $Y$ and $l - l_s$. \fig{thi}-b shows that the contribution from the third interaction to the scattering amplitude turns out to be negligibly small ($\leq\,1\%$).

\section{ General BFKL kernel: second and third iterations }
  For the general BFKL kernel we have to take into account that  (i)  \eq{SK3} does not describe the value of $\Omega$ for $ r_{01} < r$; and (ii) for $r_{01}>r$ we have to use a more general kernel instead of $ r^2/r^4_{02}$. Bearing these in mind we can rewrite equation for $ \mathscr{N_{L}}[\Omega^{(0)}]  $ in a general form:
  \begin{subequations}
  \bea 
\mathscr{N_{L}}[\Omega^{(0)}] & =& - \!\!\!\!\!\!\!\!\!\!\!\!\intl^\xi_{\begin{subarray}{l}  r_{2} \,\gg\,1/Q_s(Y)\\
r_{1}\,\gg\,1/Q_s(Y)\end{subarray}}\!\!\!\!\!\!\!\!\!\!\!\!\frac{d^2 r_1}{2 \pi}\,\underbrace{K\Lb r; r_1,r_2\Rb \exp\Lb \,-\,\,
 \Omega^{(0)}\Lb r_1,Y\Rb\,-\,  \Omega^{(0)}\Lb r_2 ,Y\Rb \Rb - \bas(r_1) \frac{1}{r^2_1} \exp\Lb \,-\,\,
 \Omega^{(0)}\Lb r_1,Y\Rb\,-\,  \Omega^{(0)}\Lb r ,Y\Rb \Rb}_{ \mbox{Difference between the general and model BFKL kernels in the region: $r_1<r$}}\nn\\ 
&- &\!\!\!\!\!\!\!\!\!\!\!\!\intl^\infty_{\begin{subarray}{l}~~~~~~ \xi\\ r_{2} \,\gg\,1/Q_s(Y)\\
r_{1}\,\gg\,1/Q_s(Y)\end{subarray}}\!\!\!\!\!\!\!\!\!\!\!\!\frac{d^2 r_1}{2 \pi}\,\underbrace{K\Lb r; r_1,r_2\Rb \exp\Lb \,-\,\,
 \Omega^{(0)}\Lb r_1,Y\Rb\,-\,  \Omega^{(0)}\Lb r_2 ,Y\Rb \Rb}_{\mbox{Contributions of the region: $r_1> r$}} \label{SIG1}\\
 & =& - \!\!\!\!\!\!\!\!\!\!\!\!\intl^{1/\Lambda_{QCD}}_{\begin{subarray}{l} ~~~~~~ -\,\xi_s\\ r_{2} \,\gg\,1/Q_s(Y)\\
r_{1}\,\gg\,1/Q_s(Y)\end{subarray}}\!\!\!\!\!\!\frac{d^2 r_1}{2 \pi}\,K\Lb r; r_1,r_2\Rb e^{ \,-\,\Omega^{(0)}\Lb r_1,Y\Rb\,-\,  \Omega^{(0)}\Lb r_2 ,Y\Rb} 
+\, \!\!\!\!\!\!\!\!\!\!\!\! \intl^\xi_{\begin{subarray}{l} ~~~~~~ \xi_s\\ r_{2} \,\gg\,1/Q_s(Y)\\
r_{1}\,\gg\,1/Q_s(Y)\end{subarray}} \!\!\! \underbrace{ \bas(r_1) \frac{d r^2_1}{r^2_1} e^{ \,-\,\,
 \Omega^{(0)}\Lb r_1,Y\Rb\,-\,  \Omega^{(0)}\Lb r ,Y\Rb }}_{\mbox{Contribution of the first iteration}} \label{SIG2}\eea
   \end{subequations}
   
    In appendix A we discuss a simplification of \eq{SIG2}. Using formulae of this appendix we reduce \eq{SI2} for the second iteration to the form:
    \beq \label{SIG3}
 \dfrac{\pp^2 \Omega^{(1)}\Lb Y, l -l_s\Rb}{\pp Y\,\pp l}\,\,=\,\, \Lb 2 - e^{-\Omega^{(1)}\Lb Y, l-l_s\Rb}\Rb\,\,e^{ - \Omega^{(0)}\Lb Y,l-l_s\Rb} \,\,-\,\,\dfrac{\pp}{\pp\,l}\Lb  e^{ \Omega^{(0)}\Lb Y,l-l_s\Rb} \mathscr{N'_{L}}[\Omega^{(0)}] \Rb
 \eeq    
 
 However, it turns out that the formulae take the more economic form for the difference between \eq{SIG2} and \eq{SI31} :  $\Delta \mathscr{N_{L}}[\Omega^{(0)}] \,\,=\,\,\mathscr{N_{L}}[\Omega^{(0)},\eq{SIG2}] \,-\,\mathscr{N_{L}}[\Omega^{(0)},\eq{SI31}] $. Using the formulae of appendix A for $\Delta \mathscr{N_{L}}[\Omega^{(0)}] $ we have
 
 \beq \label{SIG31}
 \Delta \mathscr{N_{L}}[\Omega^{(0)}] \,\,=\,\, 2 \bas\Lb r\Rb\!\!\!\!\!\!\!\!\!\!\!\!\intl^{1/\Lambda_{QCD}}_{\begin{subarray}{l} ~~~~~~ -\,\xi_s\\ r_{2} \,\gg\,1/Q_s(Y)\\
r_{1}\,\gg\,1/Q_s(Y)\end{subarray}}\!\!\!\!\!\! \!\!\!\! \frac{d r^2_1}{ r^2_1}  \frac{ r r_1 \cos\Lb \phi\Rb- r^2_1}{ r^2_1+ r^2_2} \exp\Lb \,-\,\,
 \Omega^{(0)}\Lb r_1,Y\Rb\,-\,  \Omega^{(0)}\Lb r_2 ,Y\Rb  \Rb
 \eeq

\begin{figure}[h]
 	\begin{center}
 	\leavevmode
\begin{tabular}{c c}
		\includegraphics[width=8.5cm]{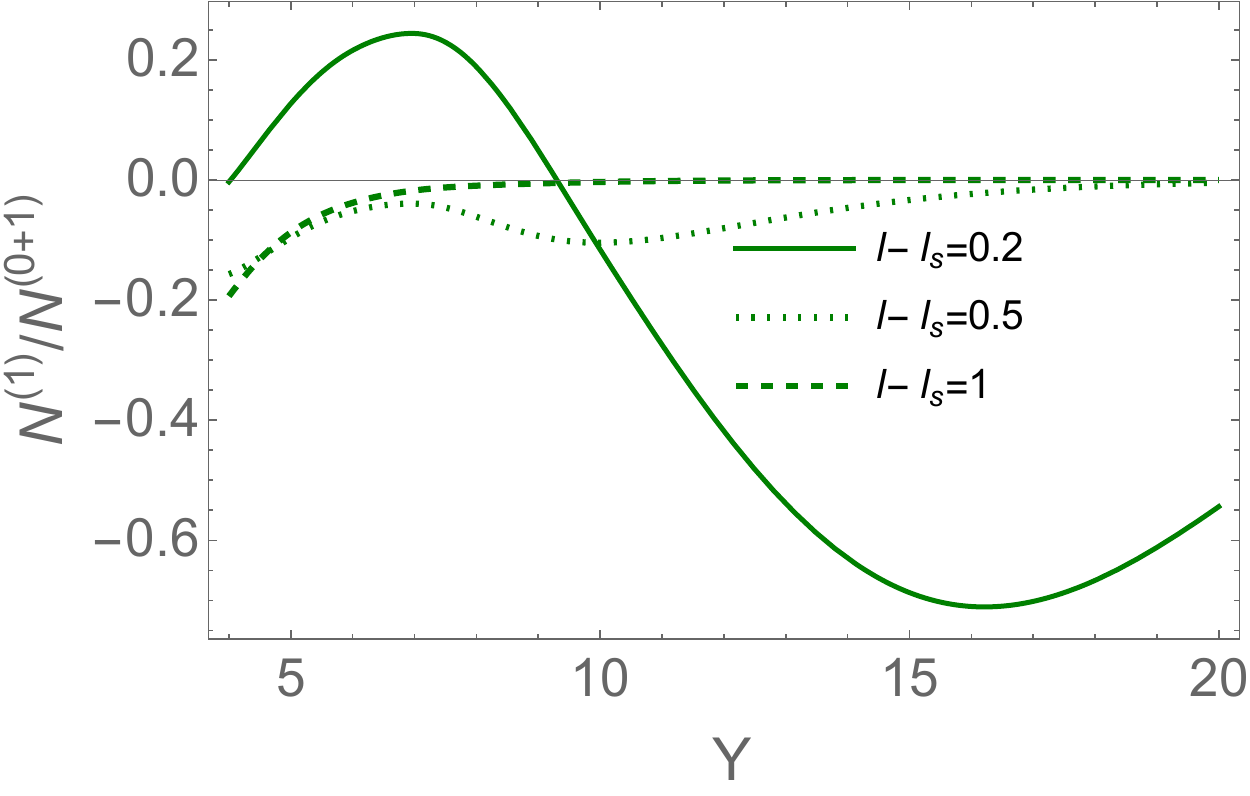}&\includegraphics[width=8.7cm]{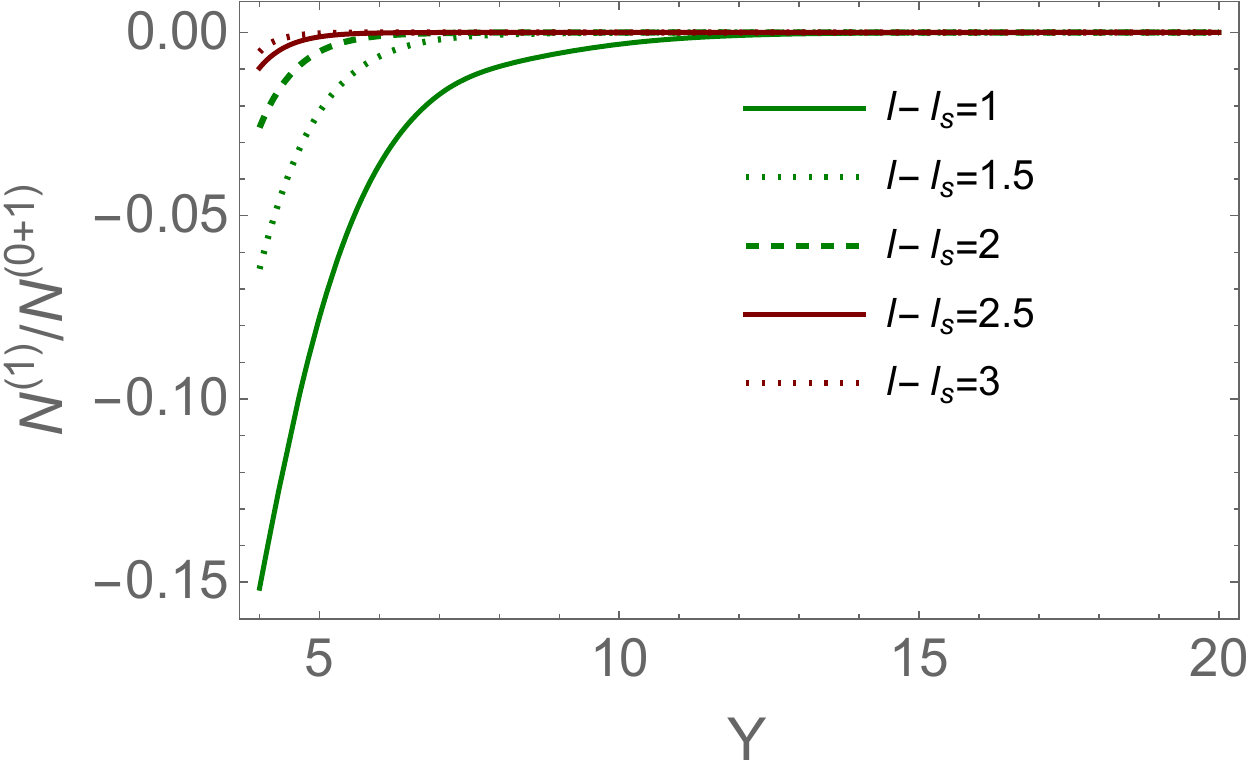} \\
		\fig{estful}-a&\fig{estful}-b\\
				\end{tabular}	
\end{center}
	
 	\caption{$\frac{N^{(1)}\Lb Y,l-l_s\Rb}{N^{(0+1)}\Lb Y,l-l_s\Rb}$ (see caption of \fig{thi}) for the general BFKL kernel versus $Y$ at fixed $l - l_s$.	  $\Omega_0 =0.25, \bar{\gamma}=0.63$.}	
	
\label{estful}
 \end{figure}

 The solution to \eq{SIG2} for the general BFKL kernel is plotted in \fig{estful}.     One can see that  the second iteration gives a small corrections for large $ l - l_s$ ($l - l_s\geq 1$) . On the other hand these corrections are large for small $l - l_s$. For large $l-l_s $ we need to calculate the third iteration to obtain corrections of the order of few percents.

  
 For the third iteration we have \eq{THI2} with

  \bea \label{SIG4}
  \mathscr{N_{L}}[\Omega^{(0)}+\Omega^{(1)}] & = & 
\!\!\!\!\!\!\!\!\!\!\!\!\intl^{1/\Lambda_{QCD}}_{\begin{subarray}{l}  r_{2} \,\gg\,1/Q_s(Y)\\
r_{1}\,\gg\,1/Q_s(Y)\end{subarray}}\,\!\!\!\!\!\!\!\!\!\!\!\!\frac{d^2 r_1}{2 \pi}\,K\Lb r; r_1,r_2\Rb \exp\Lb \,-\,\,
 \Omega^{(0)}\Lb r_1,Y;\vec{b} - \frac{1}{2}\,\vec{r}_2 \Rb\,-\,  \Omega^{(0)}\Lb r_2 ,Y;\vec{b}- \frac{1}{2} \vec{r}_1\Rb \Rb \nn\\
  &\times&~ \Lb 1\,\,-\,\,\exp\Lb - \Omega^{(1)}\Lb r_1,Y;\vec{b} - \frac{1}{2}\,\vec{r}_2 \Rb\,-\,  \Omega^{(1)}\Lb r_2 ,Y;\vec{b}- \frac{1}{2} \vec{r}_1\Rb \Rb\Rb
  \eea 
 
\eq{SIG4} is simplified in appendix B.

For small $l-l_s$ we have to calculate the fourth  iteration.  Facing such problem we decided to simplify our first iteration to obtain more transparent formulae for the higher ones. This is the goal of the next section.

\section{ General BFKL kernel: alternative approach }
In this section we are going to discuss an alternative approach for the general BFKL kernel which is based on the analytic of the BK equation solution (see Ref.\cite{LETU}) deeply in the saturation region.
 \subsection{Scattering amplitude at high energies for the general BFKL kernel:} 
 The success of our approach actually stems from the simple solution of the nonlinear equation for the general BFKL kernel in high energy kinematic region\cite{LETU}. Indeed, at high energies (at $Y \gg 1$) the scattering amplitude approaches the unitarity limit: $N\Lb r,Y;b \Rb \xrightarrow{Y \gg 1} 1$. Introducing $ N\Lb r,Y;b \Rb \,=\,1\,\,-\,\,\Delta\Lb r,Y;b \Rb$  with $\Delta\Lb r,Y;b \Rb \,\ll\,1$, 
 we can rewrite \eq{GA1} in the form  for the saturation region:
 \bea \label{C1}
 \dfrac{\pp \Delta\Lb r,Y;b \Rb}{ \pp Y}\,\,&=&\,\,  \int\,\frac{d^2 r_1}{2 \pi}\,K\Lb r; r_1,r_2\Rb \Bigg\{ - \Delta\Lb r,Y;\vec{b}\Rb \,+\,
 \Delta \Lb r_1,Y;\vec{b} - \frac{1}{2}\,\vec{r}_2 \Rb\, \Delta\Lb r_2 ,Y;\vec{b}- \frac{1}{2} \vec{r}_1\Rb \Bigg\} \nn\\
 &\xrightarrow{Y \gg 1}& -  \Delta\Lb r,Y;\vec{b}\Rb\!\!\!\!\!\!\!\! \!\!\!\! \underbrace{\intl_{\begin{subarray}{l} ~ r_{2} \,\gg\,1/Q_s(Y)\\
~r_{1}\,\gg\,1/Q_s(Y)\end{subarray}}\!\!\!\!\!\!\!  \!\!\!\!\! \frac{d^2 r_1}{2 \pi}\,K\Lb r; r_1,r_2\Rb }_{\mbox{gluon reggeization}}
\eea
 Introducing $\Delta\Lb r,Y;\vec{b}\Rb\,=\,\,\exp\Lb - \Omega\Lb r,Y;\vec{b}\Rb \Rb$ and taking the main logarithmic contribution for gluon reggeization we reduce \eq{C1} to the following equation:
 
 \beq \label{C2}
\dfrac{ \pp \Omega\Lb r,Y;\vec{b}\Rb}{\pp\,Y}\,\,=\,\,  \intl^{r^2}_{1/Q^2_s} d r'^2\frac{\bas\Lb r'^2\Rb}{r'^2}\,\,=\,\,l - l_s
\eeq
In \eq{C2} we used the variable $l - l_s$ which has been introduced in \eq{XIL}. \eq{C2} can be rewritten as
\beq \label{C3}
\dfrac{ \pp^2 \Omega\Lb r,Y;\vec{b}\Rb}{\pp\,Y\,\pp(l - l_s)} \,\,= 1
\eeq
which coincides with \eq{PL1}. Solution to this equation is equal to
\beq \label{C4} 
\Omega\Lb Y,l-l_s\Rb\,\,=\,\,Y\,\Lb l - l_s\Rb + \tilde{\Omega}\Lb Y,l-l_s\Rb;\,\, ~~~\dfrac{ \pp^2 \tilde{\Omega}\Lb r,Y;\vec{b}\Rb}{\pp\,Y\,\pp l} \,\,= 0 \,\leftarrow \mbox{homogeneous equation}
\eeq

Therefore, we demonstrated that at large values of $Y$ we obtain the same solution as for our nonlinear equation with the leading twist BFKL kernel. For matching the initial condition we use the solution of   \eq{PL5}. 

We suggest to use \eq{C3} as the first homotopy iteration and build the iterative procedure repeating all steps in finding the solution given in section III.


\subsection{Second iteration }
The second iteration to \eq{C1} we can write as follows:

\beq \label{C4}
 \dfrac{\pp \Omega^{(1)}\Lb r,Y;b \Rb}{ \pp Y}\,\,=\,\,   
  (l - l_s)    -  e^{\Omega^{(0)}\Lb Y, l - l_s \Rb} \int\,\frac{d^2 r_1}{2 \pi}\,K\Lb r; r_1,r_2\Rb \exp\Lb -
 \Omega^{(0)}  \Lb r_1,Y;\vec{b} - \frac{1}{2}\,\vec{r}_2 \Rb\,- \Omega^{(0)}\Lb r_2 ,Y;\vec{b}- \frac{1}{2} \vec{r}_1\Rb \Rb\eeq
 where $\Omega^{(0)}\Lb Y, l - l_s \Rb = Y ( l - l_s) \,-\,\bar{\gamma}\,\Omega_0\, \Lb e^{-\frac{b_0}{4 N_c}  l} \,-\,e^{-\frac{b_0}{4 N_c} l_s}\Rb\,+\,\frac{1}{4\,\kappa} \Lb e^{-\frac{b_0}{2 N_c} l}\,-\,e^{-\frac{b_0}{2 N_c}l_s}\Rb\,\,+\,\,\Omega_0
$  (see \eq{PL5}).

 Using \eq{A1} of appendix A we can rewrite \eq{C4} in the form:
  \beq \label{C5}
  \dfrac{\pp \Omega^{(1)}\Lb r,Y;b \Rb}{ \pp Y}\,\,=\,\,   
  (l - l_s)     -\bas\Lb r^2\Rb e^{\Omega^{(0)}\Lb Y, l - l_s \Rb}\!\!\!\!\!\!\!\!\!\!\!\!\intl^{1/\Lambda_{QCD}}_{\begin{subarray}{l}  r_{2} \,\gg\,1/Q_s(Y)\\
r_{1}\,\gg\,1/Q_s(Y)\end{subarray}}\!\!\!\!\!\!\!\!\!\!\!\! \frac{d \phi d r_1^2}{2 \pi}\,\Bigg\{\frac{r^2}{r^2_1\,(r^2_1+r^2_2)} \,\,+\,\,\frac{1}{r^2_1} \frac{\ln\Lb r^2_2/r^2_1\Rb}{\ln r^2_1} \Bigg\}  \exp\Lb \,-\,\,
 \Omega^{(0)}\Lb r_1,Y\Rb\,
 -\,  \Omega^{(0)}\Lb r_2 ,Y\Rb \Rb
 \eeq
Hence,
\bea \label{C6}
&&\Omega^{(1)}\Lb Y, l-l_s \Rb\,\,=\,\,\Omega^{(0)}\Lb Y, l-l_s \Rb\\
 &&- \intl^Y_0 d Y'
\bas\Lb r^2\Rb e^{\Omega^{(0)}\Lb Y', l - l_s \Rb}\!\!\!\!\!\!\!\!\!\!\!\!\intl^{1/\Lambda_{QCD}}_{\begin{subarray}{l}  r_{2} \,\gg\,1/Q_s(Y')\\
r_{1}\,\gg\,1/Q_s(Y')\end{subarray}}\!\!\!\!\!\!\!\!\!\!\!\! \frac{d \phi d r_1^2}{2 \pi}\,\Bigg\{\frac{r^2}{r^2_1\,(r^2_1+r^2_2)} \,\,+\,\,\frac{1}{r^2_1} \frac{\ln\Lb r^2_2/r^2_1\Rb}{\ln r^2_1} \Bigg\}  \exp\Lb \,-\,\,
 \Omega^{(0)}\Lb r_1,Y'\Rb\,
 -\,  \Omega^{(0)}\Lb r_2 ,Y'\Rb \Rb\nn
 \eea
 \begin{figure}[h]
 	\begin{center}
 	\leavevmode
\begin{tabular}{c c}
		\includegraphics[width=8.5cm]{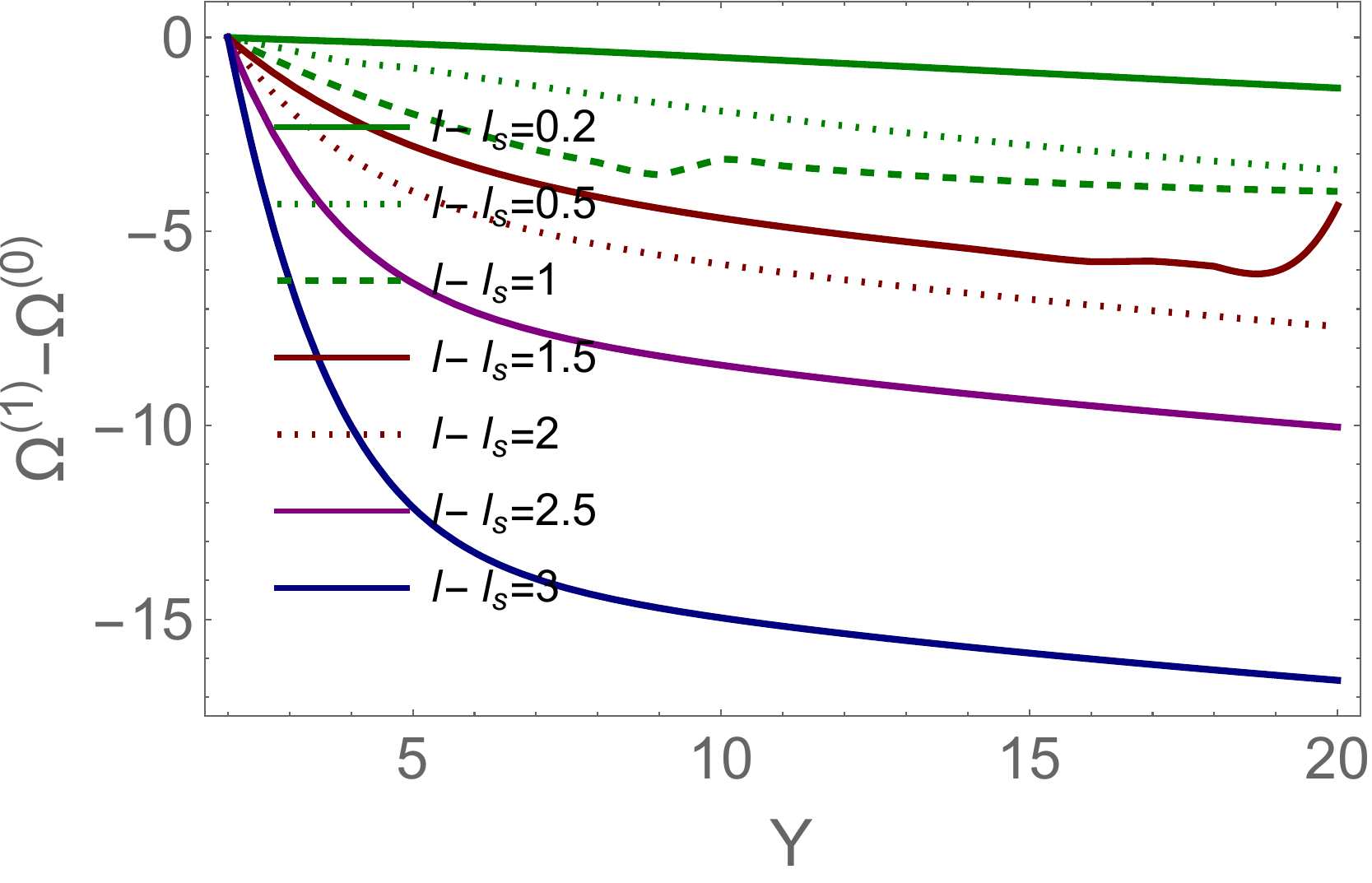}&\includegraphics[width=8.7cm]{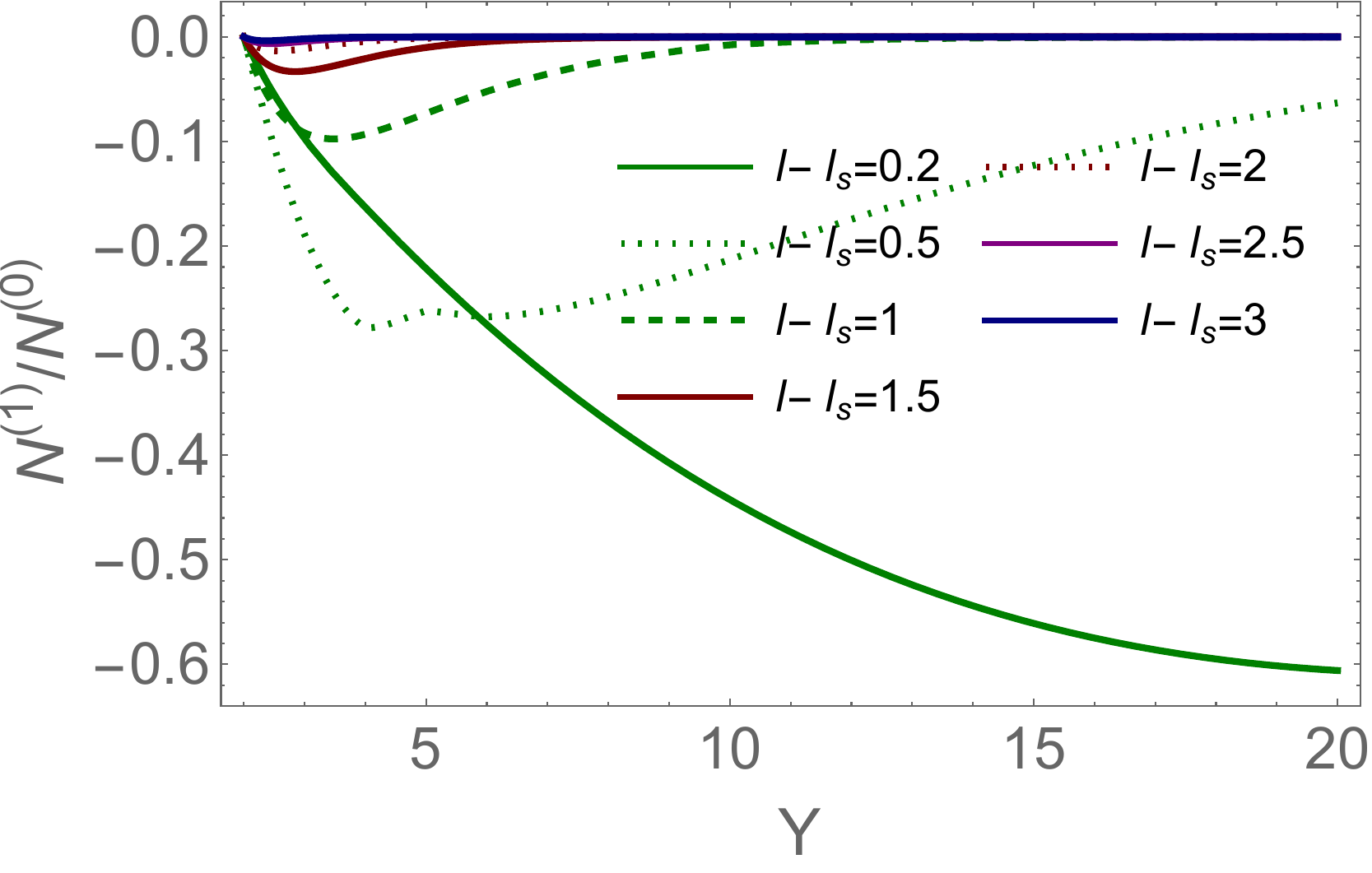} \\
		\fig{newit}-a&\fig{newit}-b\\
				\end{tabular}	
\end{center}
	
 	\caption{\fig{newit}-a: The difference $\Omega^{(1)}\Lb Y', l - l_s \Rb - \Omega^{(0)}\Lb Y', l - l_s \Rb$ (see \eq{C7})	
	 for the general BFKL kernel in the alternative approach versus $Y$ at fixed $l - l_s$.
\fig{newit}-b:$\frac{N^{(1)}\Lb Y,l-l_s\Rb}{N^{(0+1)}\Lb Y,l-l_s\Rb}$ (see caption of \fig{thi}) for the general BFKL kernel in the alternative approach (see \eq{C7}) versus $Y$ at fixed $l - l_s$.	  $\Omega_0 =0.25, \bar{\gamma}=0.63$.}	
	
\label{newit}
 \end{figure} 
 As we have discussed in \eq{C4}  we need to add to this equation the solution of homogeneous equation (see \eq{C4}) to satisfy the initial condition: $$
 \Omega^{(1)}\Lb Y, l-l_s=0 \Rb\,\,=\,\,\Omega^{(0)}\Lb Y, l-l_s=0 \Rb;
 ~~~\frac{\pp}{\pp l}\Omega^{(1)}\Lb Y, l-l_s\Rb\Bigg{|}_{l = ls}\,\,=\,\,0 $$

  Hence, finally we have
 \bea \label{C7}
 &&\Omega^{(1)}\Lb Y, l-l_s \Rb   \,\,=\,\,\Omega^{(1)}\Lb Y, l-l_s; \eq{C6} \Rb \\,\,&&-\,\,
 \Omega^{(1)}\Lb Y, l-l_s=0; \eq{C6} \Rb \,\,-\,\,\intl^l_{l_s} d l' \frac{ \pp\Omega^{(1)}\Lb\frac{b_0 \xi^2_s}{8 \,N_C\,\kappa}e^{ - 
 \frac{b_0}{2\,N_c} l'} , l-l_s; \eq{C6} \Rb}{\pp l}\Bigg{|}_{ l = l_s}\nn
  \eea

 It is instructive to note that \eq{C6} gives  an analytic solution for $\Omega^{(1)}$. From \fig{newit}-a one can  see that $\Omega^{(1)} - \Omega^{(0)}$ tuns out to be smaller than $\Omega^{(0)}$ at least for $l - l_s \geq 0.5$.   \fig{newit} shows that the second iteration, which led to \eq{C6} , can be treated as corrections to $\Omega^{(0)}$.  One can see that for $l - l_s > 1$ the first  iteration gives less that $10 \%$ accuracy. Including the second iteration, which can be calculated analytically using \eq{C6}, we can reach accuracy $\leq 1\%$. Therefore, we demonstrated that our procedure works.

\subsection{Third iteration }
Our second iteration was solution to the equation:
\bea \label{C8}
\frac{\pp \Delta^{(1)}\Lb Y, l-l_s \Rb  }{\pp \,Y}\,\,&=&\,\,-\,\Lb  l-l_s \Rb\,\Delta^{(1)}\Lb Y, l-l_s \Rb\,+\,\,\mathscr{N_{L}}[\Delta^{(0)}]\nn\\~\mbox{with}~~
\mathscr{N_{L}}[\Delta^{(0)}] \,&=&\,\!\!\!\!\!\!\!\!\!\!\!\!\intl^{1/\Lambda_{QCD}}_{\begin{subarray}{l}  r_{2} \,\gg\,1/Q_s(Y)\\
r_{1}\,\gg\,1/Q_s(Y)\end{subarray}}\!\!\!\!\!\!\!\!\!\!\!\!\frac{d^2 r_1}{2 \pi}\,K\Lb r; r_1,r_2\Rb \exp\Lb \,-\,\,
 \Omega^{(0)}\Lb r_1,Y\Rb\,-\,  \Omega^{(0)}\Lb r_2 ,Y\Rb \Rb
 \eea
 
 For the third iteration the equation takes the form:
 \bea \label{C9}
\frac{\pp \Delta^{(2)}\Lb Y, l-l_s \Rb  }{\pp \,Y}\,\,&=&\,\,-\,\Lb  l-l_s \Rb\,\Delta^{(2)}\Lb Y, l-l_s \Rb\,+\,\,\mathscr{N_{L}}[\Delta^{(1)}]\nn\\~\mbox{with}~~
\mathscr{N_{L}}[\Delta^{(1)}] \,&=&\,\!\!\!\!\!\!\!\!\!\!\!\!\intl^{1/\Lambda_{QCD}}_{\begin{subarray}{l}  r_{2} \,\gg\,1/Q_s(Y)\\
r_{1}\,\gg\,1/Q_s(Y)\end{subarray}}\!\!\!\!\!\!\!\!\!\!\!\!\frac{d^2 r_1}{2 \pi}\,K\Lb r; r_1,r_2\Rb \exp\Lb \,-\,\,
 \Omega^{(1)}\Lb r_1,Y\Rb\,-\,  \Omega^{(1)}\Lb r_2 ,Y\Rb \Rb
 \eea 
 One can see from \fig{newit}-a that $\mathscr{N_{L}}[\Delta^{(1)}] \,<\,\mathscr{N_{L}}[\Delta^{(0)}] $ and, therefore, we can hope that the solution to this equation will lead to the better approximation for non-linear BK equation.
  
 Introducing $\Delta^{(2)}\Lb Y, l-l_s \Rb\,\,=\,\,\exp\Lb - \Omega^{(2)}\Lb Y, l-l_s \Rb\Rb$ we can rewrite the solution to \eq{C9} as follows:
 \bea \label{C10}
&&\Omega^{(2)}\Lb Y, l-l_s \Rb\,\,=\,\,\Omega^{(0)}\Lb Y, l-l_s \Rb\\
 &&- \intl^Y_0 d Y'
\bas\Lb r^2\Rb e^{\Omega^{(0)}\Lb Y',l - l_s \Rb}\!\!\!\!\!\!\!\!\!\!\!\!\intl^{1/\Lambda_{QCD}}_{\begin{subarray}{l} r_{2} \,\gg\,1/Q_s(Y')\\
r_{1}\,\gg\,1/Q_s(Y')\end{subarray}}\!\!\!\!\!\!\!\!\!\!\!\! \frac{d \phi d r_1^2}{2 \pi}\,\Bigg\{\frac{r^2}{r^2_1\,(r^2_1+r^2_2)} \,\,+\,\,\frac{1}{r^2_1} \frac{\ln\Lb r^2_2/r^2_1\Rb}{\ln r^2_1} \Bigg\}  \exp\Lb \,-\,\,
 \Omega^{(1)}\Lb r_1,Y'\Rb\,
 -\,  \Omega^{(1)}\Lb r_2 ,Y'\Rb \Rb\nn
 \eea 
 
\begin{figure}[h]
 	\begin{center}
 	\leavevmode
\begin{tabular}{c c}
		\includegraphics[width=9.1cm]{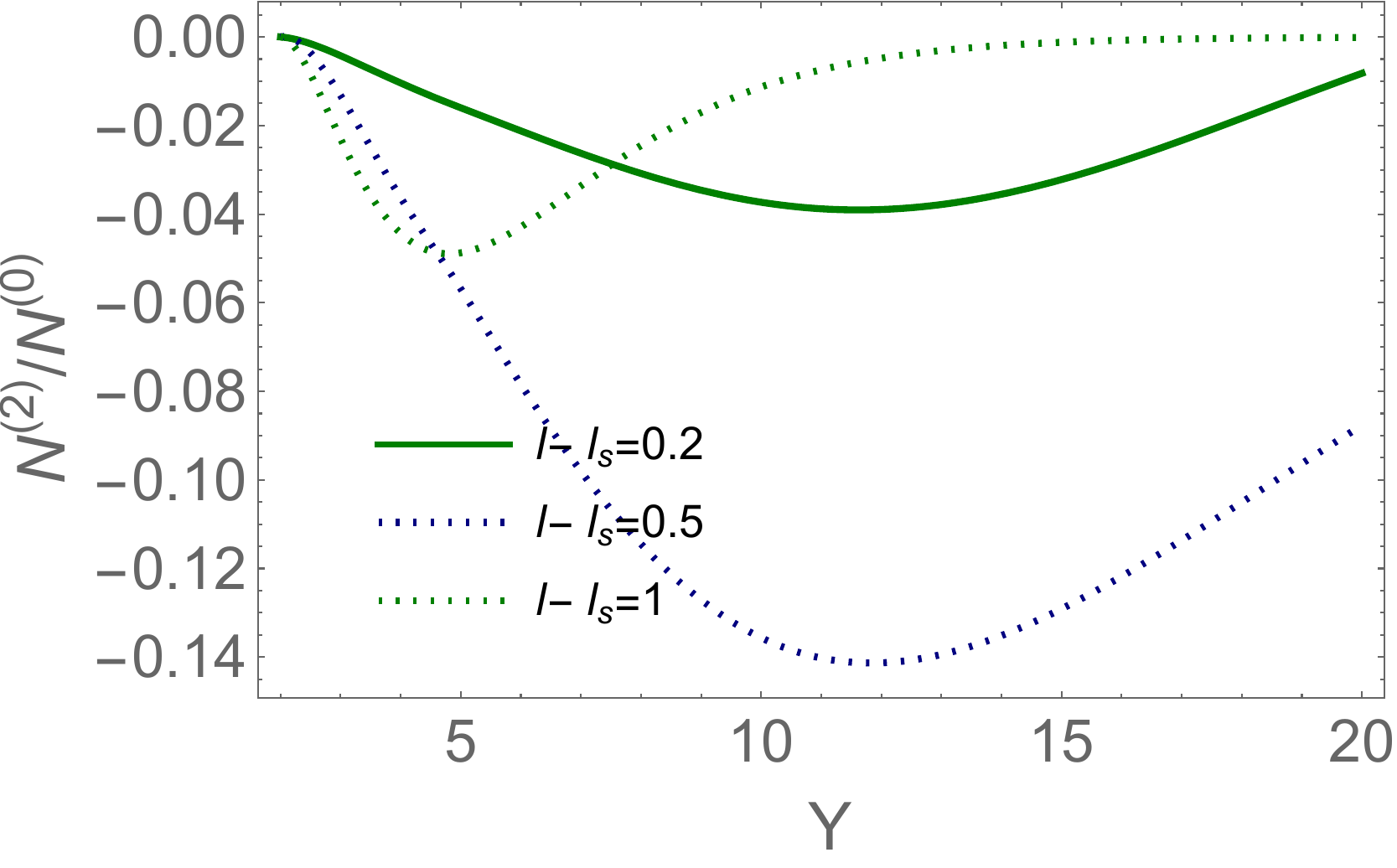}&\includegraphics[width=9.5cm]{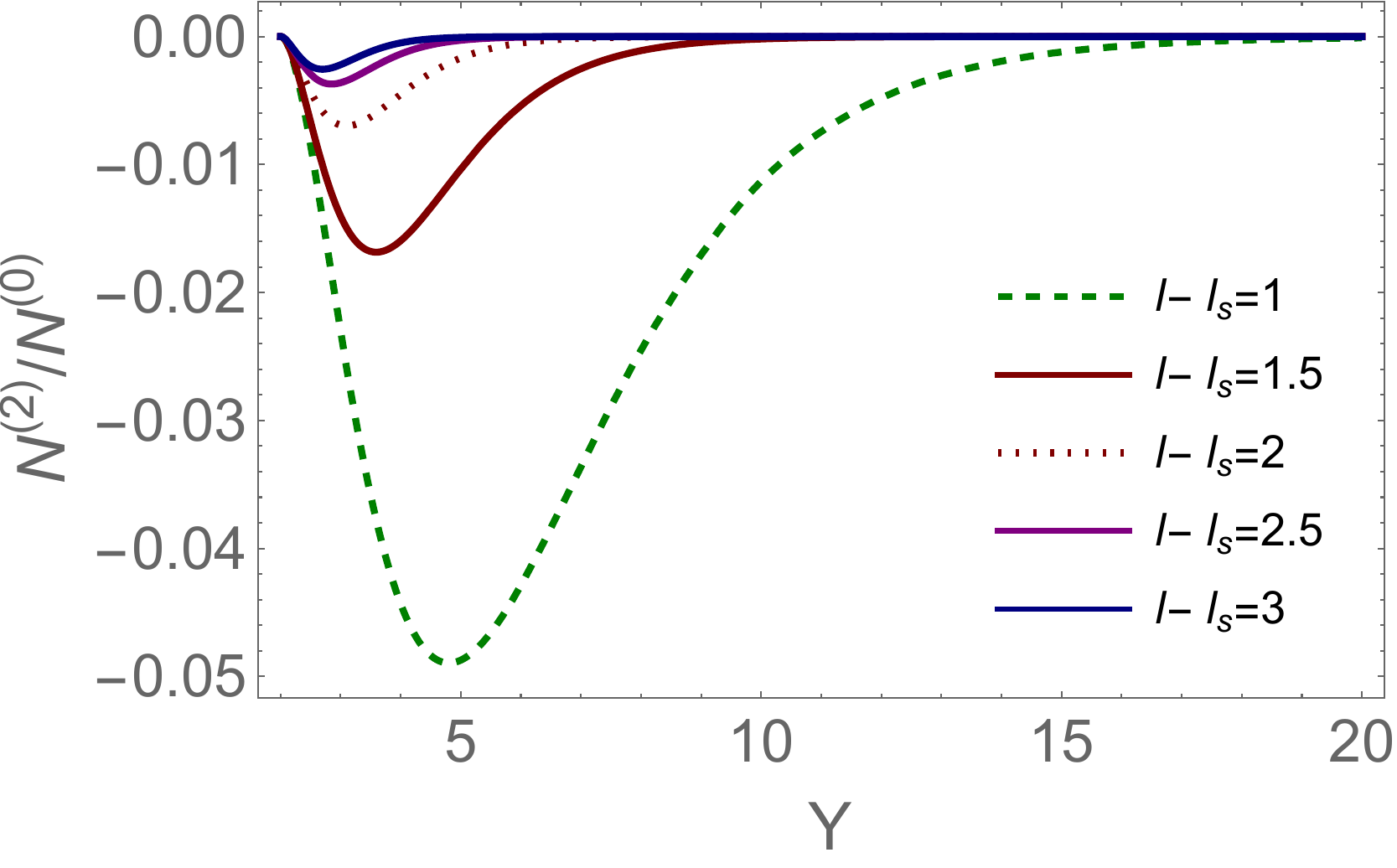} \\
		\fig{newit1}-a&\fig{newit1}-b\\
				\end{tabular}	
\end{center}
	
 	\caption{\fig{newit1}-a:$\frac{N^{(2)}\Lb Y,l-l_s\Rb}{N^{(0)}\Lb Y,l-l_s\Rb}$ (see caption of \fig{thi}) for the general BFKL kernel in the alternative approach (see \eq{C10}) versus $Y$ at fixed small values of $l - l_s$. \fig{newit1}-b: The same  at large values of $l - l_s$.
$\Omega_0 =0.25, \bar{\gamma}=0.63$.}		
\label{newit1}
 \end{figure}

\begin{figure}[h]
 	\begin{center}
 	\leavevmode
\begin{tabular}{c c}
		\includegraphics[width=9.1cm]{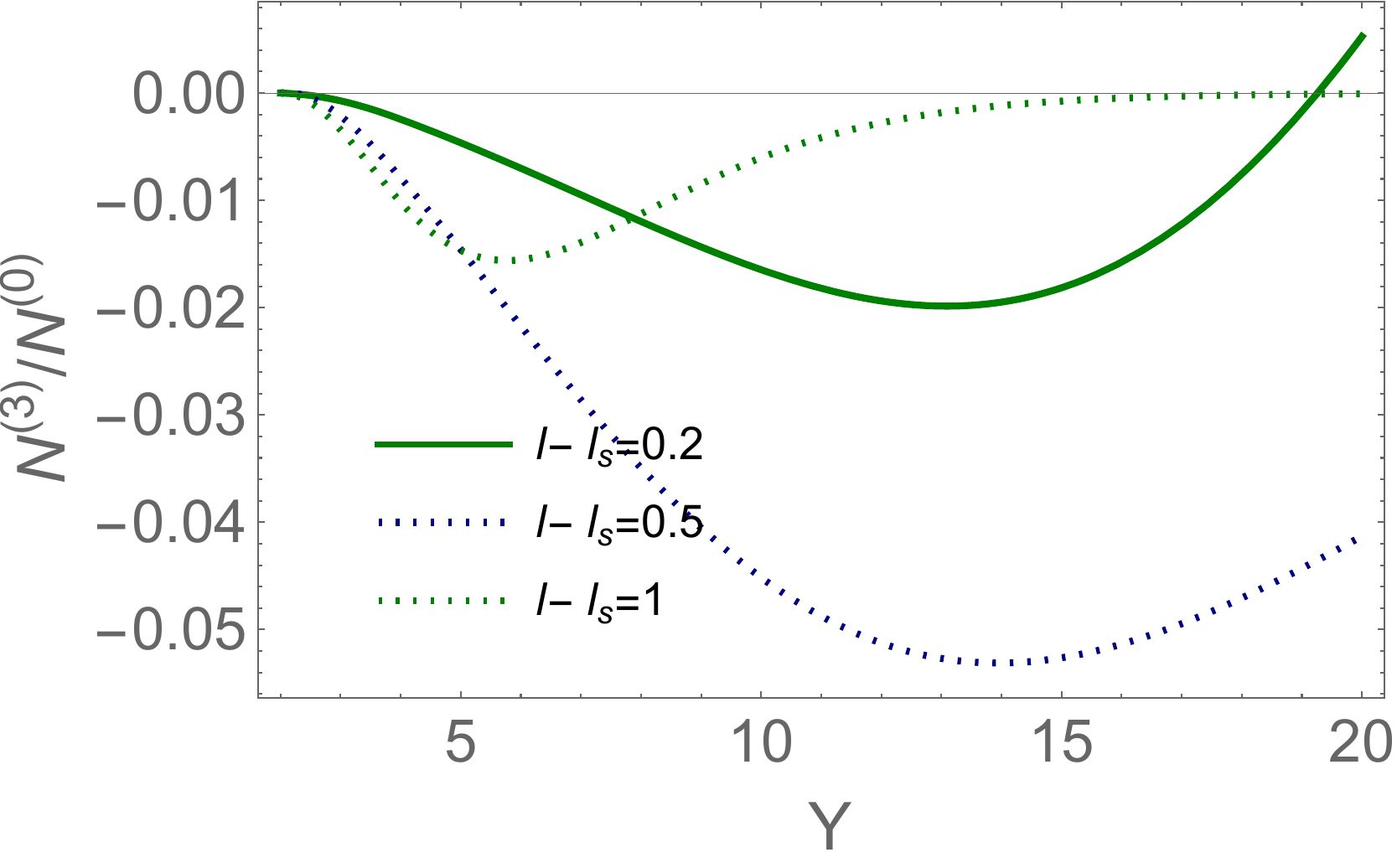}&\includegraphics[width=9.5cm]{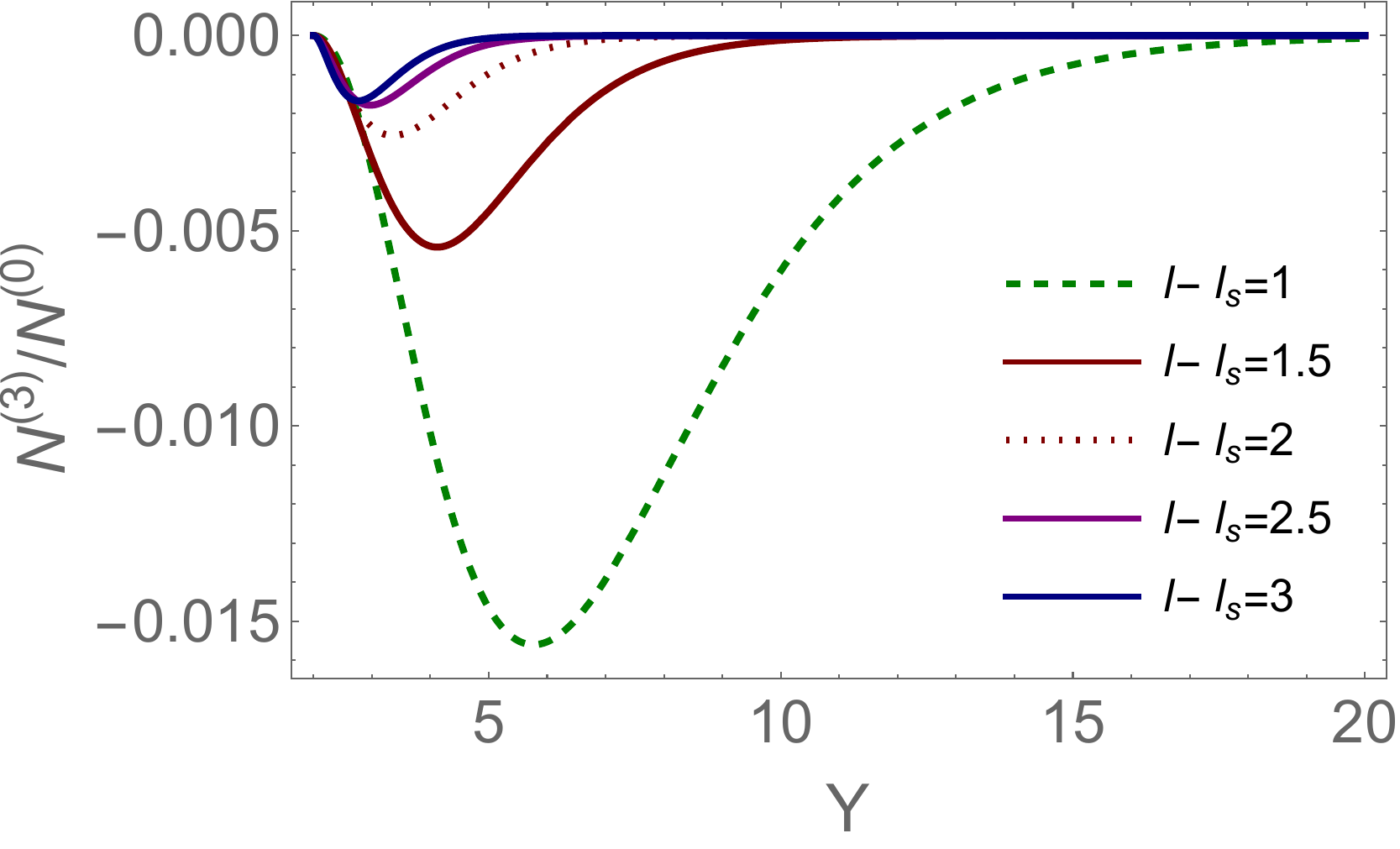} \\
		\fig{newit2}-a&\fig{newit2}-b\\
				\end{tabular}	
\end{center}
	
 	\caption{\fig{newit2}-a:$\frac{N^{(3)}\Lb Y,l-l_s\Rb}{N^{(0)}\Lb Y,l-l_s\Rb}$ (see caption of \fig{thi}) for the general BFKL kernel in the alternative approach (see \eq{C10}) versus $Y$ at fixed small values of $l - l_s$. \fig{newit2}-b: The same  at large values of $l - l_s$.
$\Omega_0 =0.25, \bar{\gamma}=0.63$.}		
\label{newit2}
 \end{figure}

\begin{figure}[h]
 	\begin{center}
 	\leavevmode
\begin{tabular}{c c}
		\includegraphics[width=9.1cm]{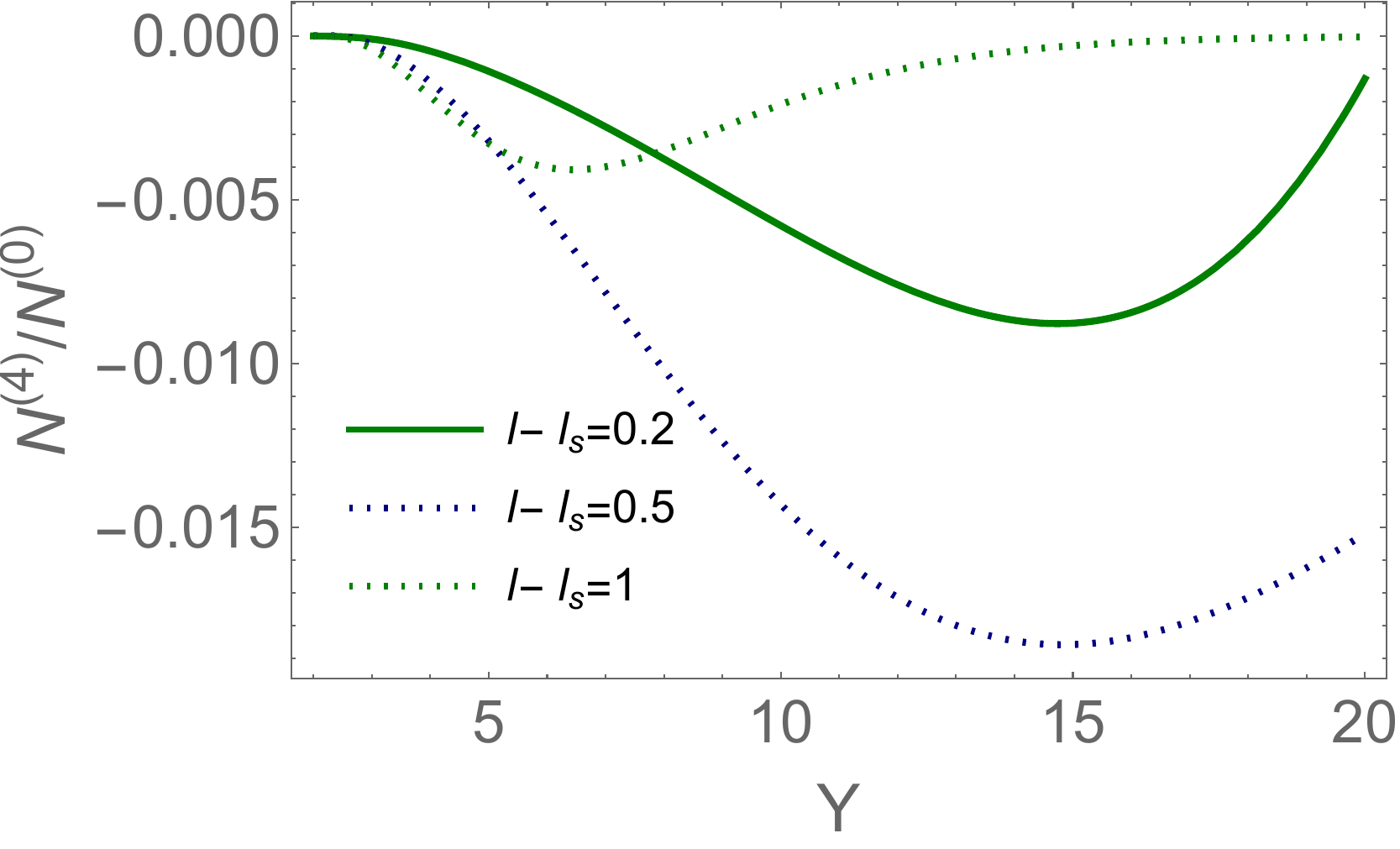}&\includegraphics[width=9.5cm]{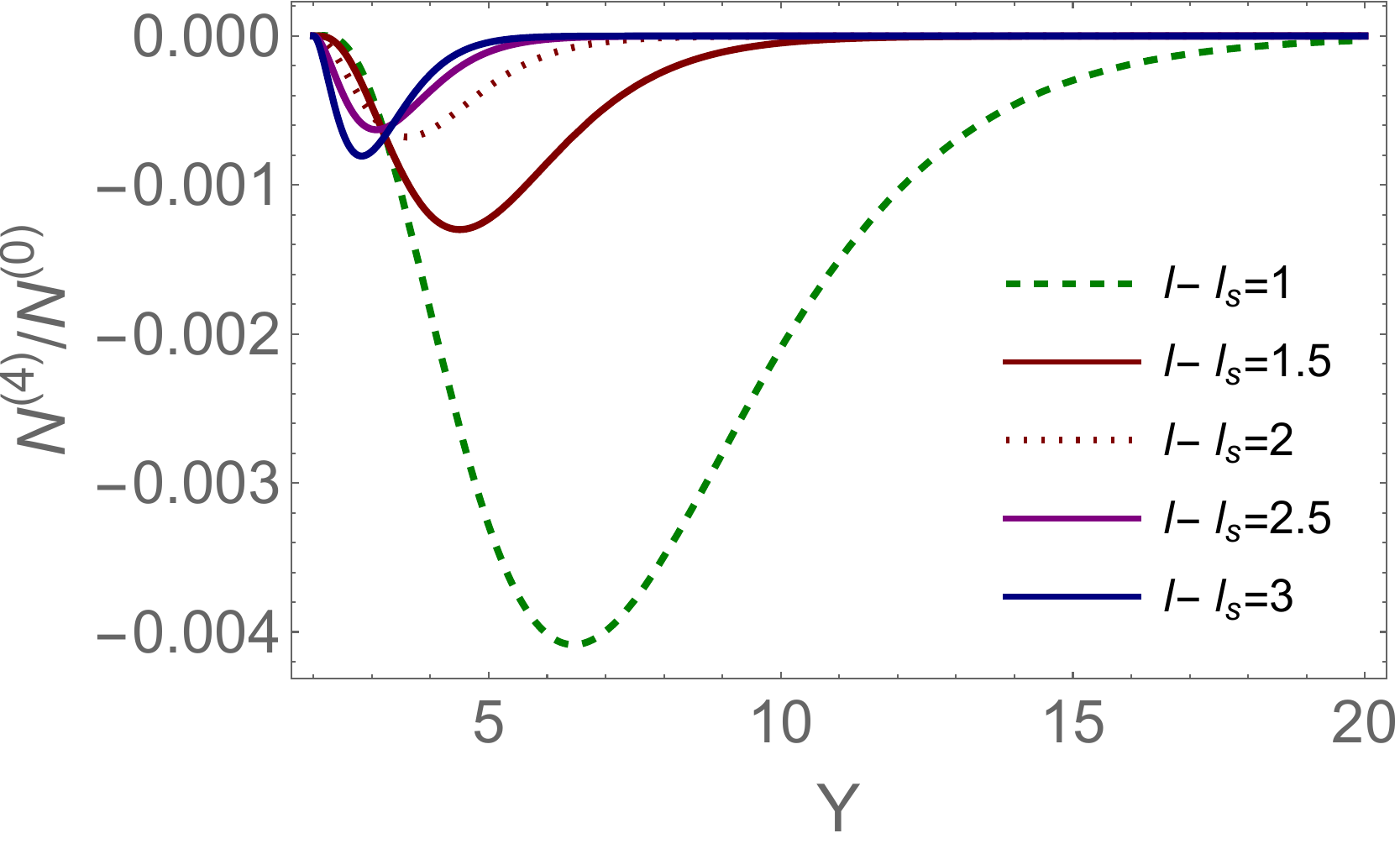} \\
		\fig{newit3}-a&\fig{newit3}-b\\
				\end{tabular}	
\end{center}
	
 	\caption{\fig{newit3}-a:$\frac{N^{(4)}\Lb Y,l-l_s\Rb}{N^{(0)}\Lb Y,l-l_s\Rb}$ (see caption of \fig{thi}) for the general BFKL kernel in the alternative approach (see \eq{C10}) versus $Y$ at fixed small values of $l - l_s$. \fig{newit3}-b: The same  at large values of $l - l_s$.
$\Omega_0 =0.25, \bar{\gamma}=0.63$.}		
\label{newit3}
 \end{figure}

In \fig{newit1} we plot the contribution of the third iteration. One can see that our procedure works and  the third iteration improves the accuracy.  However, for small values of $l - l_s \leq 1$ we certainly need  to calculate the third iteration and,   perhaps, the fourth iteration has to be included.  In \fig{newit2} and \fig{newit3} we plotted the contribution of the third and fouth iterations. One can see that our procedure works.

For large values of $l-l_s\geq 1$ the third iteration leads to the scattering amplitude within several percent accuracy ($ \leq 4\%$). At first sight it looks as we need more iterations than for the leading twist BFKL kernel. However, this is misleading impression since for the leading twist BFKL kernel we also need  four iteration as it has been discussed in section III-C (see \fig{omegazata}).

\section{Conclusions }
 \paragraph{Main results:}
In the paper we developed the homotopy approach to the nonlinear Balitsky-Kovchegov equation for the running QCD coupling. As has been mentioned in the introduction our approach consists of two stages. The first one is the analytic solution to the nonlinear equation with a simplified BFKL kernel which has only the leading twist contributions. We found the analytic function for the scattering amplitude which satisfies the initial and boundary conditions. The second step was to find the corrections to this solution. The paper contains two investigation of these corrections: the detailed analysis of them in the simplified leading twist BFKL kernel and the development of the iteration procedure for the general BFKL kernel.

\subsection{Leading twist BFKL kernel}

The substantial part of this paper is devoted to the investigation of the leading twist BFKL. The motivation for such approach is very simple: this kernel shows the largest difference between the fixed and running QCD coupling.

 For the leading twist BFKL kernel we gave the detailed description of the  corrections to our analytic solution. 
It is instructive to note that we calculate  them using  numerical estimates but the equation, that we solved numerically, depends only on analytic solution and did not bring any uncertainties. We demonstrated that the analytic first iteration leads the  solution within 5\% accuracy and calculating the second iteration we increased the accuracy to $\leq 1\%$.

\begin{figure}[h]
 	\begin{center}
 	\leavevmode
 		\includegraphics[width=11cm]{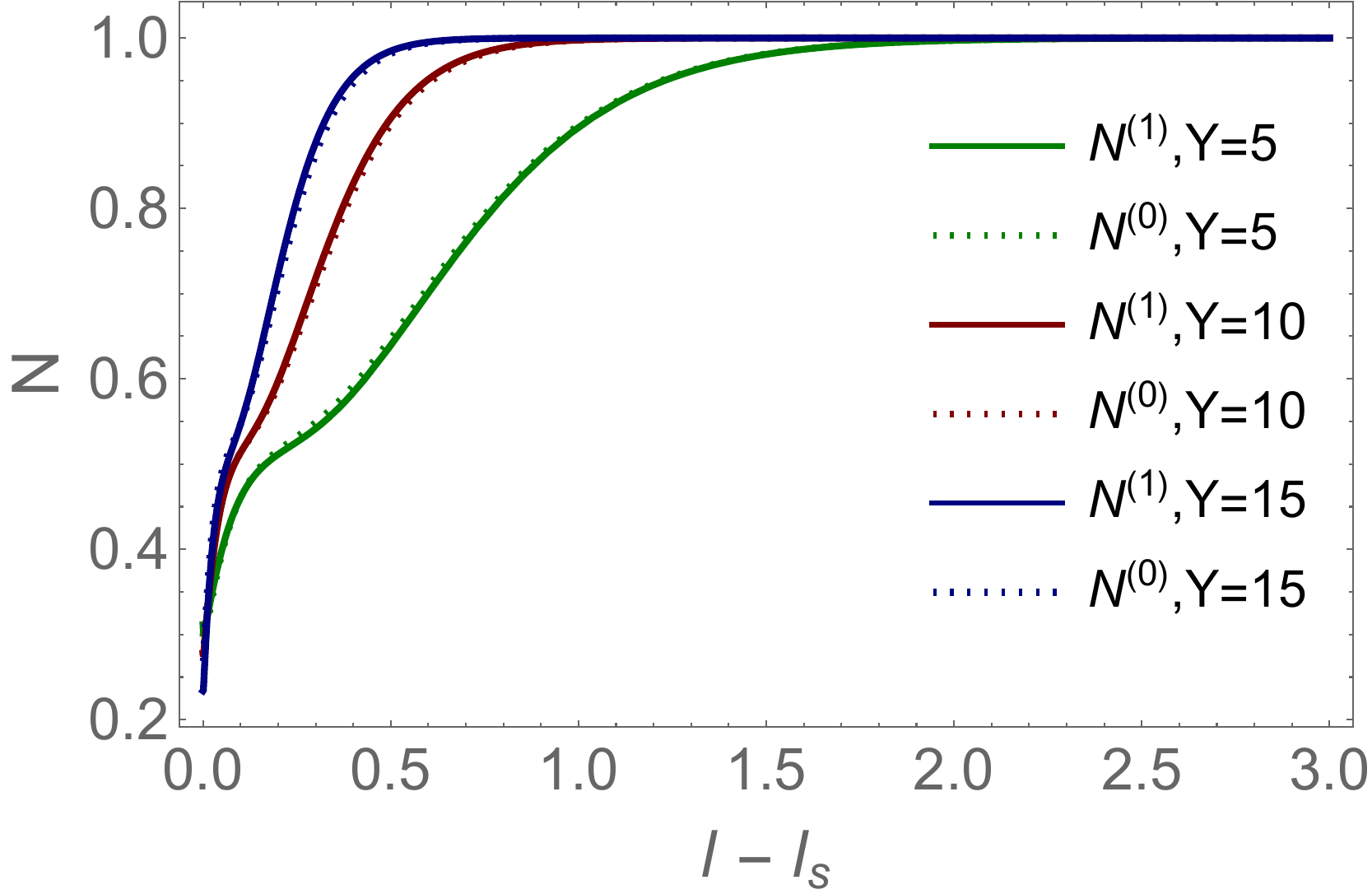}
 	\end{center}
 	\caption{The scattering amplitude in the first and second iterations for the leading twist BFKL kernel.}
 	\label{omega}
 \end{figure}

\fig{omega}  shows the scattering amplitude in the first and second iterations. One can see that we need to account for the second iteration for small $Y$ and $l - l_s$.

  
\begin{figure}[h]
 	\begin{center}
 	\leavevmode
	\begin{tabular}{c c }
 		\includegraphics[width=8.5cm]{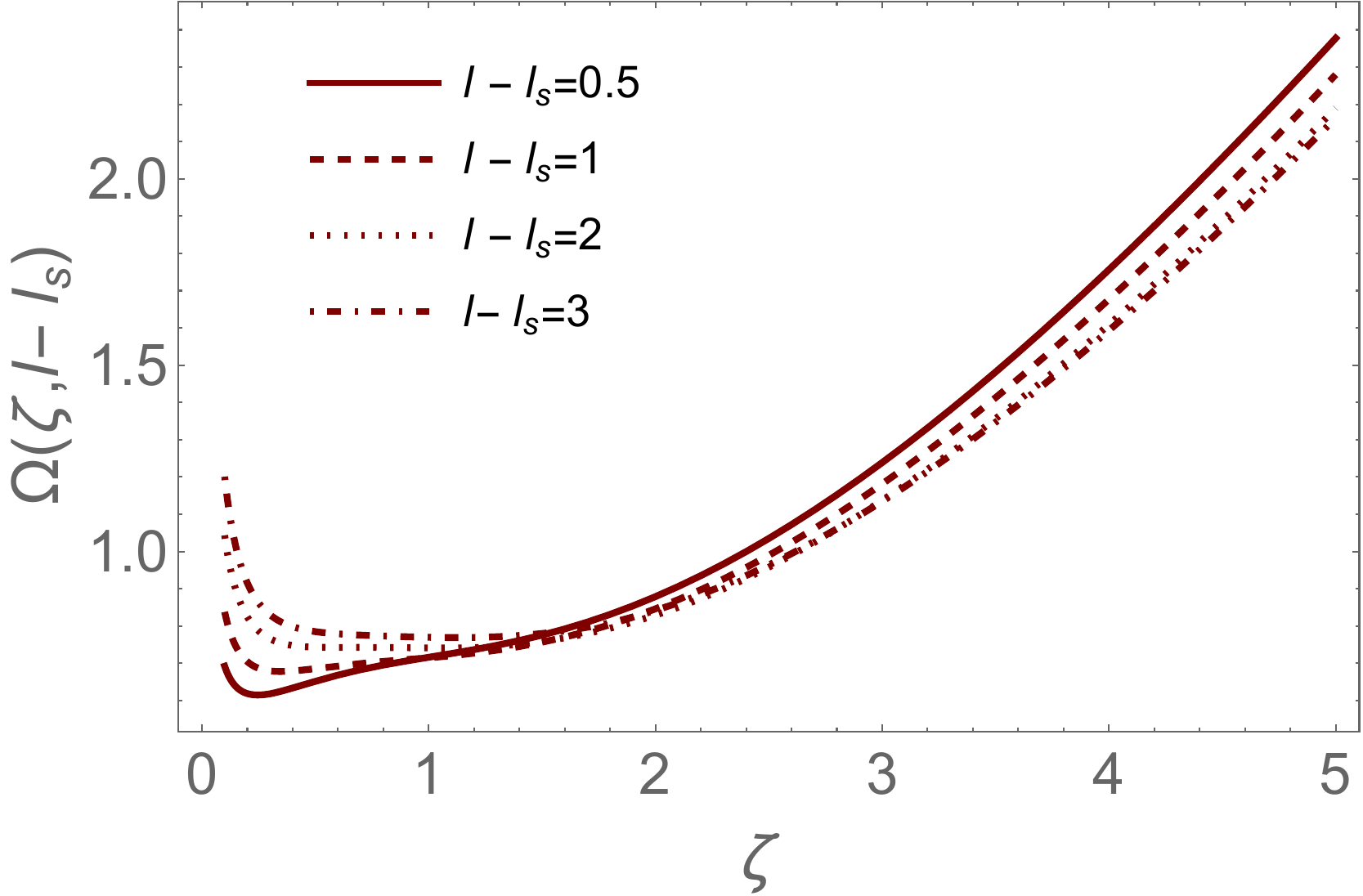}&\includegraphics[width=8.7cm]{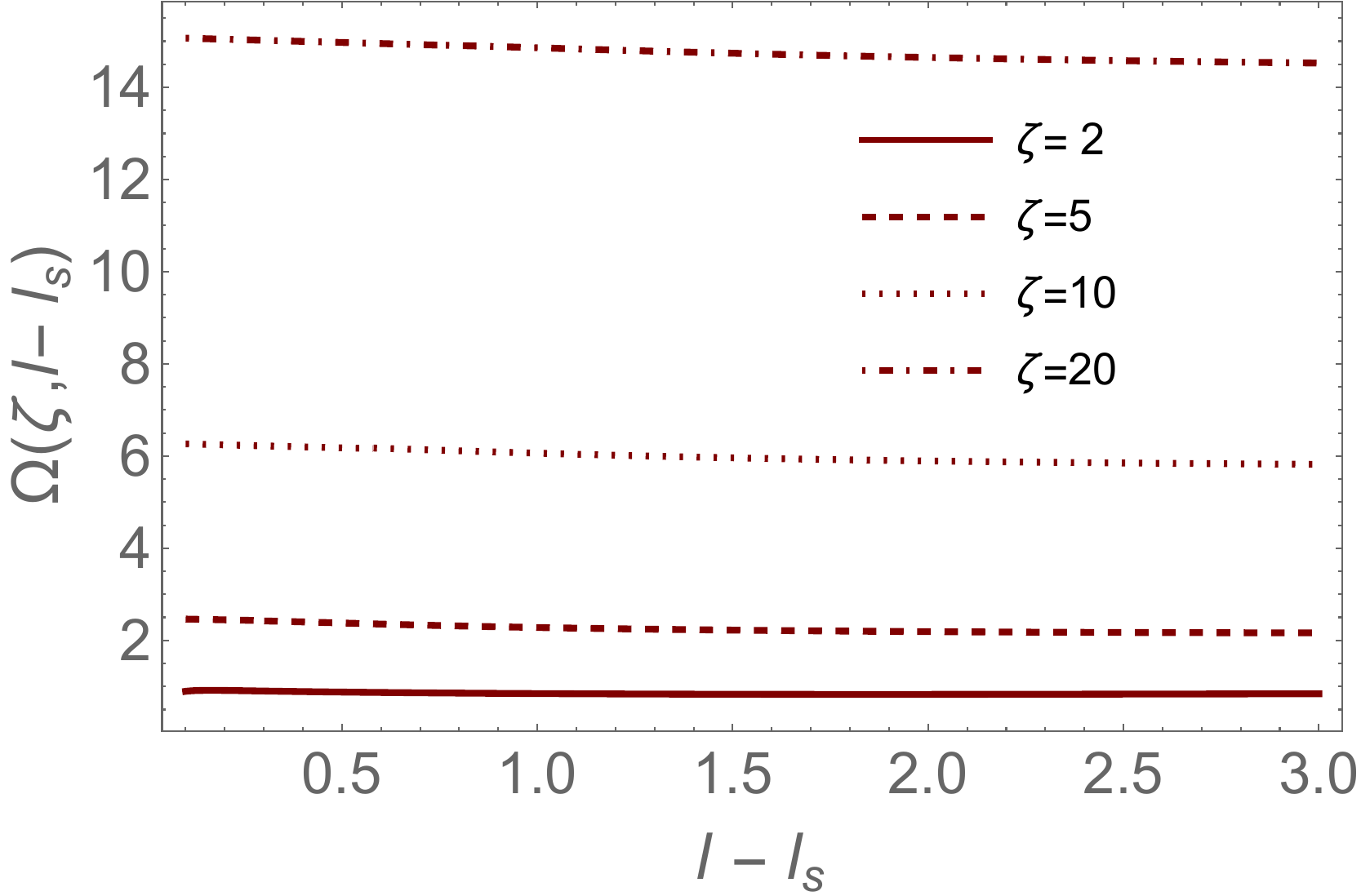} \\
		\fig{zsc}-a & \fig{zsc}-b\\
		\end{tabular}
			\end{center}
 	\caption{ $\zeta = Y (l - l_s)$ scaling behaviour for $\Omega\Lb \zeta, l-l_s\Rb = \,\Omega^{(0)}\Lb Y, l-l_s\Rb + \Omega^{(1)}\Lb Y, l-l_s\Rb  + \Omega^{(2)}\Lb Y, l-l_s\Rb	$. 
}		
\label{zsc}
 \end{figure}

      \paragraph{$\mathbf{\zeta}$ scaling:} 
     \eq{C4} as well as the solutions that have been discussed in sections I II-C and D,   depend on one variable $\zeta = Y (l - l_s)$ and therefore, show the $\zeta$ scaling behaviour. However this behaviour cannot  satisfy the initial and boundary conditions and we have to add some violation of $\zeta$ scaling behaviour at low $l - l_s$. In   \fig{zsc} the $\Omega\Lb \zeta, l-l_s\Rb $ is plotted. One can see that (i) at large $\zeta$ and $l - l_s$ we have $\zeta$ scaling behaviour, but (ii) at low $\zeta$ and $l - l_s$ this behaviour is violated.
Note, that this violation is larger at lower values of $\zeta$.
 

      \paragraph{Geometric scaling:} 
     
\begin{figure}[h]
 	\begin{center}
 	\leavevmode
	\begin{tabular}{c c }
 		\includegraphics[width=8.5cm]{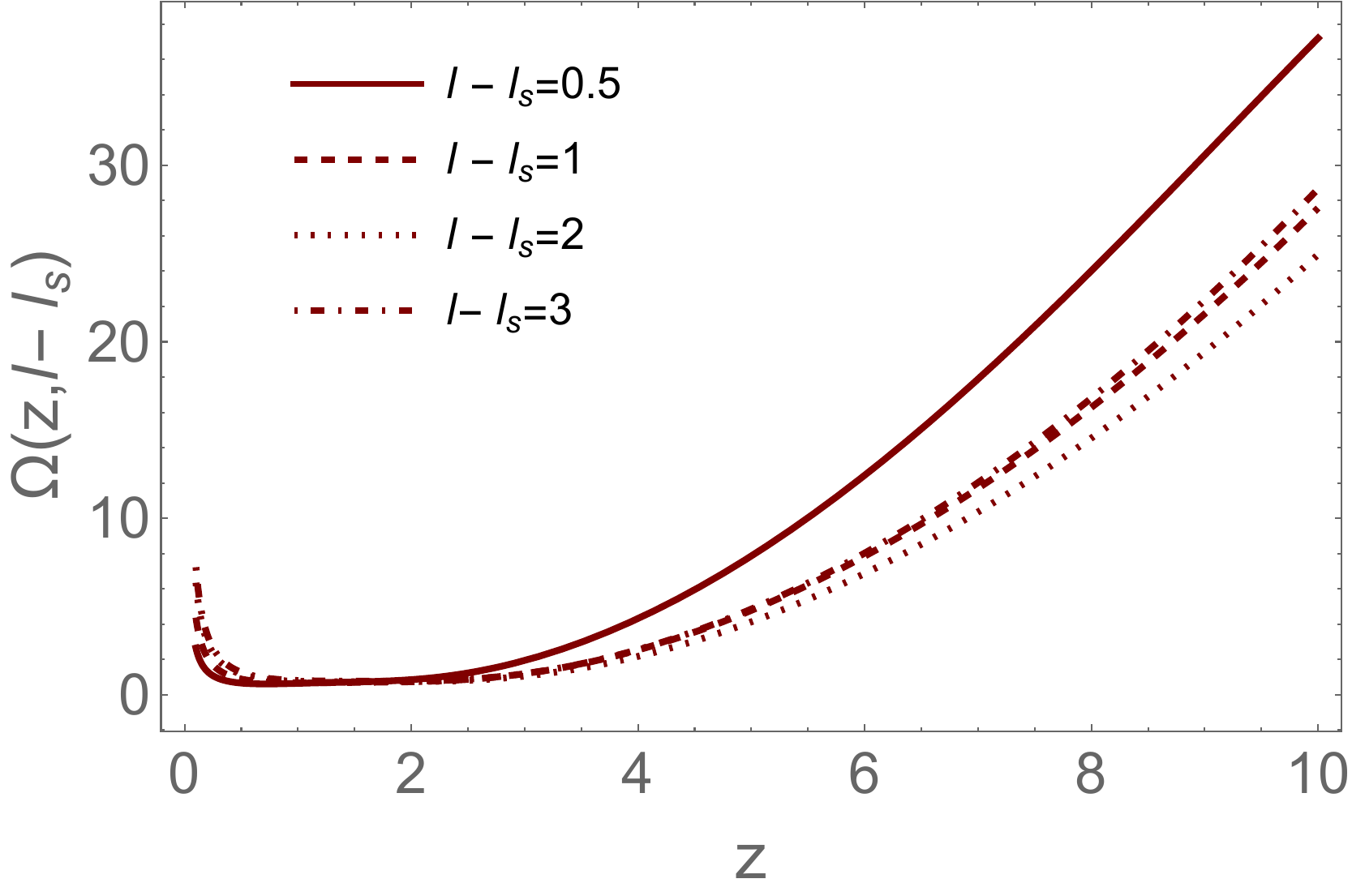}&\includegraphics[width=9.3cm]{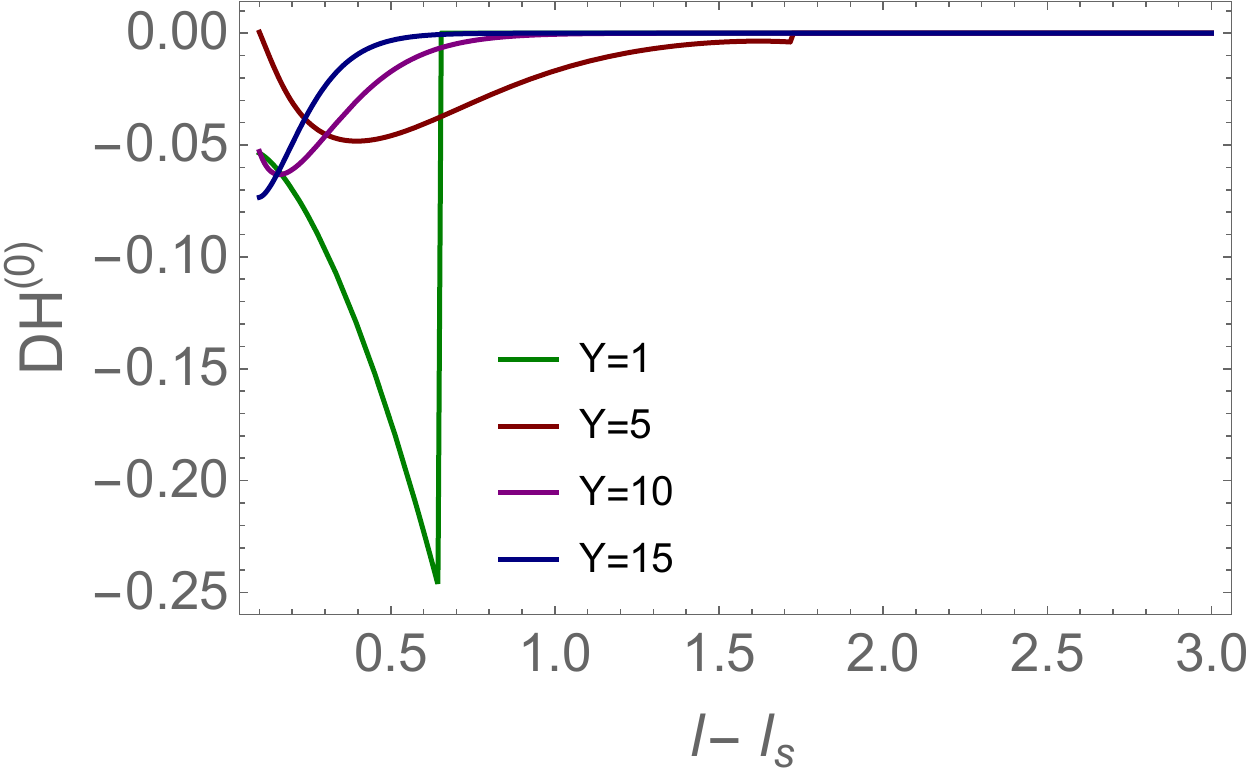} \\
		\fig{zetasc}-a & \fig{zetasc}-b\\
		\end{tabular}
			\end{center}
 	\caption{\fig{zetasc}-a: $z = \xi_s + \xi$  scaling behaviour for $\Omega\Lb \zeta, l-l_s\Rb = \,\Omega^{(0)}\Lb Y, l-l_s\Rb + \Omega^{(1)}\Lb Y, l-l_s\Rb + \Omega^{(2)}\Lb Y, l-l_s\Rb	$. \fig{zetasc}-b: The nonhomogeneous term of \eq{SI2} $DH^{(0)}\Lb Y,l-l_s\Rb$ versus $l - l_s$ at fixed $Y$ with the cut.}	
\label{zetasc}
 \end{figure}

     We wish to emphasize that    we do not see any reason for the geometric scaling behaviour of $\Omega$ for the running QCD coupling.  On the other hand,  we have argued in section III-A  (see \eq{PL5} - \eq{PL7})  that the geometric scaling behaviour could be in the vicinity of the saturation scale.   
Indeed, \fig{zetasc} show an approximate $z$-scaling behaviour in the limited  region of $z$. In particular, one can see that the geometric scaling behaviour holds for small $l-l_s$  in rather wide region of $z = \xi_s + \xi$.

      We wish to emphasize that  the leading twist kernel demonstrates the most pronounced    difference between the fixed QCD coupling with $z$ scaling  and the running QCD coupling with $\zeta$ scaling behaviours.
             
        \paragraph{Infrared cutoff:} 
 As we have discussed, we need to introduce an infrared cutoff to avoid the Landau pole in the running QCD coupling, since the nonperturbative methods of QCD have not solved the problem of  long distance behaviour of the QCD coupling. In this paper we used this cutoff $r_{max} = 1/\Lambda_{QCD}$. We demonstrated that the contributions of the distances of the order of $1/\Lambda_{QCD}$ in our equations can be neglected. We wish only to note that we can introduce any long distance behaviour of the running QCD coupling which could come from nonperturbative QCD or related phenomenology. In particular, there is a phenomenological reason\cite{IFRCUT,KOVCUT}  to cut long distances at $\bas\Lb r_{max}\Rb = \h$. In \fig{zetasc}-b one can see the nonhomogeneous term in \eq{SI2}  with this cut. Comparing this figure with \fig{dh}-a we see that the difference in the infrared cut off change $DH^{(0)}\Lb Y,l-l_s\Rb$ only at small values of $Y$.

\subsection{Solution to the equation with the general BFKL kernel}

\begin{figure}[h]
 
 	\begin{center}
 	\leavevmode
	\begin{tabular}{c c }
 		\includegraphics[width=8.5cm]{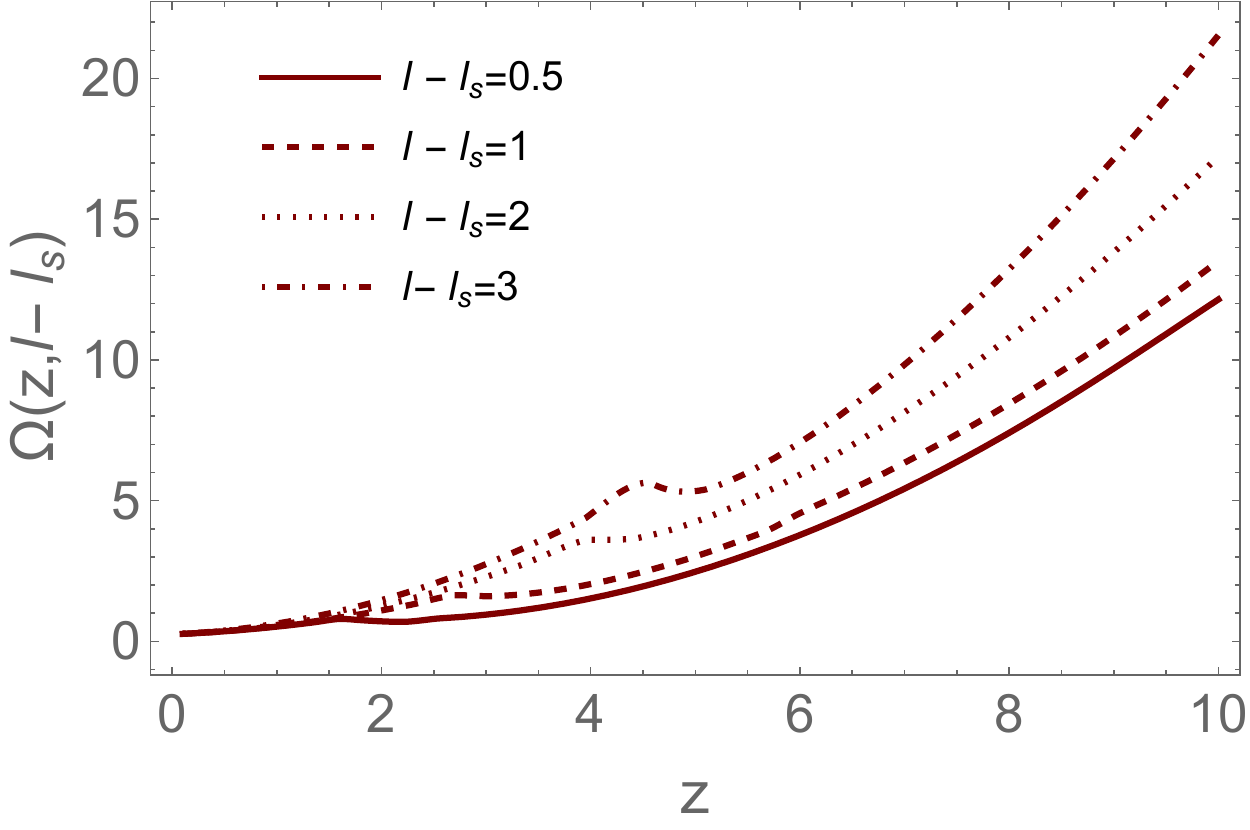}&\includegraphics[width=8.5cm]{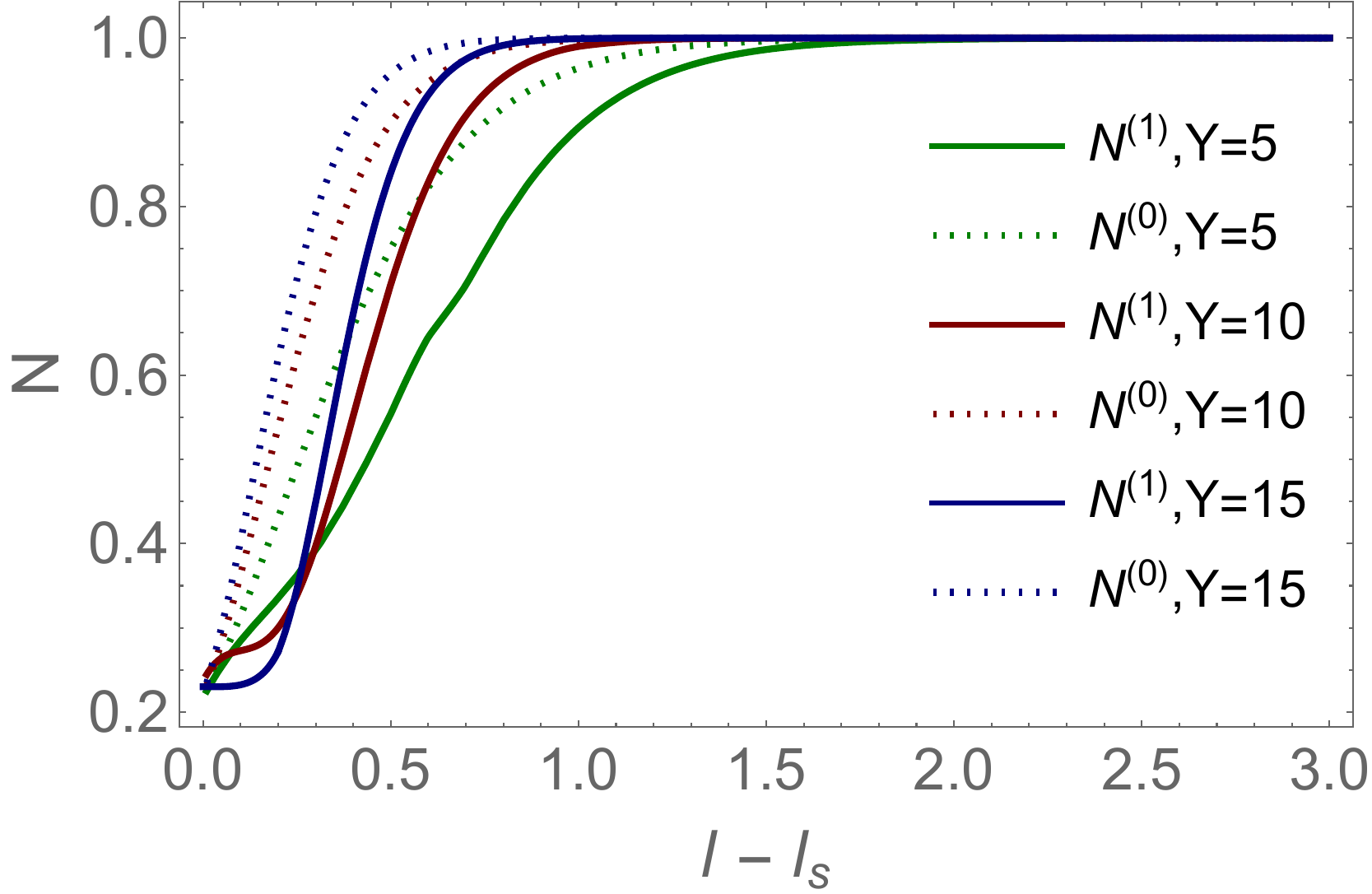} \\
		\fig{zetascf}-a & \fig{zetascf}-b\\
		\end{tabular}
			\end{center}
 	\caption{\fig{zetascf}-a: $z = \xi_s + \xi$  scaling behaviour of $\Omega\Lb \zeta, l-l_s\Rb = \,\Omega^{(0)}\Lb Y, l-l_s\Rb + \Omega^{(1)}\Lb Y, l-l_s\Rb $  for the general BFKL kernel. \fig{zetascf}-b: The scattering amplitude with the general BFKL kernel in the first and second iterations}
 	\label{zetascf}
 \end{figure}

 The major part of this paper is devoted to the homotopy approach with the simplified, leading twist, BFKL kernel, since the main features of the running QCD coupling can be seen for this model in the clearest way.
 For the general BFKL kernel we demonstrated that our approach works well, considering two cases: (i) the first homotopy iteration is the nonlinear equation with the leading twist BFKL kernel; and (ii) the first iteration is the solution to the nonlinear BK equation at large values of $Y$.  We demonstrated that for the general BFKL kernel our second approach gives the most economic way to describe large $Y$ region. Indeed, for large values of $Y$ we can use only the first iteration which has simple analytic solution. It is instructive to note that we can use the asymptotic solution of \eq{C4} in this region. However  \fig{zetascf}-a  shows that the $\zeta$-scaling behaviour is strongly violated for the general BFKL kernel. Only for large values of $\zeta$ we see that  $\Omega$ is independent of $l - l_s$. The  $z$ scaling behaviour of the scattering amplitude (see \fig{zsc}) have the same patterns as for the leading twist BFKL kernel.
 
On the other hand for moderate value of $Y$ ($Y = 1-7$) the simple approach we can trust only outside the vicinity of saturation scale at $l - l_s > 1$. For smaller $l - l_s$ we have to take into account the second and, perhaps, even the third iterations. 
The behaviour of the scattering amplitude for the general BFKL kernel one can see in 
\fig{zetascf}-b.  This figure shows that corrections related to higher iterations are rather large.

We consider as one of the results of the paper that the leading twist BFKL kernel which models sufficiently well the nonlinear evolution for the freezed QCD coupling, does not work so well for the running QCD coupling. Indeed, we have seen that the solution to the general nonlinear equation does not reproduce the $zeta$ scaling, which is the inherent feature of the leading twist BFKL kernel.

  \begin{figure}[h]
 	\begin{center}
 	\leavevmode
	\begin{tabular}{c c }
 		\includegraphics[width=8.5cm]{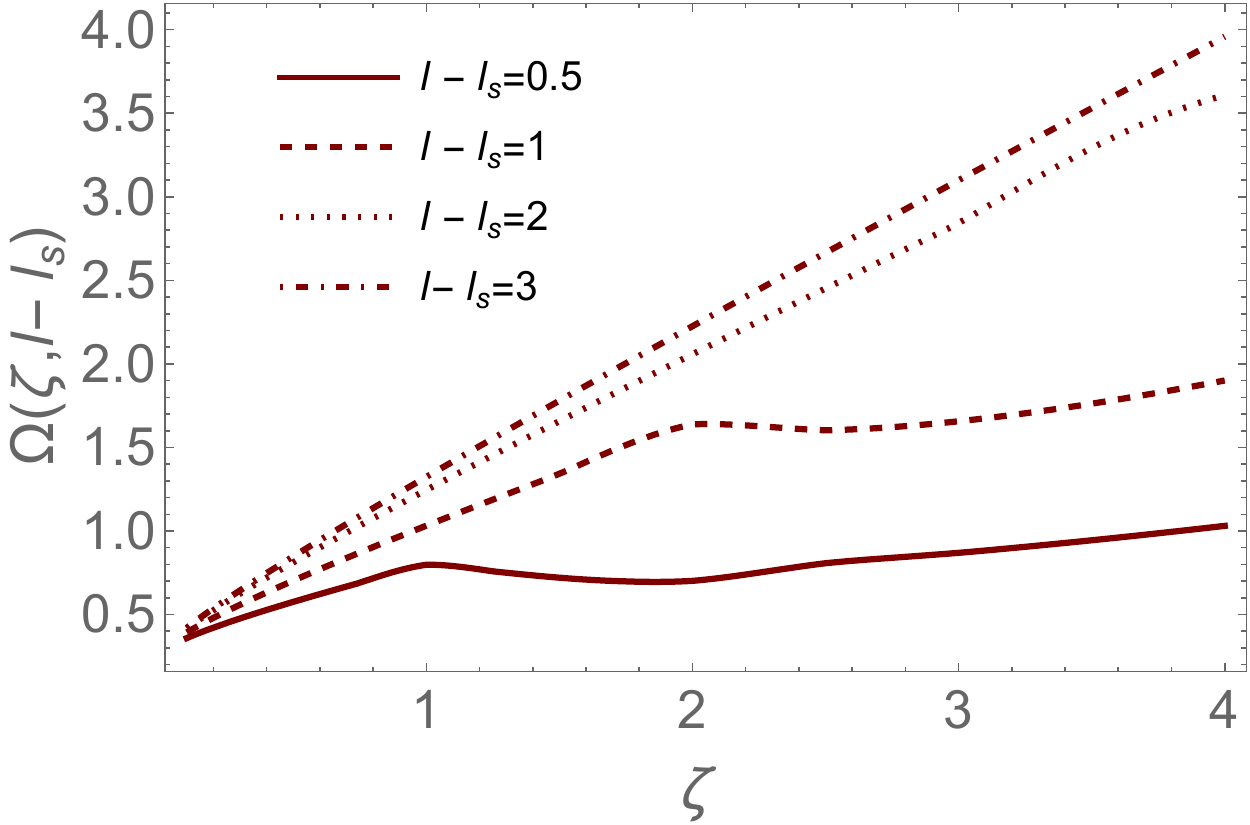}&\includegraphics[width=8.5cm]{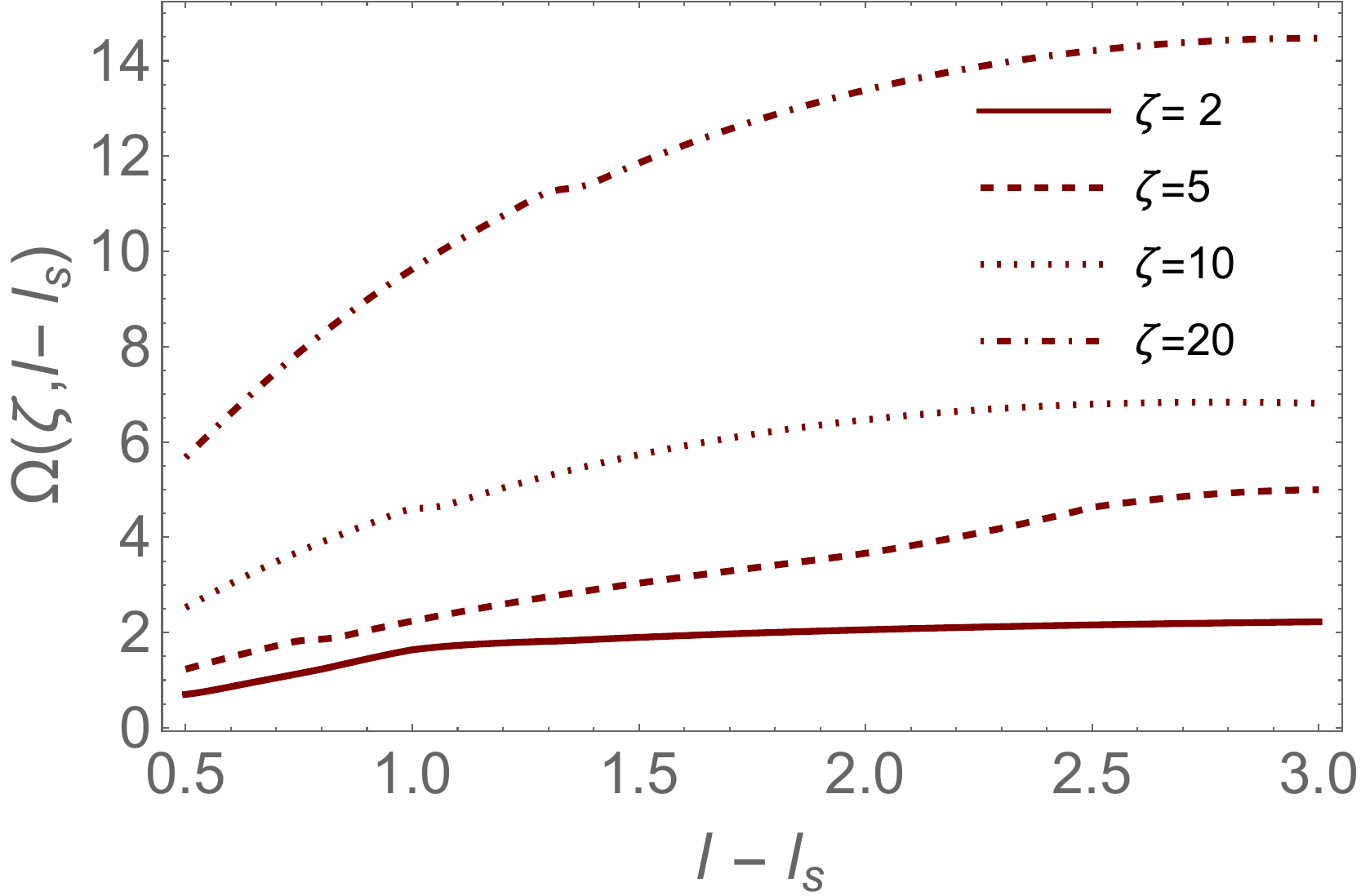} \\
		\fig{zscf}-a & \fig{zscf}-b\\
		\end{tabular}
			\end{center}
 	\caption{$\zeta = Y (l - l_s)$ scaling behaviour for $\Omega\Lb \zeta, l\Rb = \,\Omega^{(0)}\Lb Y, l-l_s\Rb + \Omega^{(1)}\Lb Y, l-l_s\Rb $ for the general BFKL kernel }	
\label{zscf}
 \end{figure} 	

      \subsection{Concluding remarks:} 
  We wish to recall that we consider the interaction of the dipole with the small size with the proton, which we model as the dipole with the size $R = 1/\Lambda_{QCD}$. Our matching with perturbative QCD is determined by the solution of the linear BFKL equation at small dipole sizes. The initial condition for the BFKL equation is the Born approximation of perturbative QCD: the exchange of two gluons. In this paper we do not consider the initial condition of 
  McLerran-Venugopalan type \cite{MV}. This our reader has to bear in mind comparing our results with the numerical estimates in other papers (see, for example, Refs.\cite{KOLEB,KOVCUT,AAMS}). The influence of the McLerran-Venugopalan initial condition, which are correct for the  dipole-nucleus scattering, we will consider in a separate paper.

 We believe that the homotopy approach could be useful for treatment of the nonlinear dynamics in QCD and for accounting the effects related to the running QCD coupling.

  \section{Acknowledgements}
We thank our colleagues at Tel Aviv University and UTFSM for encouraging discussions.  This research was supported by Fondecyt (Chile) grants No. 1231829 and  1231062.
J.G. express his gratitude to the PhD scholarship USM-DP No. 029/2024 for the financial support.  J.G. would also like to thank the organizers of the long-term workshop Hadrons and Hadron Interactions in QCD 2024 (HHIQCD 2024) at the Yukawa Institute for Theoretical Physics (YITP-T-24-02), Kyoto, Japan, where part of this work was completed.
  \appendix
  
 \section{The general BFKL kernel:  formulae for the second iteration}
Using the identity :
$$ \frac{1}{r_1^2\,r^2_2}\,=\, \frac{ 2}{r^2_1\,(r^2_1+r^2_2)} $$
we can rewrite \eq{SIG2} in the form:
\bea \label{A1}
\mathscr{N_{L}}[\Omega^{(0)}] & =&\underbrace{ -\bas\Lb r^2\Rb \!\!\!\!\!\!\!\!\!\!\!\!\intl^{1/\Lambda_{QCD}}_{\begin{subarray}{l} ~~~~~~ \xi_s\\ r_{2} \,\gg\,1/Q_s(Y)\\
r_{1}\,\gg\,1/Q_s(Y)\end{subarray}}\!\!\!\!\!\!\!\!\!\!\!\! \frac{d \phi d r^2}{2 \pi}\,\Bigg\{\frac{r^2}{r^2_1\,(r^2_1+r^2_2)} \,\,+\,\,\frac{1}{r^2_1} \frac{\ln\Lb r^2_2/r^2_1\Rb}{\ln r^2_1} \Bigg\}  \exp\Lb \,-\,\,
 \Omega^{(0)}\Lb r_1,Y\Rb\,
 -\,  \Omega^{(0)}\Lb r_2 ,Y\Rb \Rb}_{\mbox{$\mathscr{N'_{L}}[\Omega^{(0)}]$}} 
  \nn\\
 &+ &\underbrace{\intl^\xi_{\begin{subarray}{l} ~~~~~~ \xi_s\\ r_{2} \,\gg\,1/Q_s(Y)\\
r_{1}\,\gg\,1/Q_s(Y)\end{subarray}}\!\!\!\!\!\!\!\!\!\!\!\! \bas(r_1^2) \frac{d r^2_1}{r^2_1} \exp\Lb \,-\,\,
 \Omega^{(0)}\Lb r_1,Y\Rb\,-\,  \Omega^{(0)}\Lb r ,Y\Rb \Rb}_{\mbox{$\mathscr{N''_{L}}[\Omega^{(0)}]$}} \eea

Note that $r^2_2 = r^2 + r^2_1 - 2  r_1 r \cos\phi$ and the identity used the symmetry between $r_1$ and $r_2$ in the integration.

The equation for $\Omega^{(1)}$ has the same form as \eq{SI2}. One can see that $\dfrac{\pp}{\pp\,l}\Lb  e^{ \Omega^{(0)}\Lb Y,l-l_s\Rb} \mathscr{N_{L}}[\Omega^{(0)}] \Rb$ from the second term in \eq{A1} can be found and it is equal to
\beq \label{A2}
\dfrac{\pp}{\pp\,l}\Lb  e^{ \Omega^{(0)}\Lb Y,l-l_s\Rb} \mathscr{N''_{L}}[\Omega^{(0)}] \Rb\,\, =\,\bas(\xi)  \exp\Lb -\,  \Omega^{(0)}\Lb r ,Y\Rb \Rb\dfrac{d \xi}{d l} =\exp\Lb -\,  \Omega^{(0)}\Lb r ,Y\Rb \Rb
\eeq
Using \eq{A2} we obtain \eq{SIG3} for the second iteration.
 \section{The general BFKL kernel:  formulae for the third iteration}
\eq{SIG4} can be rewritten in the following form using \eq{SIG2}:
\bea \label{B1}
\mathscr{N_{L}}[\Omega^{(0)}+ \Omega^{(1)}] & =&-\bas\Lb r^2\Rb \!\!\!\!\!\!\!\!\!\!\!\!\intl^{1/\Lambda_{QCD}}_{\begin{subarray}{l} ~~~~~~ \xi_s\\ r_{2} \,\gg\,1/Q_s(Y)\\
r_{1}\,\gg\,1/Q_s(Y)\end{subarray}}\!\!\!\!\!\!\!\!\!\!\!\! \frac{d \phi d r^2}{2 \pi}\,\Bigg\{\frac{r^2}{r^2_1\,(r^2_1+r^2_2)} \,\,+\,\,\frac{1}{r^2_1} \frac{\ln\Lb r^2_2/r^2_1\Rb}{\ln r^2_1} \Bigg\}  \exp\Lb \,-\,\,
 \Omega^{(0)}\Lb r_1,Y\Rb\,
 -\,  \Omega^{(0)}\Lb r_2 ,Y\Rb \Rb\nn\\
 &\times& \Lb 1-\exp \left(-\Omega^{(1)}\left(Y,\frac{4}{3} \log \left(\frac{\ln (r^2_2}{\xi_s}\right)\right)-\Omega^{(1)}\left(Y,\frac{4}{3} \ln \left(\frac{\ln \left(r_1^2\right)}{\xi_s}\right)\right)\right)\Rb\eea
Note, that $ \xi_s=\sqrt{ \frac{8 N_c}{b_0}\,\,\frac{\chi\Lb \gamma_{cr} \Rb}{ 1 \,-\,\gamma_{cr}}\,Y}$ and $r^2_2 = r^2 + r^2_1 - 2  r_1 r \cos\phi$.


\begin{thebibliography}{99} \frenchspacing

\bibitem{HE1}
	J.H. He, 
	Comput. Methods Appl. Mech. Engrg. {\bf 173} (1999) 257.  
	
	\bibitem{HE2}
	J.H. He, 
	 Int. J. Nonlinear Mech. {\bf 35} (2000) 37.
 
 
 
 

\bibitem{BK}
I.~Balitsky,
{Phys.\ Rev.} {\bf D60}, 014020 (1999);
Y.~V.~Kovchegov,
{Phys.\ Rev.}  {\bf D60}, 034008  (1999).


\bibitem{CLMNEW}
C.~Contreras, E.~Levin and R.~Meneses,
Phys. Rev. D \textbf{107} (2023) no.9, 094030
doi:10.1103/PhysRevD.107.094030
[arXiv:2302.10497 [hep-ph]].
\bibitem{KOLE}
  Y.~V.~Kovchegov and E.~Levin,
  Nucl.\ Phys.\ B {\bf 577} (2000) 221.

\bibitem{CGLM}
C.~Contreras, J.~Garrido, E.~Levin and R.~Meneses,
Phys. Rev. {\bf D 110}  (2024)  054045
    \bibitem{BFKL}
   V.~S. Fadin, E.~A. Kuraev and L.~N. Lipatov,
\newblock Phys. Lett. {\bf B60}, 50 (1975);\,\,\,
E.~A. Kuraev, L.~N. Lipatov and V.~S. Fadin,
\newblock Sov. Phys. JETP {\bf 45}, 199 (1977),
\newblock [Zh. Eksp. Teor. Fiz.72,377(1977)];\,\,\,
I.~I. Balitsky and L.~N. Lipatov,
\newblock Sov. J. Nucl. Phys. {\bf 28}, 822 (1978),
\newblock [Yad. Fiz.28,1597(1978)].

\bibitem{RUNAL}
B.~Diaz Saez and E.~Levin,
Nucl. Phys. A \textbf{870-871} (2011), 83-93
doi:10.1016/j.nuclphysa.2011.09.003
[arXiv:1106.6257 [hep-ph]].
\bibitem{RAL}
I.Balitsky, Phys. Rev. {\bf D 75} (2007)  014001 
\bibitem{RAL2}
 Y.V. Kovchegov and H. Weigert, Nucl. Phys. {\bf A 784} (2007), {\bf 789} (2007) 260. 

\bibitem{KOVCUT}
J.~L.~Albacete and Y.~V.~Kovchegov,
Phys. Rev. D \textbf{75}, 125021 (2007)
doi:10.1103/PhysRevD.75.125021
[arXiv:0704.0612 [hep-ph]].

\bibitem{GLR} 
L.~V.~Gribov, E.~M.~Levin and M.~G.~Ryskin,
  Phys.\ Rept.\  {\bf 100}, 1 (1983).
 
    \bibitem{MUT}
A.~H.~Mueller and D.~N.~Triantafyllopoulos,
{\it Nucl.\ Phys.} \, {\bf B640} (2002) 331
[arXiv:hep-ph/0205167];\,\,D.~N.~Triantafyllopoulos,
{\it Nucl.\ Phys.}\,  {\bf B648} (2003) 293
[arXiv:hep-ph/0209121].
    
\bibitem{MUPE}
S.~Munier and R.~B.~Peschanski,
  Phys.\ Rev.\  D {\bf 70} (2004) 077503
  [arXiv:hep-ph/0401215];\,\,
Phys.\ Rev.\  D {\bf 69} (2004) 034008
  [arXiv:hep-ph/0310357];\,\,
  Phys.\ Rev.\ Lett.\  {\bf 91} (2003) 232001
  [arXiv:hep-ph/0309177].
    
       \bibitem{KOLEB}
Yuri V. Kovchegov and Eugene Levin, {\it `` Quantum Chromodynamics at High Energies"}, Cambridge Monographs on Particle Physics, Nuclear Physics and Cosmology, Cambridge University Press, 2012 . 





 \bibitem{IIM}
 E.~Iancu, K.~Itakura, L.~McLerran,
  Nucl.\ Phys.\  {\bf A708 } (2002)  327-352.
  [hep-ph/0203137]   
  
  \bibitem{GOST}
K.~J.~Golec-Biernat and A.~M.~Stasto,
Nucl. Phys. B \textbf{668} (2003), 345-363
[arXiv:hep-ph/0306279 [hep-ph]].
\bibitem{BEST}
J.~Berger and A.~Stasto,
Phys. Rev. D \textbf{83} (2011), 034015
[arXiv:1010.0671 [hep-ph]].


   \bibitem{LIP}
 L.~N.~Lipatov,
  Sov.\ Phys.\ JETP {\bf 63}, 904 (1986)
  [Zh.\ Eksp.\ Teor.\ Fiz.\  {\bf 90}, 1536 (1986)].
 \bibitem{LIREV}
                L.~N.~Lipatov,
  Phys.\ Rept.\  {\bf 286} (1997) 131.
      
 
 


\bibitem{LETU}
E.~Levin and K.~Tuchin,
  Nucl.\ Phys.\ B {\bf 573}, 833 (2000)
  [hep-ph/9908317];\,\,\,
  Nucl.\ Phys.\ A {\bf 691}, 779 (2001)
  [hep-ph/0012167]; 
   {\bf 693}, 787 (2001)
  [hep-ph/0101275].
   \bibitem{CLMS}
C.~Contreras, E.~Levin, R.~Meneses and M.~Sanhueza,
Eur. Phys. J. C \textbf{80}, no.11, 1029 (2020)
doi:10.1140/epjc/s10052-020-08580-w
[arXiv:2007.06214 [hep-ph]].
           
   \bibitem{MURC}
A.H. ~Mueller, Nucl.Phys. {\bf B643} (2002)  501

 \bibitem{RY}
I. Gradstein and I. Ryzhik, {\it `` Table of Integrals, Series, and Products''},
Fifth Edition, Academic Press, London, 1994.	
 	

\bibitem{MATH}
Andrei D. Polyanin and Valentin F. Zaitsev,
{\it `` Handbook of nonlinear Partial Differential Equations"},  Chapman $\&$ Hall/CRC, 2004.
\bibitem{IFRCUT}
S.~Godfrey and N.~Isgur,
Phys. Rev. D \textbf{32}, 189-231 (1985)
doi:10.1103/PhysRevD.32.189


 \bibitem{MV}
L. McLerran and R. Venugopalan, 
Phys. Rev. {\bf D49} (1994) 2233,
Phys. Rev. {\bf D49} (1994), 3352;   
{\bf D50} (1994) 2225;
 {\bf D59} (1999) 09400. 

\bibitem{AAMS}
  Javier L. Albacete, N\'estor  Armesto, N\'estor,Jos\'e Guilherme  Milhano and Carlos A.Salgado,
  Phys. Rev.   {\bf D  80} (2009)  034031, [arXiv:0902.1112 [hep-ph]], and reference therein.

   \end{thebibliography}
\end{document}